\documentclass[]{aa}  

\usepackage{graphicx}
\usepackage{txfonts}
\bibpunct[; ]{(}{)}{;}{a}{}{;}

\usepackage{color}
\usepackage{xcolor}
\usepackage[colorlinks=true,
            linkcolor=blue,
            citecolor=blue,
            urlcolor=blue,
            breaklinks=True]{hyperref}

\newcommand{\eg}{e.g.,~}

\begin{document} 

   \title{Black hole parameter estimation with synthetic Very Long Baseline Interferometry data from the ground and from space}

\author{Freek Roelofs\inst{\ref{inst1},\ref{inst1_1}}\and
Christian M. Fromm\inst{\ref{inst2_0},\ref{inst2}, \ref{inst3}}\and Yosuke Mizuno\inst{\ref{inst4},\ref{inst2}}\and Jordy Davelaar\inst{\ref{inst7},\ref{inst8},\ref{inst1}}\and Michael Janssen\inst{\ref{inst1}}\and Ziri Younsi\inst{\ref{inst5},\ref{inst2}}\and Luciano Rezzolla\inst{\ref{inst2},\ref{inst6}}\and Heino Falcke\inst{\ref{inst1}}}

\institute{
Department of Astrophysics, Institute for Mathematics, Astrophysics and Particle Physics (IMAPP), Radboud University, P.O. Box 9010, 6500 GL Nijmegen, The Netherlands\label{inst1}\and
Now at: Center for Astrophysics $|$ Harvard \& Smithsonian,  60 Garden Street, Cambridge, MA 02138, USA\label{inst1_1} \\ \email{freek.roelofs@cfa.harvard.edu}\and
Black Hole Initiative at Harvard University, 20 Garden Street, Cambridge, MA 02138, USA\label{inst2_0}\and
Institut f\"ur Theoretische Physik, Goethe-Universit\"at Frankfurt, Max-von-Laue-Stra{\ss}e 1, D-60438 Frankfurt am Main, Germany\label{inst2}\and
Max-Planck-Institut f\"{u}r Radioastronomie, Auf dem H\"{u}gel 69, D-53121 Bonn, Germany\label{inst3}\and
Tsung-Dao Lee Institute and School of Physics and Astronomy, Shanghai Jiao Tong University, 800 Dongchuan Road, Shanghai, 200240, People's republic of China\label{inst4}\and
Department of Astronomy and Columbia Astrophysics Laboratory, Columbia University, 550 W 120th St, New York, NY 10027, USA\label{inst7}\and
Center for Computational Astrophysics, Flatiron Institute, 162 Fifth Avenue, New York, NY 10010, USA \label{inst8}\and
Mullard Space Science Laboratory, University College London, Holmbury St. Mary, Dorking, Surrey, RH5 6NT, UK\label{inst5}\and
School of Mathematics, Trinity College, Dublin 2, Ireland\label{inst6}
}  

   \date{\today}

% \abstract{}{}{}{}{} 
% 5 {} token are mandatory
 
  \abstract
   {
  The Event Horizon Telescope (EHT) has imaged the shadow of the supermassive black hole in M87. A library of general relativistic magnetohydrodynamics (GMRHD) models was fit to the observational data, providing constraints on black hole parameters. 
   }
   %  aims heading (mandatory)
   {
We investigate how much better future experiments can realistically constrain these parameters and test theories of gravity.
   }
   {
We generate realistic synthetic 230 GHz data from representative input models taken from a GRMHD image library for M87, using the 2017, 2021, and an expanded EHT array. The synthetic data are run through an automated data reduction pipeline used by the EHT. Additionally, we simulate observations at 230, 557, and 690 GHz with the Event Horizon Imager (EHI) Space VLBI concept. Using one of the EHT parameter estimation pipelines, we fit the GRMHD library images to the synthetic data and investigate how the black hole parameter estimations are affected by different arrays and repeated observations.
   }
   {
Repeated observations play an important role in constraining black hole and accretion parameters as the varying source structure is averaged out. A modest expansion of the EHT already leads to stronger parameter constraints in our simulations. High-frequency observations from space with the EHI rule out all but $\sim$\,15\% of the GRMHD models in our library, strongly constraining the magnetic flux and black hole spin. The 1$\sigma$ constraints on the black hole mass improve by a factor of five with repeated high-frequency space array observations as compared to observations with the current ground array. If the black hole spin, magnetization, and electron temperature distribution can be independently constrained, the shadow size for a given black hole mass can be tested to $\sim\,0.5\%$ with the EHI space array, which allows tests of deviations from general relativity. With such a measurement, high-precision tests of the Kerr metric become within reach from observations of the Galactic Center black hole Sagittarius A*.
   }
   {}

   \titlerunning{Black hole parameter estimation with synthetic VLBI data}
   \maketitle
%
%-------------------------------------------------------------------

\section{Introduction}

\nocite{eht-paperI}
\nocite{eht-paperII}
\nocite{eht-paperIII}
\nocite{eht-paperIV}
\nocite{eht-paperV}
\nocite{eht-paperVI}

The Event Horizon Telescope (EHT) is a mm Very Long Baseline Interferometry (VLBI) array imaging supermassive black holes on event horizon scales \citep{eht-paperII}. In April 2019, the Event Horizon Telescope Collaboration (EHTC) published its first results from 230 GHz observations of the supermassive black hole in M87 conducted in 2017 \citep{eht-paperI,eht-paperII,eht-paperIII,eht-paperIV,eht-paperV,eht-paperVI}. The source was imaged as an asymmetric ring with a diameter of $42\pm 3$ $\mu$as that is brighter in the south than in the north. This ring structure is interpreted as the black hole ``shadow'' \citep{Falcke2000}, which is formed by gravitational lensing of synchrotron photons emitted by the near-horizon accretion flow plasma or relativistic jet. The observed shadow is largely dominated by properties of the black hole itself, but also affected by the astrophysics of the emitting region. Hence, simulations are used to disentangle these effects. The structure is asymmetric due to Doppler boosting of emission from plasma moving towards us on the southern side of the ring, in agreement with the orientation of the jet seen on larger scales for a clockwise rotation of the accretion flow. Through the combination of fitting a library of general relativistic magnetohydrodynamics (GRMHD) simulations and a crescent model to the interferometric data, and fitting a ring to the reconstructed images, a mass measurement of $M=6.5\pm 0.2|_{\mathrm{stat}}\pm 0.7|_{\mathrm{sys}}\times 10^9 M_{\odot}$ could be made, which is consistent with earlier measurements from stellar dynamics in M87 \citep{Gebhardt2011}.

Constraining black hole and accretion parameters is important for several reasons. Accurate measurements of the mass and spin of the black hole would together determine the Kerr metric, describing the spacetime near the black hole in general relativity. In the Kerr metric, the size and shape of the black hole shadow are completely determined by the mass and spin combined with the distance to the source. If an independent mass measurement is made, the accuracy of deriving the black hole mass from VLBI imaging data can be used as a proxy for testing general relativity. Especially high-resolution VLBI observations of the Galactic Center black hole Sagittarius A* (Sgr A*) could provide a unique opportunity to test the Kerr metric \citep[e.g.][]{Psaltis2015}, since its mass is known accurately (at sub-percent level) and independently, from monitoring of stellar orbits. The mass of Sgr A* was measured to be $(4.148\pm0.014)\times 10^6 M_{\odot}$ by \citet{Do2019} and $(3.964 \pm 0.047_{\mathrm{stat}} \pm 0.0264_{\mathrm{sys}}) \times 10^6 M_{\odot}$ by \citet{Gravity2019}. With an expected shadow size of 51 $\mu$as, Sgr A* is the other important target for black hole imaging by the EHT.

Apart from measuring and testing the Kerr metric, high-resolution VLBI observations also provide the opportunity to establish the near-horizon plasma properties and behaviour. Potentially measurable quantities include the plasma magnetization and the electron temperature distribution. The black hole spin is important here as well, as it determines the plasma orbits. The black hole spin and plasma properties play an important role in the launching and collimation of relativistic jets as seen in M87 and potentially in Sgr A*, which is currently not clearly understood. For instance, in the Blandford-Znajek mechanism \citep{Blandford1977} the jet launching energy is extracted from the black hole as its spin twists magnetic field lines, while in the Blandford-Payne mechanism \citep{Blandford1982} the accretion disk is the dominant source of jet launching energy. While the black hole spin, plasma magnetization, and electron temperature distribution could not be determined from the EHT2017 observations, an enhanced EHT array may provide the necessary resolution and fidelity to start constraining these parameters.

Apart from extending the EHT on the ground, several studies have been done into employing space-based antennas for high-resolution observations. The baseline lengths attainable from Earth are limited by its size. Also, a ground-based imaging array at frequencies higher than about 345 GHz is extremely challenging due to strong attenuation and turbulence introduced by water vapor in the troposphere. A space-based array could overcome these limitations and potentially increase the array resolution by an order of magnitude \citep{Roelofs2019, Fish2020}, or provide fast $uv$-coverage suitable for dynamical imaging of variable sources like Sgr A* \citep{Palumbo2019}. For this work, we consider the Event Horizon Imager (EHI) concept \citep{Martin2017, Kudriashov2019, Roelofs2019}, which is a purely space-based interferometer consisting of two or three satellites in Medium Earth Orbits. It is suitable for high-resolution (with a nominal beam size down to about 4 $\mu$as) and high-fidelity imaging at high frequencies up to $\sim690$ GHz \citep[see][for EHI imaging simulations of Sgr A*] {Roelofs2019}.

Synthetic data tools have been developed and used to predict the imaging performance of the EHT \citep[e.g.][]{Fish2010, Falcke2010, Lu2014, Lu2016, Chael2016, Chael2018, Johnson2017, Bouman2018, RoelofsJanssen2020} and Space VLBI arrays \citep{Roelofs2019, Palumbo2019, Fish2020}. However, image comparisons are challenging to quantify except for simple metrics evaluating pixel-by-pixel differences or cross-correlations, which do not always match a quality assessment by eye where a person looks at the reconstruction of certain model features, like a photon ring or extended jet. Furthermore, because of the sparsely sampled $uv$-plane, image reconstruction generally requires additional assumptions in the form of, e.g., regularization by image smoothness or sparsity \citep{eht-paperIV}. In this work, we aim to overcome both these limitations by evaluating the current and future EHT and EHI array performance based on model parameters that are estimated from synthetic visibility data directly. 

There are two main motivations for testing array performance under varying observational circumstances. For the near future, studies like these are required for setting a set of triggering requirements for the EHT. During an EHT campaign, which typically lasts two weeks, a few observing nights can be triggered based on the technical readiness of the telescopes and weather conditions at the different sites. These triggering decisions could be optimized by quantifying the effect of losing certain stations or varying weather conditions on the science output. 

The second motivation for this project and main focus of this paper is for the longer term: by adding new stations to the synthetic observations at different sites, their effect on the science output can be measured directly. Such a procedure can help optimizing the design of the future EHT and other VLBI arrays, and establish how accurately parameters can be measured and ultimately how accurately general relativity and models for black hole accretion and jet launching can be tested.

In this paper, we investigate the potential of current and future EHT and EHI arrays to constrain black hole and accretion parameters. To this end, we employ the GRMHD library model fitting framework of \citet{eht-paperV, eht-paperVI} to fit the black hole mass and spin, electron temperature distribution, plasma magnetization, and sky orientation for a suite of realistic EHT and EHI synthetic datasets. In Section \ref{sec:methods}, we present our model image generation, synthetic data generation, and model fitting framework. In Section \ref{sec:results}, we show how the recovered parameter constraints are affected by the array and observing strategy used. We summarize and provide suggestions for next steps in Section \ref{sec:summary}.

\section{Simulation and parameter estimation methods}
\label{sec:methods}
\subsection{Model image generation}
\label{sec:grmhd}

\subsubsection{The EHT GRMHD model image library}
The data and images from the EHT 2017 campaign were compared to a library of ray-traced general relativistic magnetohydrodynamics (GRMHD) simulations \citep{eht-paperV, eht-paperVI}. In a GRMHD simulation, the plasma dynamics near the event horizon are simulated from a set of initial conditions by evolving the plasma according to the laws of magnetohydrodynamics embedded in, e.g., the Kerr metric as described by general relativity \citep[e.g.][]{Gammie2003, Moscibrodzka2009, Porth2017, Porth2019}. The resulting plasma quantities such as the density, magnetic field, and temperature are then used to calculate synchrotron emissivities and absorptivities, and photons are ray-traced along the geodesics through the simulation domain to a distant observer to form an image at a specific frequency. 

The images in the EHT M87 GRMHD library depend on several parameters. At the GRMHD stage, there is a distinction between Standard And Normal Evolution \citep[SANE,][]{Narayan2012,Sadowski2013} models which have a relatively low magnetic flux crossing a hemisphere of the event horizon, and Magnetically Arrested Disk \citep[MAD,][]{Igumenshchev2003, Narayan2003} models which are characterized by a high magnetic flux. Intermediate magnetic fluxes (INSANE) are possible as well, but these were not included in the analysis by \citet{eht-paperV, eht-paperVI}. The dimensionless black hole spin parameter $a_*$ can be set to values between -1 and 1, where negative and positive value represent a retrograde and prograde spin, respectively, with respect to the rotation of the outer large-scale accretion flow.

At the ray-tracing stage, a choice needs to be made for the ratio of the ion temperature $T_i$, which follows from the GRMHD simulations, and the electron temperature $T_e$, which characterizes the emission. In \citet{eht-paperV}, the model from \citet{Moscibrodzka2016} was adopted, where 
\begin{equation}
    \frac{T_i}{T_e}=R_{\mathrm{high}}\frac{\beta_p^2}{1+\beta_p^2} + \frac{1}{1+\beta_p^2}.
\end{equation}
Here, $\beta_p$ is the gas-to-magnetic pressure ratio, which is large in the disk and small in the strongly magnetized jet. It follows from the GRMHD simulation, so that the free parameter $R_{\mathrm{high}}$ fixes the electron temperature. An increased value of $R_{\mathrm{high}}$ represents a larger ion-to-electron temperature ratio in the disk, so that emission in that region is suppressed and the resulting ray-traced image becomes more jet dominated. A thermal electron temperature parametrization is justified here, as the inclusion of non-thermal particles is mostly relevant for high-energy emission. The appearance of the black hole shadow depends only weakly on the inclusion of non-thermal particles \citep{Davelaar2019}. Other ray tracing parameters that need to be set are the inclination angle of the line of sight with respect to the black hole spin axis, the plasma mass unit, which sets the average total flux density of the model, the black hole mass, and the emission frequency.

\subsubsection{GRMHD images used in this work}
Our ray-traced images, which were used as input for the synthetic data generation and for fitting the synthetic data against, were generated from the \texttt{BHAC} \citep{Porth2017} EHT GRMHD library using the general-relativistic ray-tracing radiative transfer code \texttt{BHOSS} \citep{Younsi2012,Younsi2016,Younsi2020}. We used MAD and SANE models with spins -0.94, -0.5, 0, 0.5, 0.94 (MAD), and 0.97 (SANE). Once the accretion rate had reached a steady state at $t>6000\,GM/c^3$ for SANE models and $t>12000\,GM/c^3$ for MAD models, we computed the emission. Following \citet{eht-paperV}, $R_{\mathrm{high}}$ was set to 1, 10, 20, 40, 80, or 160. During the radiative transfer calculations we excluded regions in the GRMHD simulations with magnetization (ratio of magnetic and kinetic energy density) $\sigma>1$. These regions, mainly found in the funnel, are typically subject to numerical uncertainties which could lead to spurious features in the final images. We use a black hole mass $6.2\times 10^9M_{\odot}$ and a distance of 16.9 Mpc. For the flux density normalization we selected the last GRMHD snapshot of each model and iterated the accretion rate until a flux density of 0.8 Jy at 230 GHz was obtained.
The inclination angle was fixed at 163 degrees, as motivated by the large scale orientation of the M87 jet and the fact that the 230 GHz data favor an image that is brighter in the south\footnote{Due to frame dragging, the plasma in the innermost accretion flow is forced to rotate with the black hole spin, even if the orbits are retrograde further out. Given that the approaching jet of M87 extends towards the northwest, the Doppler-boosted bright region in the south thus implies that the black hole spin axis is pointed away from us.} \citep{Walker2018, eht-paperIV, eht-paperV}. If the inclination angle is not known a priori from other measurements, which is the case for Sgr A*, it should be added as an additional GRMHD parameter.

For each combination of magnetic flux, spin, and $R_{\mathrm{high}}$, 100 frames were generated at a cadence of 10~$GM/c^3$, at frequencies of 230, 557, and 690 GHz. The image field of view was set to 160 $\mu$as, with a pixel size of 1 $\mu$as. At the higher frequencies, the emission is optically thin and the average total flux density decreases from 0.8 Jy at 230 GHz to 0.42 and 0.33 Jy at 557 and 690 GHz, respectively.

\subsection{Synthetic data generation}
\label{sec:synthdata}
\subsubsection{EHT simulations} 
The EHT synthetic data generation for this work was performed with \texttt{SYMBA} \citep{RoelofsJanssen2020}. This pipeline mimics the full VLBI observation process, including realistic observation and calibration effects. Raw frequency-resolved synthetic data is generated following real observation schedules with \texttt{MeqSilhouette} \citep{Blecher2017, Natarajan2019}. Added data corruptions include an atmospheric model which introduces delays, phase turbulence, amplitude attenuation, and sky noise. Furthermore, the effects of antenna pointing offsets due to atmospheric seeing, wind wobbling the dish, and inaccurate pointing solutions are simulated. Fixed antenna gain offsets and polarimetric leakage are added as well. The raw synthetic data is then passed through the VLBI data calibration pipeline \texttt{rPICARD} \citep{Janssen2019}, which is used to calibrate real EHT data as well \citep{eht-paperIII}. The calibrated datasets can then be used for further analysis like image reconstruction or parameter estimation.

The synthetic data generation setup, which includes the locations and properties of the antennas and the local weather conditions during the observations, was set equal to the setup used for the EHT2017 and EHT2020 arrays in \citet{RoelofsJanssen2020}, where the latter was renamed to the EHT2021 array as the 2020 observations were cancelled. The EHT2017 coverage is identical to that of 11 April, where scans that were scheduled but not observed were flagged. For the simulation of atmospheric corruptions, \texttt{SYMBA} takes the ground pressure, ground temperature, precipitable water vapor, and coherence time at each site as input. The input antenna and weather parameters were based on site measurements taken during the EHT2017 campaign or estimated from the EHT2017 data. For new stations that did not participate in 2017, weather parameters were estimated using climatological modelling \citep{Paine2019}, assuming decent observing conditions in April. See \citet{RoelofsJanssen2020} for more details and used values for the different observation parameters. 

\begin{figure}[h]
\centering
\includegraphics[width=0.48\textwidth]{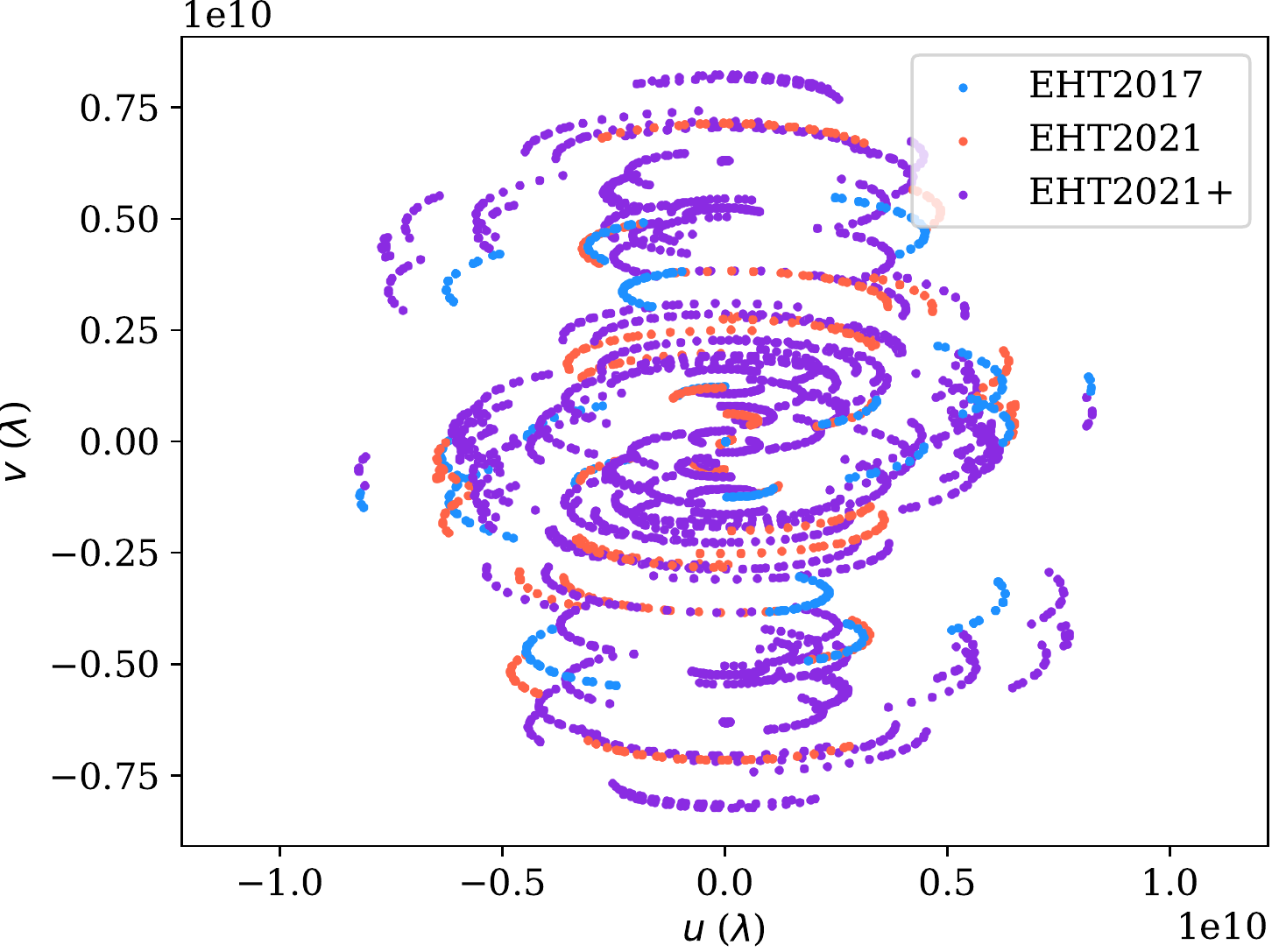}
  \caption{$uv$-coverage for the scan-averaged ground-based simulations with different EHT arrays. The EHT2021 coverage includes the points labeled as EHT2017, and the EHT2021+ coverage includes the points labeled as EHT2021 and EHT2017.}
     \label{fig:uv_ngeht}
\end{figure}

We also simulated an array configuration with six additional stations added to the EHT2021 array. These are the Africa Millimeter Telescope, which is planned to be built on the Gamsberg in Namibia \citep{Backes2016}, Haystack Observatory in Massachusetts, and potential new sites on the summit of the Greenland ice sheet, in La Palma in Spain, in R\'io Negra in Argentina, and on Pikes Peak in Colorado. These sites would be suitable for M87 observations based on the $uv$-coverage they add to the array, taking into account station dropouts due to bad weather conditions in a Monte Carlo analysis by \citet{Raymond2020}. This expanded array is referred to as EHT2021+. There are no plans to build this particular array: it is included in this work as a preliminary example. An effort to investigate the possibilities of a more strongly expanded and enhanced array is ongoing in the context of the next-generation EHT \citep[ngEHT,][]{Doeleman2019}. The $uv$-coverage of the EHT2017 array and the additional EHT2021 and EHT2021+ baselines are shown in Figure \ref{fig:uv_ngeht}.

\begin{figure}[h]
\centering
\includegraphics[width=0.48\textwidth]{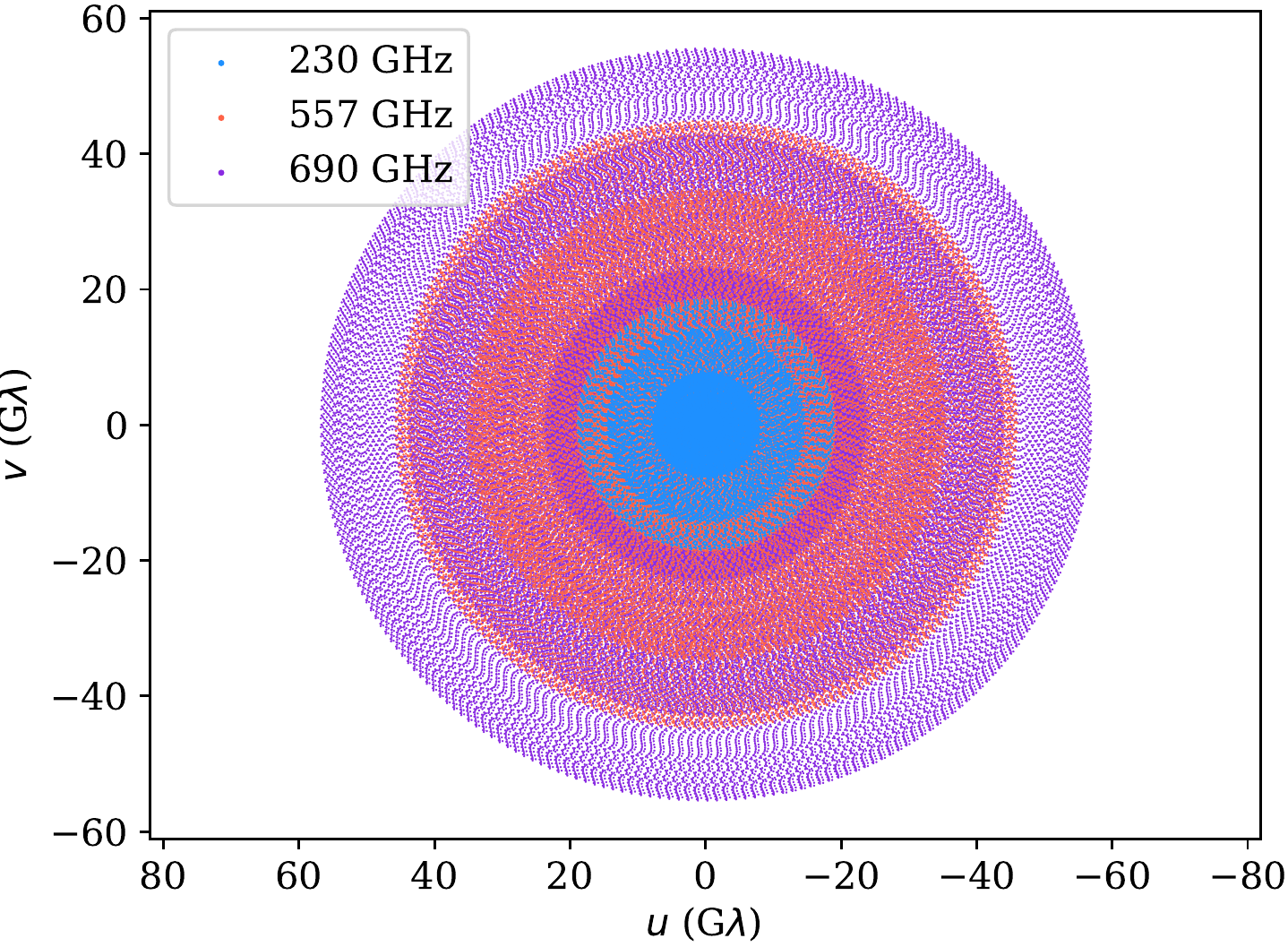}
  \caption{$uv$-coverage for the space-based simulations with the three-satellite EHI at different frequencies. The spacing between the points is set by the $uv$-smearing limit \citep{TMS2017, Roelofs2019}.}
     \label{fig:uv_ehi}
\end{figure}

\subsubsection{EHI simulations}
The EHI Space VLBI mission concept consists of two or three satellites in circular polar orbits around Earth, with radii of $\sim14,000$ km \citep{Martin2017,Kudriashov2019,Roelofs2019}. Because of a small difference in the orbit radii, the inner satellite orbits faster than the outer one, and the baseline between them slowly grows as the satellites move from their initial positions that are aligned with the center of the Earth. This setup then results in dense spiral-shaped $uv$-coverage. For the orbit radius difference of 21 km adopted for the simulations in \citet{Roelofs2019} and in this work, it takes 29 days to reach the longest baseline of 25,000 km. At this distance, the intersatellite laser link, which is required for on-board data correlation and intersatellite distance measurements, is obscured by the Earth's atmosphere. Longer baselines have thus not been included in the simulations, even though the satellites can be further apart in these orbits. For synthetic data generated for this work, we adopted the three-satellite setup with 4-meter antenna diameters and system noise parameters from \citet{Roelofs2019}. The model visibilities were calculated with the \texttt{eht-imaging} library \citep{Chael2016, Chael2018}. The $uv$-coverage of the EHI at 230, 557, and 690 GHz is shown in Figure \ref{fig:uv_ehi}.

\subsection{Parameter estimation}

\subsubsection{The EHT GRMHD library parameter estimation framework}
In \citet{eht-paperV, eht-paperVI}, the GRMHD library was scored against the EHT 2017 data by rotating and scaling (in size and total flux density) each model image frame to provide the best possible fit to the EHT data. The size scaling is of particular importance here, since it sets the angular size of one gravitational radius $GM/Dc^2$. Combined with distance measurements of M87, the size scaling then gives a best-fit black hole mass for each model frame. The scoring was performed and vetted with two independent pipelines: the MCMC parameter estimation framework \texttt{Themis} \citep{Broderick2020}, and the genetic algorithm \texttt{GENA} \citep{Fromm2019}, which is used in this work. 

Due to variability from plasma orbits and turbulence, different GRMHD frames of the same movie may appear significantly different on the scales probed by the EHT. Therefore, synthetic EHT data generated from a frame of that model will not necessarily give a formally acceptable fit ($\chi^2 \approx 1$) to the other movie frames. To account for this model variance, \citet{eht-paperV, eht-paperVI} developed the method of average image scoring (AIS). In this procedure, synthetic EHT data generated from the average image of a GRMHD movie is scored against the individual frames of the same model. The resulting $\chi^2$ distribution is then compared to the $\chi^2$ obtained from fitting the observed data to the average model image. A $p$-value is then computed as
\begin{equation}
\label{eq:ais}
    p_{\mathrm{AIS}}(\mathcal M|\boldsymbol{D}) = \frac{N_{>|\chi^2-\chi^2_{\mathrm{med}}|}}{N}.
\end{equation}
Here, $\mathcal M$ is a GRMHD model, $\boldsymbol{D}$ is the observed data, and $N$ is the number of frames in the model. $N_{>|\chi^2-\chi^2_{\mathrm{med}}|}$ is the number of frames in the model for which the difference between the reduced $\chi^2$ of the frame scored against synthetic data from the average model and the median $\chi^2$ of all frames is larger than the difference between the reduced $\chi^2$ of the average model scored against the observed data and the median $\chi^2$ of the average model. Equation \ref{eq:ais} is effectively computing a two-sided p-value \citep[see for details][]{eht-paperVI}. The model is rejected when $p_{\mathrm{AIS}}<0.01$.

Using this method, MAD models with $a_*=-0.94$ formed the only class of models that could be ruled out based on the EHT2017 data alone. With the addition of other criteria like the X-ray flux and jet power as predicted by the models compared to other observational data, the number of fitting models was reduced significantly. The remaining models, combined with a distance measurement of $D=16.8_{-0.7}^{+0.8}$ Mpc and mass estimates from a crescent model fit to the data and from a ring fit to the reconstructed image, gave the measured mass of $M=6.5\pm 0.2|_{\mathrm{stat}}\pm 0.7|_{\mathrm{sys}}\times 10^9 M_{\odot}$.

\subsubsection{GENA pipeline description}
In the following we provide a short description of the mathematical and numerical methods used within \texttt{GENA}, which is the genetic algorithm pipeline we used for the fitting of the GRMHD model library to our input synthetic data.
\paragraph{Constrained non-linear optimization:}
The extraction of physical parameters from observational 
data via numerical, semi-analytical or geometrical models can be considered as a constrained non-linear optimization problem. Its mathematical formulation is given by:

\begin{eqnarray}
\begin{array}{ll@{}ll}
\mathrm{minimize}  & \displaystyle  f(\vec{x}) &\\
\mathrm{subject\, to}& \displaystyle g_{j}(\vec{x})\leq 0, &   &j=1 ,..., n,\\
                 &  \displaystyle        x_{L,i}\leq x_{i}\leq x_{R,i},  & &i=1 ,..., m,
\end{array}
\end{eqnarray}

where $\vec{x}$ is an $m$-dimensional vector including the model parameters, $f(\vec{x})$ is the cost function (also objective or minimisation function), $g_j(\vec{x})$ are the 
constraints, and
$x_{L,i}$ and $x_{R,i}$ are lower and upper boundaries for the model parameters.

The cost function and the constraints should be constructed in such a way that key 
properties of the data and prior knowledge of the source, such as the mass of the black hole, are used to guide and speed up the 
the convergence of the optimization process.

\paragraph{Optimization algorithm:}
The above stated optimization problem can be solved by several kinds of algorithms: gradient-based, gradient-free or MCMC algorithms \citep[see \texttt{Themis},][]{Broderick2020}. Given the high computational effort in scoring of the GRRT images, which includes Fourier transforms and gain optimizations, as well as the fast convergence of gradient-free in contrast to gradient-based algorithms (a gradient-based algorithm requires a lot of computational resources for mapping out the gradient with sufficient resolution), we decided to apply a gradient-free search algorithm in the form of a population-based evolutionary algorithm. The basic idea and steps of an evolutionary algorithm (EA) are described in the next paragraph. 

\paragraph{Evolutionary algorithm (EA):}
EAs are motivated by biological evolution and follow the principle of survival of the fittest. The main steps of EAs are selection, crossover (mating), and mutation of the individuals for several generations. An individual can be seen as set of parameters, in our case the orientation of the image in the sky, $\phi$, the flux scaling, $f$ and the mass of the black hole, $M$. $\vec{x}$ can thus be written as $\vec{x}=\left[\phi,f,M\right]^{T}$. Each entry, \eg $\phi$, is typically labelled as a gene. During the initial step, $N$ random individuals are created in such a way that they cover the parameter space (\eg using the latin hypercube sampling), and their fitness is computed using $f\left(\vec{x}\right)$. 

For the creation of the next generation, offsprings are created from the current population, including random mutations to ensure diversity in the population so that the algorithm does not get stuck in local minima. The details of this step depend on the implementation of the EA. In this work, we use the Non Sorting Genetic Algorithm II (\texttt{NSGA2}) \citep{Deb:2002} and the differential evolution algorithm following \citet{Storn1997}. In this algorithm, one member of the population, $\vec{x_1}$, is selected together with three additional randomly selected population members labelled $\vec{x_{\mathrm{r}1}}$, $\vec{x_{\mathrm{r}2}}$, and $\vec{x_{\mathrm{r}3}}$. From these three random members, a new member is created, which is referred to as the mutated member $\vec{x_{\rm mut}}=\vec{x_{\mathrm{r}1}}+ F(\vec{x_{\rm r2}}-\vec{x_{\rm r3}})$, where $F$ is a constant factor in the interval [0,2]. A new trial population member $\vec{x_{\rm trial}}$ is then created from crossover between $\vec{x_1}$ and $\vec{x_{\rm mut}}$. During the crossover, a random number in the interval [0,1] is drawn for each component ($\phi$, $f$ and $M$). If this number is larger than the crossover probability, the component from $\vec{x_{\rm mut}}$ is copied over to $\vec{x_{\rm trial}}$. If not, the component from $\vec{x_1}$ is taken as new component for $\vec{x_{\rm trial}}$. After the crossover step,  $\vec{x_{\rm trial}}$ thus includes components from the initial member  $\vec{x_1}$ and from the mutated member $\vec{x_{\rm mut}}$. In the last step, $\vec{x_{\rm trial}}$ is evaluated with respect to the cost function $f(\vec{x})$. If $f(\vec{x_{\rm trial}}) < f(\vec{x_1})$, $\vec{x_1}$ is replaced by $\vec{x_{\rm trial}}$. If not, $\vec{x_1}$ is kept in the next generation. This procedure is repeated for each member of the population so that finally a new generation is created.

New generations are created up to a maximum number of iterations or until a convergence criterion is reached, i.e. when the standard deviation $\sigma_f$ of $f(\vec{x})$ across the population is smaller than a specified fraction of the mean $\overline{f(\vec{x})}$. The final fitted parameters are then the components of the $\vec{x}$ for which $f(\vec{x})$ is smallest. In this work, we use a population size of 20, a crossover probability of 0.6, we vary $F$ randomly between different generations in the interval [0.7, 1] (a process called dithering, which speeds up the convergence), and the process is stopped after 100 generations or until $\sigma_f<0.02\overline{f(\vec{x})}$.

\paragraph{Application to VLBI observations:}
Here we apply the concept of numerical optimization via EAs to the fitting of radio astronomical observations obtained via VLBI. Interferometers sample the brightness distribution $I(l,m)$ of an astronomical object in Fourier-space via its projected baseline between antenna $i$ and antenna $j$. The observed visibilities $V_{ij}$ can be written as
\begin{equation}
V_{ij}(l,m) =\int\int dl dm I(l,m) e^{-2\pi \imath (ul +vm)},
\end{equation}
where $l$ and $m$ are angular coordinates on the sky, and $u$ and $v$ are the projected baseline components measured in wavelengths, which act like spatial frequencies. The complex visibility can be converted into the visibility amplitude $\left|V_{ij}\right|$ and the visibility phase $\Phi_{ij}$. The latter is severely affected by atmospheric fluctuations. The number of complex visibility data points of an interferometer is given by $N(N-1)/2$ where N is the total number of antennas forming the interferometer, assuming that they are all able to observe the source at the same time.

A secondary observable quantity is the so-called closure phase $\Phi_{ijk}$ computed as the sum of the visibility phases on a closed triangle of baselines \citep{Jennison1958, Rogers1974}. The advantage of closure phases is that station-dependent phase variations cancel out. Closure phases are thus of great importance to measure the structure of the observed source. The total number of independent closure phases computed from an interferometer consisting of $N$ individual antennas is given by $(N-1)(N-2)/2$ \citep{TMS2017}. If prior information about the station gain phases is available, the dependence on closure phases comes with a loss of information as fewer data are available. If no prior information about the gain phases is available, as is usually the case for our observations, the information content of the visibility phases and closure phases is the same \citep{Blackburn2020}, except if there are times during which the source can only be detected on a single baseline.

Throughout this paper, we will use the visibility amplitude $\left|V_{ij}\right|$ and the closure phase $\Phi_{ijk}$ to fit models to VLBI observations. The visibility amplitude may still be corrupted by, e.g., gain offsets due to antenna mispointings that cannot be calibrated a priori. However, these gains are usually limited to $\sim$ a few tens of percents, while the visibility phases are randomized due to the atmospheric corruptions. Note that there are additional quantities which could be computed from the complex visibilities, such as the closure amplitudes \citep[\eg][]{TMS2017}, which are also free of station-based gain errors. For the minimization function, we use the least squares computed from the visibility amplitudes, $\chi^2_{\rm{VA}}$, and the closure phase, $\chi^2_{\rm{CP}}$, \citep[we follow the convention given in ][for the calculation of the $\chi^2$]{Chael2018}. Finally, the cost function for the optimization process can be written as: 
\begin{equation}
f\left(\vec{x}\right)=\chi^2_{\rm{VA}}+w\chi^2_{\rm{CP}},
\label{minfunc}
\end{equation}
where $w$ is a weighting factor. The information on the structure of the source is mostly encoded in the closure phases. Moreover, these quantities are free of gain corruptions and therefore more reliable than visibility amplitudes. Using a weighting factor $w<1$ enforces the optimization to prioritize the reduction of the $\chi^2_{\rm CP}$. This setup typically increases the convergence of the optimisation procedure.

\paragraph{Station-dependent gain factors:}
Besides the free parameters of our model ($\phi$, $f$, and $M$), the antenna gains $g_i$ can be considered as free parameters \citep[see][]{TMS2017}. Within \texttt{GENA} we use a numerical minimization as implemented in \texttt{eht-imaging} to obtain the antenna gains following \citet{Chael2018} and \texttt{Themis} \citep{Broderick2020} in a self-calibration procedure minimizing
\begin{equation}
\sum_{\rm time}\sum_{i\neq j}\left(V_{ij,\,\mathrm{obs}}-g_ig_j^\star V_{ij,\,\mathrm{model}}\right)^2.
\label{gains}
\end{equation}
During the minimisation of Eq. \ref{gains}, we assume that the gains are constant during a scan, which speeds up the calculations. To avoid compensation of large $\chi^2 $ on the visibility amplitude and/or closure phase by strong gain variations, we include a flat prior on the individual antenna gains $g_{i}$, disfavoring station gain solutions larger than 0.2 (see below).

\subsubsection{Changes made to GENA}
For this work, several changes were made to the \texttt{GENA} pipeline as compared to the version used in \citet{eht-paperV, eht-paperVI}. First, self calibration, which solves for the station gains due to, e.g., unknown pointing offsets, has been included in the process of optimizing the rotation and scaling for each frame. Previously, self calibration was only performed for the best fitting rotation and scaling parameters at the end of the optimization process to determine the eventual goodness-of-fit. Because the station gains can be significant, they strongly increase the $\chi^2_{\mathrm{VA}}$ of non self-calibrated data. In the optimization process, the weight $w$ of the $\chi^2_{\mathrm{VA}}$ therefore needed to be set low ($\sim 0.01$) with respect to the weight of the $\chi^2_{\mathrm{CP}}$ for the optimization to converge. 

However, the visibility amplitudes contain valuable information on the source morphology that can be included in the optimization process. We found that including self calibration in the optimization process and setting $w$ to 0.1 resulted in a better goodness-of-fit and more narrow (by $\sim$ 10\%) mass distributions in the end. A $w$ larger than 0.1 gave a small but significant increase in the best-fit $\chi^2_{\mathrm{CP}}$. Since closure phase remains the most reliable data product because of its robustness against station gains, we adopted $w=0.1$, which is still significantly larger than the $w=0.01$ that needed to be set previously, when self calibration was not included in the optimization process. In order to prevent the self calibration from overfitting the visibility amplitudes, station gain offset solutions larger than 0.2 were disfavored. Due to low S/N decoherence and randomized pointing offsets, the gain offsets may exceed this value in some cases despite the fact that the input station gain offsets did not exceed 0.1. Including self calibration in the optimization process increases the runtime of \texttt{GENA} by a factor $\sim$ 10-20. In order to mitigate this effect, we performed self calibration only when the closure phase fit was reasonable, i.e. $\chi^2_{\mathrm{CP}}<10$.

The second change in the \texttt{GENA} pipeline is the treatment of stations with intra-site baselines. For the current EHT, these are ALMA and APEX in Chile and JCMT and SMA in Hawaii. If the total flux density of the source is known, gains of these stations can be solved for by fixing the visibility amplitudes on the intra-site baselines and comparing the amplitudes on the baselines to a third station \citep[network calibration,][]{Blackburn2019}, and no further self calibration is required. In the EHT2017 model fitting, these stations were included in the self calibration nevertheless because the total compact flux density was uncertain due to a lack of short baselines, and the contribution of the large-scale (jet) structure to the intra-baseline amplitudes was not strongly constrained. Since there is no large-scale jet component in the GRMHD library from which we take our input models, and the input total flux density is known, our input synthetic data is network-calibrated using the implementation in \texttt{eht-imaging} \citep{Chael2016, Chael2018, Blackburn2019}, and the gains for ALMA, APEX, JCMT, and SMA are fixed throughout the fitting procedure.

\section{Parameter estimation results}
\label{sec:results}
In this section, we show our scoring results for a range of observing conditions. We start by investigating the best fit parameter distributions as a function of stations included in the past, current, and potential future EHT array. We also include results from simulating Space VLBI observations of the Event Horizon Imager (EHI), and investigate the effect of repeated observations of the same source with a fixed array.

As input model for our synthetic observations, we use the 20th frame of the SANE model with $a_*=0.5$ and $R_{\mathrm{high}}=80$, shown in Figure \ref{fig:models}. Fits to the EHT2017 data were not considered in the choice for this particular model, but it was chosen because its appearance is fairly representative for most library frames in that it has no strong jet footprint or emission far from the horizon. No specific preference is involved with choosing the 20th frame as the input model for our simulated ground truth data. The typical correlation time between GRRT images created from BHAC data is around 50M. Thus, images separated by 5 frames can be regarded as uncorrelated and independent realizations of the variability. The model was observed with \texttt{SYMBA} and the EHI simulator using the setups described in Section \ref{sec:synthdata}.

\subsection{General trends}
\label{sec:gentrends}
\begin{figure*}[ht!]
\centering
\includegraphics[width=0.43\textwidth]{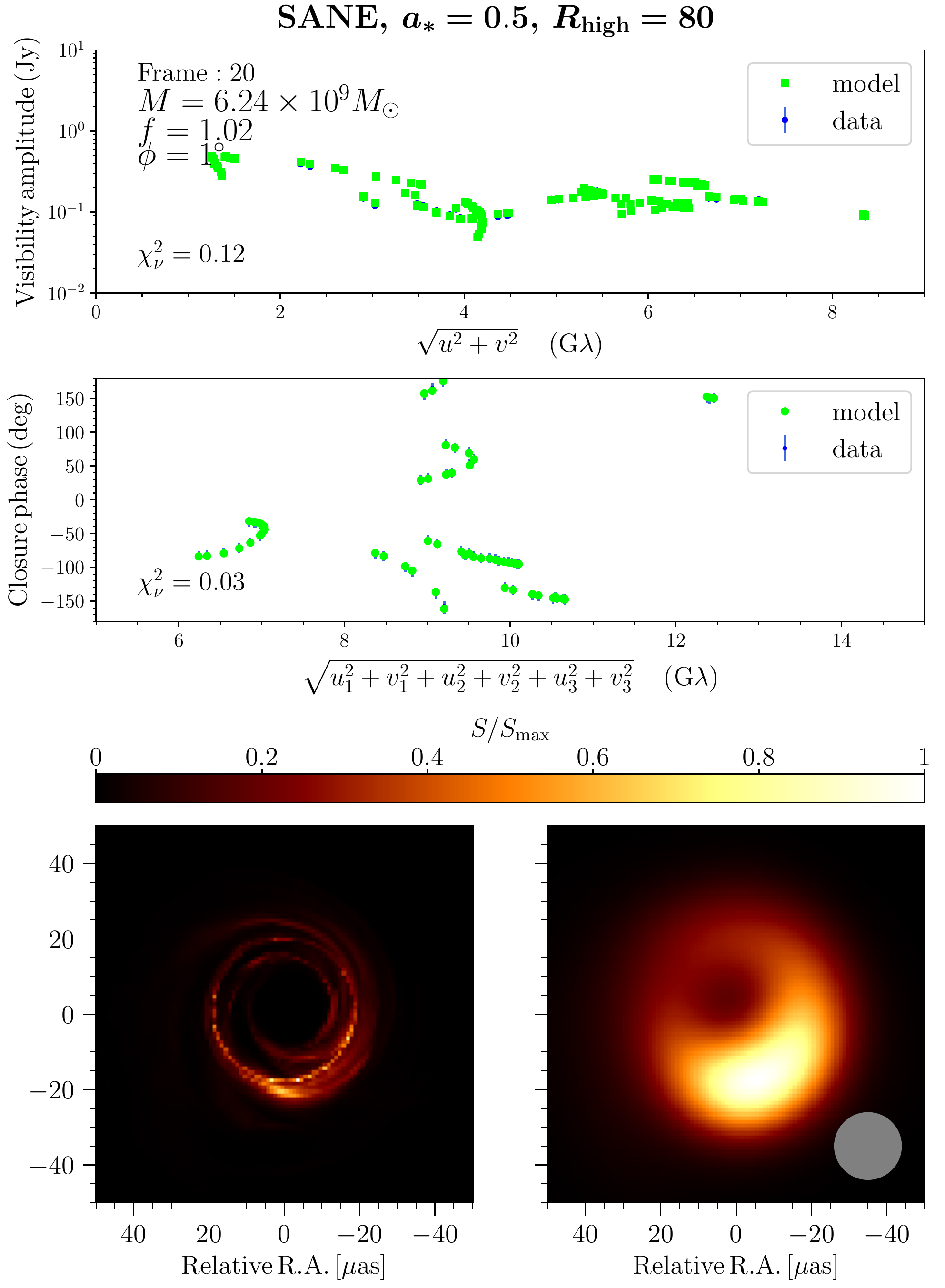}
\includegraphics[width=0.43\textwidth]{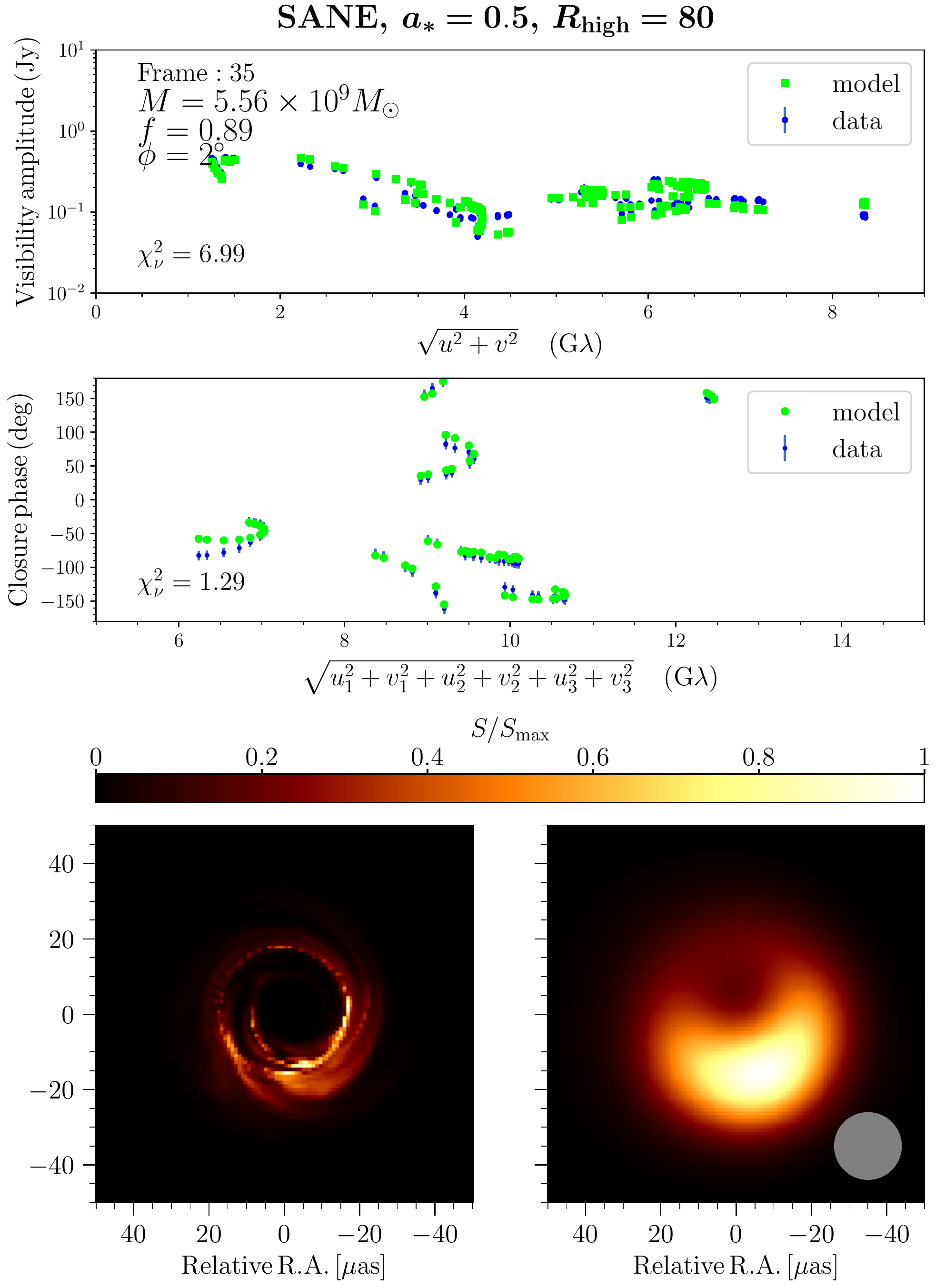}\\
\includegraphics[width=0.43\textwidth]{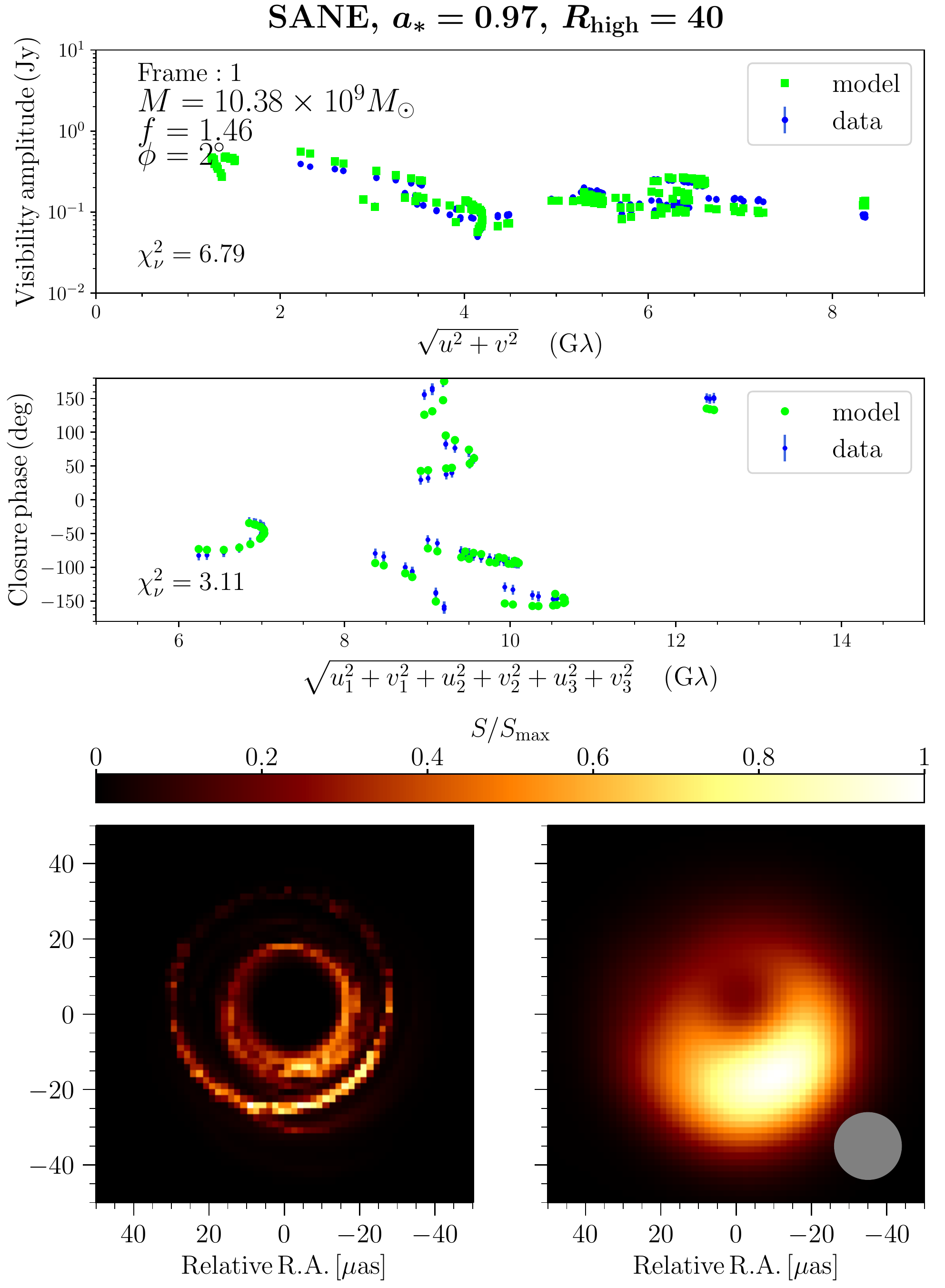}
\includegraphics[width=0.43\textwidth]{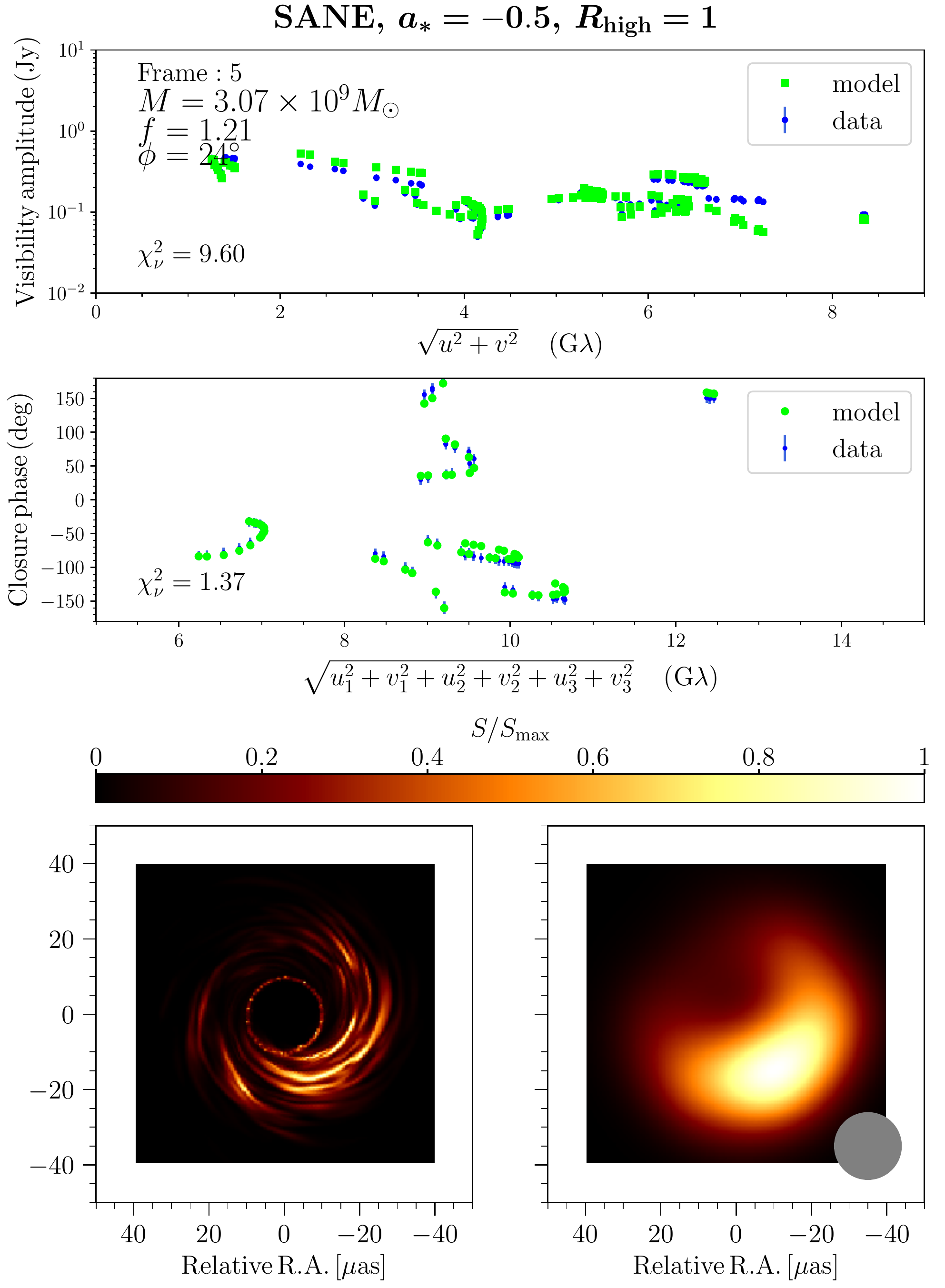}
  \caption{Four example fits to visibility amplitudes (upper panels) and closure phases (middle panels) of the EHT2017 synthetic data generated from frame 20 of the SANE model with $a_*=0.5$ and $R_{\mathrm{high}}=80$. The lower panels show the model frames that were scaled and rotated by the best-fit values indicated in the upper panels, without (left) and with (right) blurring by a 20 $\mu$as Gaussian beam.
  }
     \label{fig:models}
\end{figure*}

Figure \ref{fig:models} shows the visibility amplitude and closure phase fits for four different library frames that were fit to the EHT2017 synthetic data. The best fit parameters (mass, amplitude, sky orientation) for the input model frame (top left) are recovered at the $\sim$percent level. 
As illustrated by the fit to a different frame from the same model (top right), the best fit parameters and goodness-of-fit can indeed vary considerably due to the variability within the model (model variance).

\begin{figure*}[h]
\centering
\includegraphics[width=\textwidth]{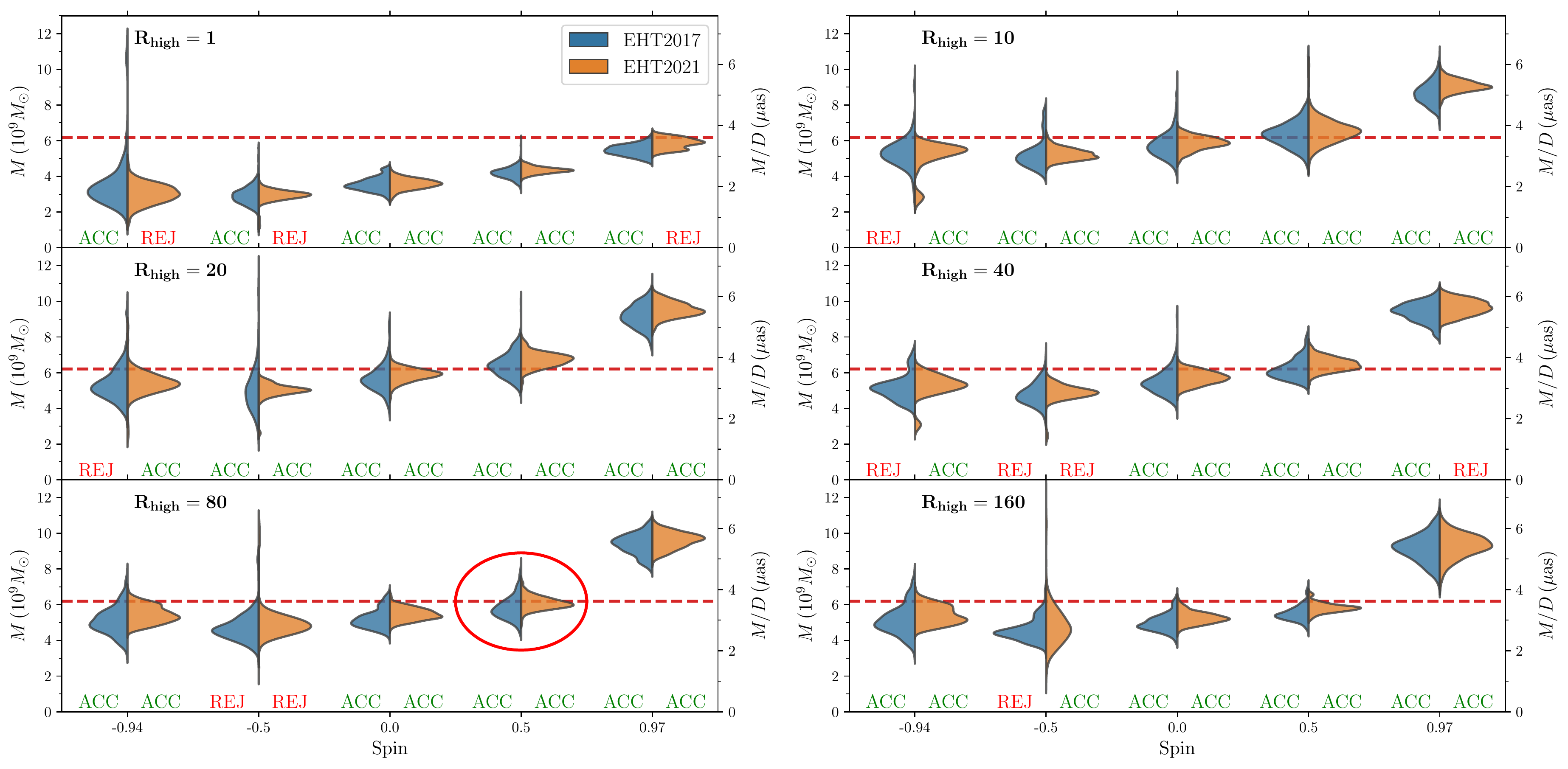}
  \caption{Mass distributions with best-fit values of all SANE model frames fit to the EHT2017 (blue) and EHT2021 (orange) synthetic data generated from frame 20 of the SANE model with $a_*=0.5$ and $R_{\mathrm{high}}=80$. The six panels correspond to the six values of $R_{\mathrm{high}}$ in the model library, and the black hole spin is on the horizontal axis. The red dashed line indicates the true mass of the input model (encircled). The models corresponding to the distributions with annotation ``ACC'' and ``REJ'' were accepted and rejected, respectively, by the average image scoring procedure.}
     \label{fig:sanemassdist}
\end{figure*}

\begin{figure*}[h]
\centering
\includegraphics[width=\textwidth]{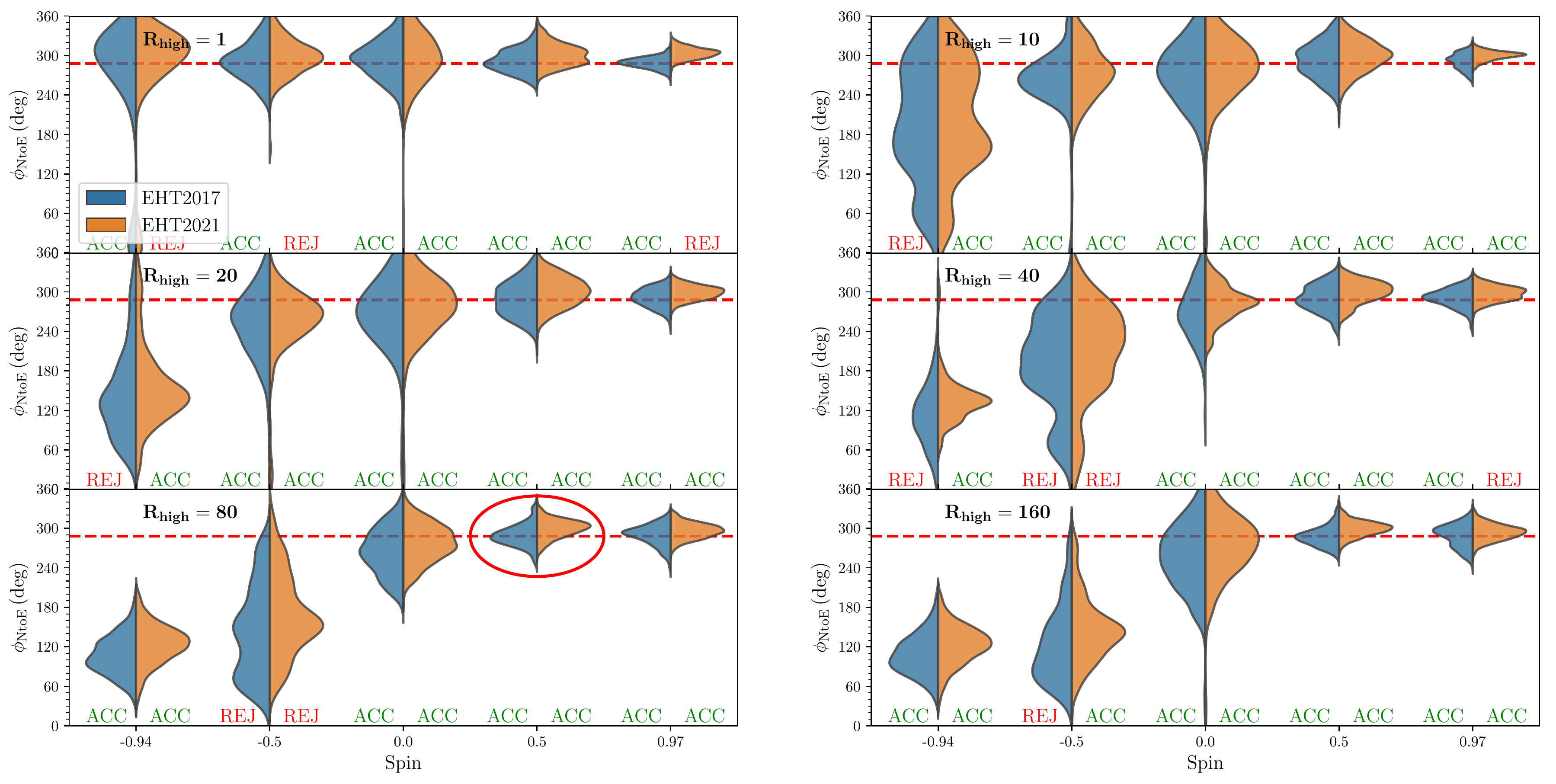}
  \caption{Same as Fig. \ref{fig:sanemassdist}, but for the best-fit sky orientations.}
     \label{fig:sanephidist}
\end{figure*}

Larger variations in mass estimates may occur where the image is not dominated by flux on or close to the photon ring. The bottom row of Figure \ref{fig:models} shows two examples. On the left is a model with strong emission from the jet footprint, which appears as a small ring in front of the shadow. To optimally fit the EHT data, the model was scaled up in size, resulting in a large mass estimate of $10.38\times10^9M_{\odot}$. On the right is a model dominated by emission further away from the photon ring, which was scaled down in size to fit the data, resulting in a small mass estimate of $3.07\times10^9M_{\odot}$. 

Figure \ref{fig:sanemassdist} shows the best fit masses for all SANE models for synthetic data generated with the EHT2017 (left, blue distributions) and EHT2021 (right, orange distributions) arrays. All model frames have been included in the distributions, i.e. no selection was made based on the goodness-of-fit. The plotted curves were generated with a kernel density estimator. The annotation below each distribution indicates whether the model was accepted of rejected based on the average image scoring procedure. See Appendix \ref{app:mass_scaling} for an analysis and discussion on the validity of the mass scaling in the context of this work.

The estimated mass is generally lower than the true value for models with low $R_{\mathrm{high}}$ and/or retrograde spin. The models with $R_{\mathrm{high}}=1$ are disk-dominated, resulting in emission further away from the horizon as the observer is looking down the jet and viewing the disk face-on (see also Fig. \ref{fig:models}, bottom right panel). A retrograde spin pushes the orbits further out, resulting in lower mass estimates. The models with high prograde spin ($a_*=0.97$) and $R_{\mathrm{high}}\geq10$ show a strong jet footprint in front of the shadow (see also Fig. \ref{fig:models}, bottom left panel), and result in a large mass estimate. Similar trends were observed in the comparison of the real EHT2017 data to the GRMHD image library \citep{eht-paperV}.

The inclusion of the GLT, KP, and PDB in the EHT2021 array generally narrows the individual mass distributions. Some models with mass estimates that are substantially lower (most $R_{\mathrm{high}}=1$ models) or higher (the $R_{\mathrm{high}}=40$, $a_*=0.97$ model) were accepted when observed with the EHT2017 array but rejected when observed with the EHT2021 array, indicating an increased ability to constrain the black hole mass when more stations are added to the array. 

A few models with retrograde spin are rejected with the EHT2017 array but accepted with the EHT2021 array (the $a_*=-0.94$ models with $R_{\mathrm{high}}=10-40$, and the $a_*=-0.5$ model with $R_{\mathrm{high}}=160$), which is counter-intuitive.  
This effect can be understood from the average image scoring procedure and investigating the $\chi^2$ distributions obtained for these cases. For example, the median $\chi^2_{\mathrm{CP}}$ from fitting the model frames to synthetic data from the average frame of the same model increases from 5.5 to 15 for the SANE, $a_*=-0.94$, $R_{\mathrm{high}}=20$ model between EHT2017 and EHT2021, and the standard deviation increases from to 3.9 to 8.3. The overall fit quality thus decreases and the spread across the frames increases as more baselines are added, which is indeed plausible if there are large variations between the different model frames. The additional and longer baselines, which allow to see more detail in the source structure, make it more difficult for \texttt{GENA} to fit the varying substructure to data obtained from the average model, where the detailed structure has been washed out. In other words, the additional baselines of the EHT2021 array make the model variance more apparent. In the average image scoring procedure, the acceptance of a model is determined by the goodness-of-fit of the average model image to the input synthetic data (which is generated from a frame of a different model) as compared to the goodness-of-fit of the model frames to synthetic data generated from the average image of the same model. The $\chi^2_{\mathrm{CP}}$ for the former decreases from 32 to 29 between EHT2017 and EHT2021 for the SANE, $a_*=-0.94$, $R_{\mathrm{high}}=20$ model, where the input synthetic data was generated from our fixed SANE, $a_*=-0.94$, $R_{\mathrm{high}}=80$, frame 20 input model. So, the average of the model can be fit slightly better to the input synthetic data when more baselines are added (although the overall fit quality remains poor), while the $\chi^2$ variations become larger within the model due to its strong variability, and some $\chi^2$-values now exceed that for the average model image fit to the input synthetic data. The rejection criterion was thus reached for EHT2017, but not for EHT2021.  
This phenomenon is thus a consequence of not setting an absolute $\chi^2$ criterion for the acceptance of a model, but comparing the fit to the model variance. It is not expected to occur for models which do not show strong variability and fit the input synthetic data well, as the additional baselines will then allow the fits to improve. It should not occur for the true input model as the input synthetic data is then from a sample of the model variance. Once rejected for a particular array, it should therefore in principle be safe to consider the model rejected for other arrays as well, provided that the model variance is well sampled by the model frames.

Apart from the mass, the sky orientation $\phi$ is optimized for each frame in the fitting procedure as well. Figure \ref{fig:sanephidist} shows the recovered distribution in $\phi$ for all SANE models. Just like the mass, these distributions show a general trend with black hole spin. The sky orientation is near the true value for models that, like the input model from which synthetic data was generated, have a prograde spin. However, the preferred orientation flips when the spin changes direction. The plasma is forced to rotate with the black hole spin due to frame dragging, and gets boosted in the opposite direction when the spin changes sign, hence changing the image asymmetry. The width of the distributions is generally smaller for models with a high prograde spin, which appear most asymmetric. For the more jet-dominated models ($R_{\mathrm{high}}\geq 10$), many retrograde models show a particularly wide distribution, most of which are rejected by the average image scoring procedure. These models indeed appear symmetric, with most emission outside the black hole shadow. For jet-dominated models ($R_{\mathrm{high}}=40-160$) with strongly retrograde spins ($a_*=-0.94$), Doppler boosting becomes significant and the images appear asymmetric, which can be made consistent with the input model synthetic data by rotating the image by $\sim 180^{\circ}$.

For the MAD models (not plotted), the results are different in that the optimal mass is less dependent on spin and usually lower than the true mass as the emission is mostly outside the black hole shadow. The trends in $\phi$ are similar to those of the SANE models.

\begin{figure*}[ht]
\centering
\includegraphics[width=.48\textwidth]{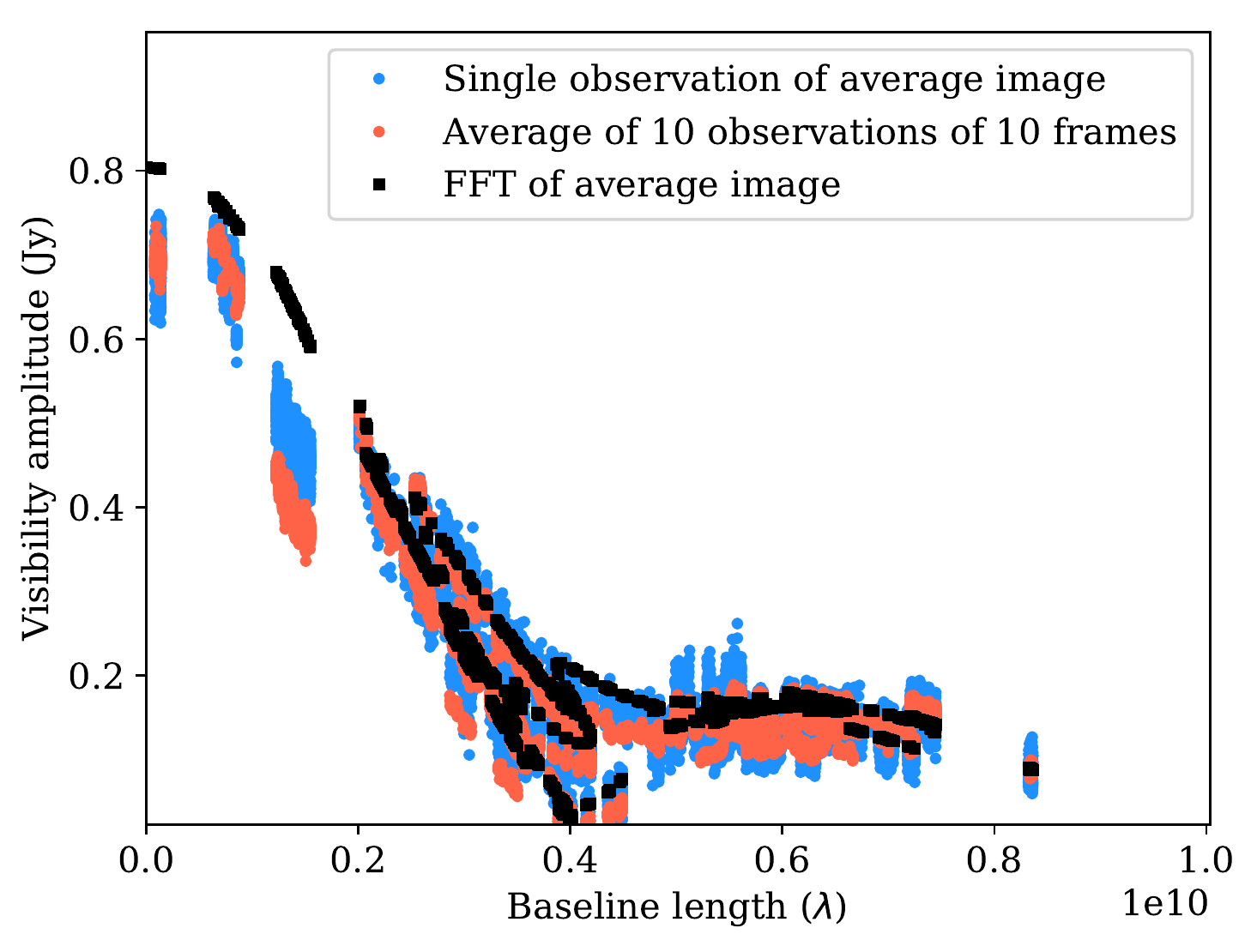}
\includegraphics[width=.48\textwidth]{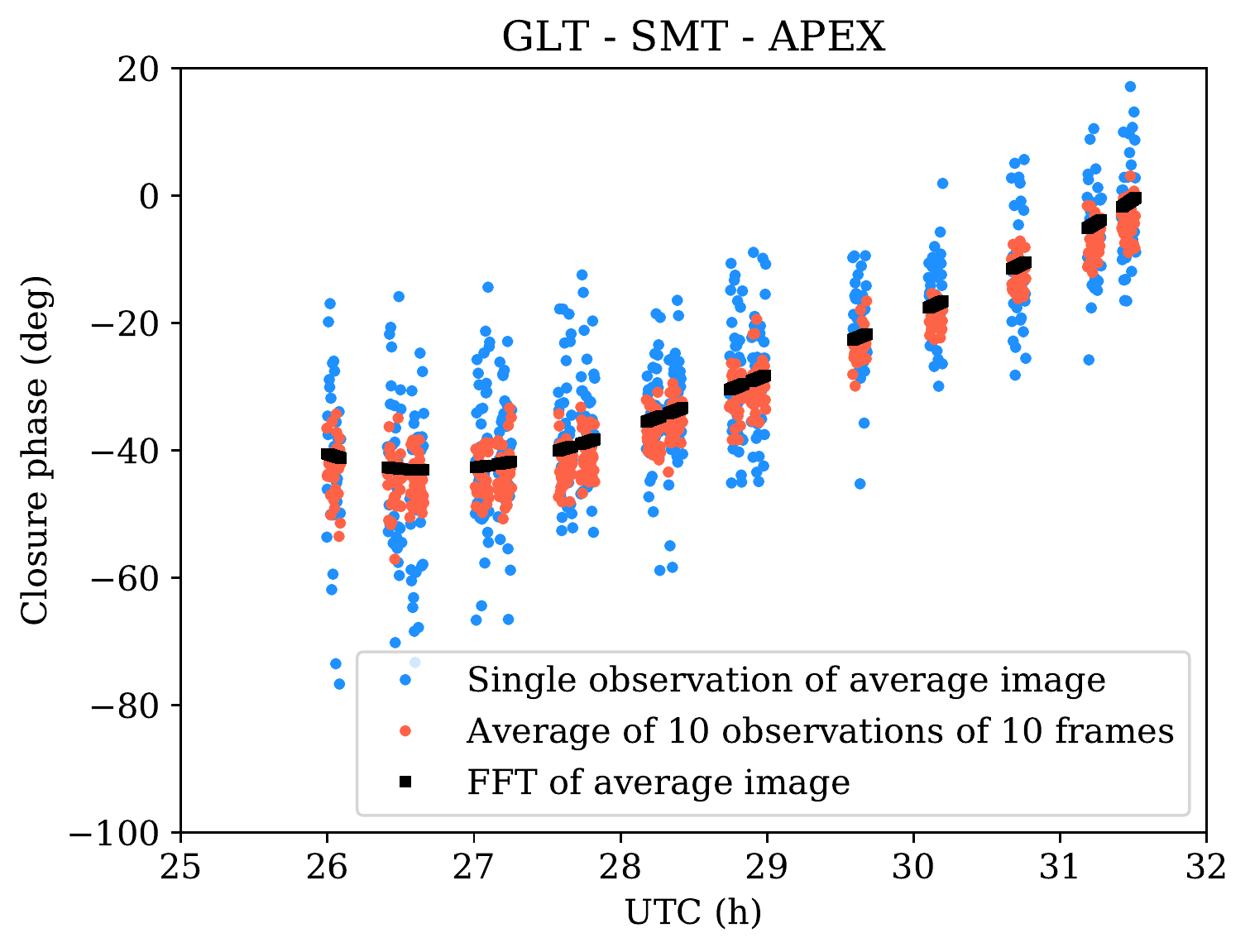} \\
\includegraphics[width=.48\textwidth]{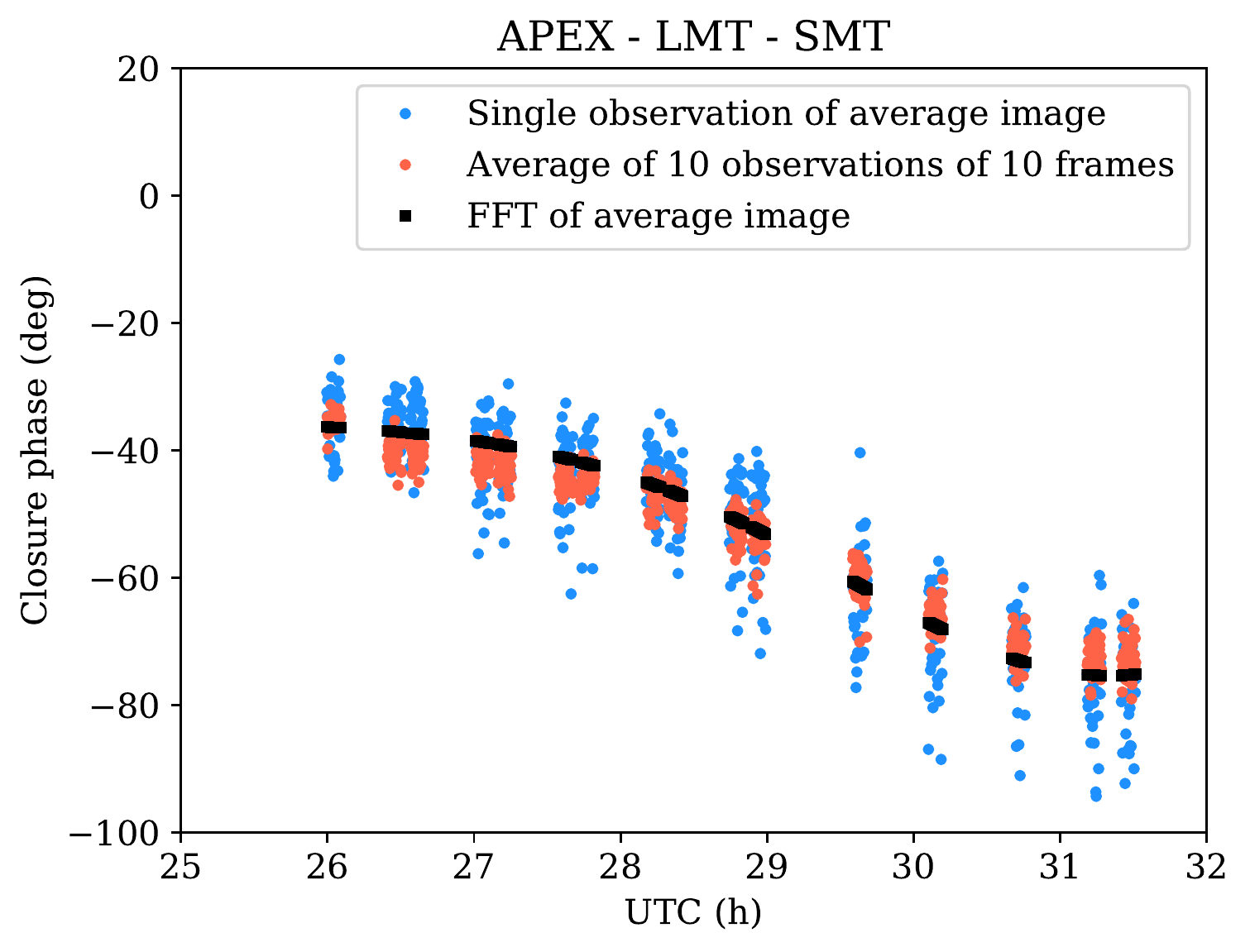}
\includegraphics[width=.48\textwidth]{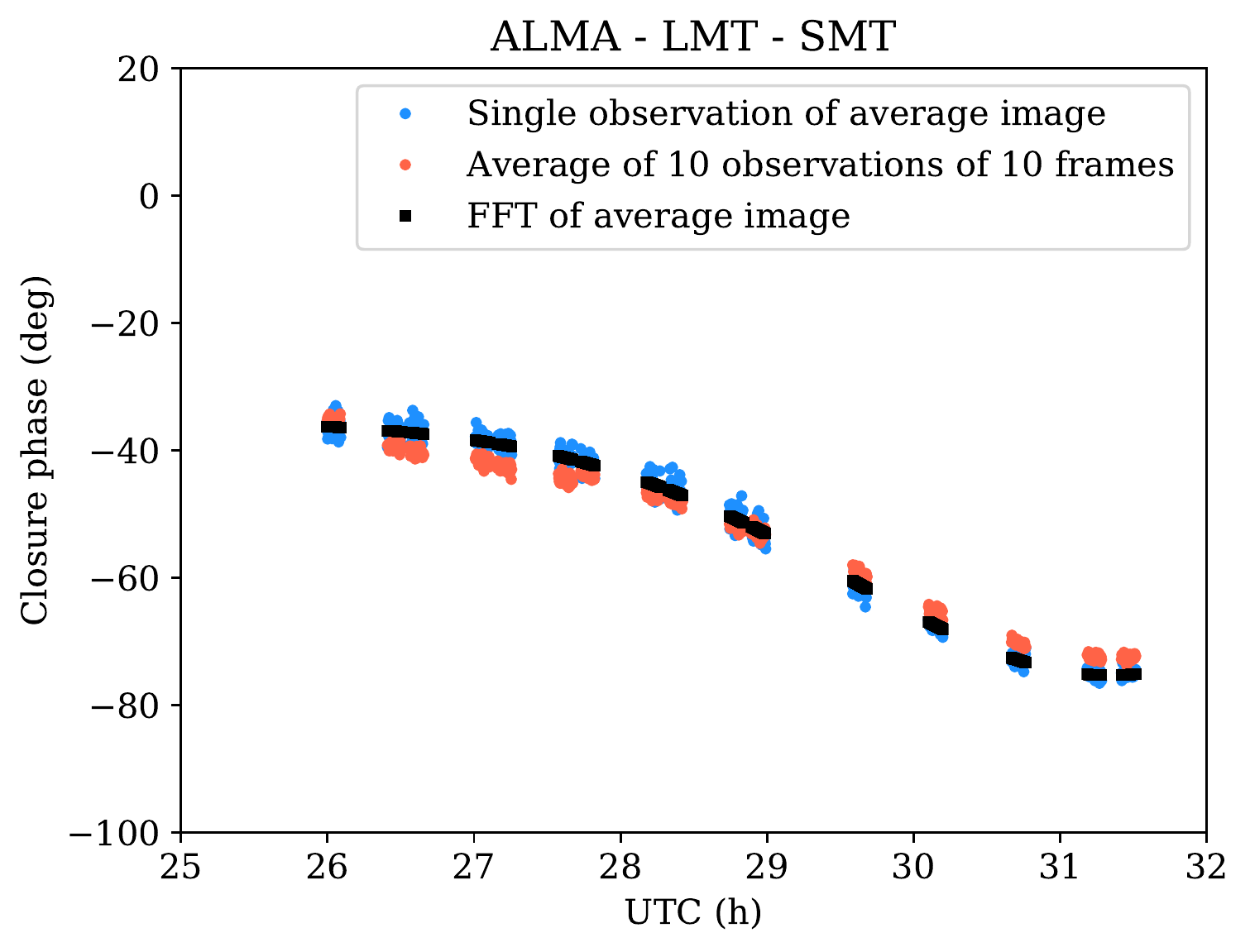}
  \caption{Visibility data from a single observation of an average image of ten randomly selected frames of the SANE, $a_*=0.5$, $R_{\mathrm{high}}=80$ model (blue), and from an average of ten observations of these ten frames (orange), compared to the FFT of the average image (black). The top left panel shows visibility amplitudes as a function of baseline length. The top right, bottom left, and bottom right panels show closure phases as a function of time on the GLT-SMT-APEX, APEX-LMT-SMT, and ALMA-LMT-SMT triangles, respectively.}
     \label{fig:averaging}
\end{figure*}

\begin{figure*}[ht]
\centering
\includegraphics[width=.16\textwidth]{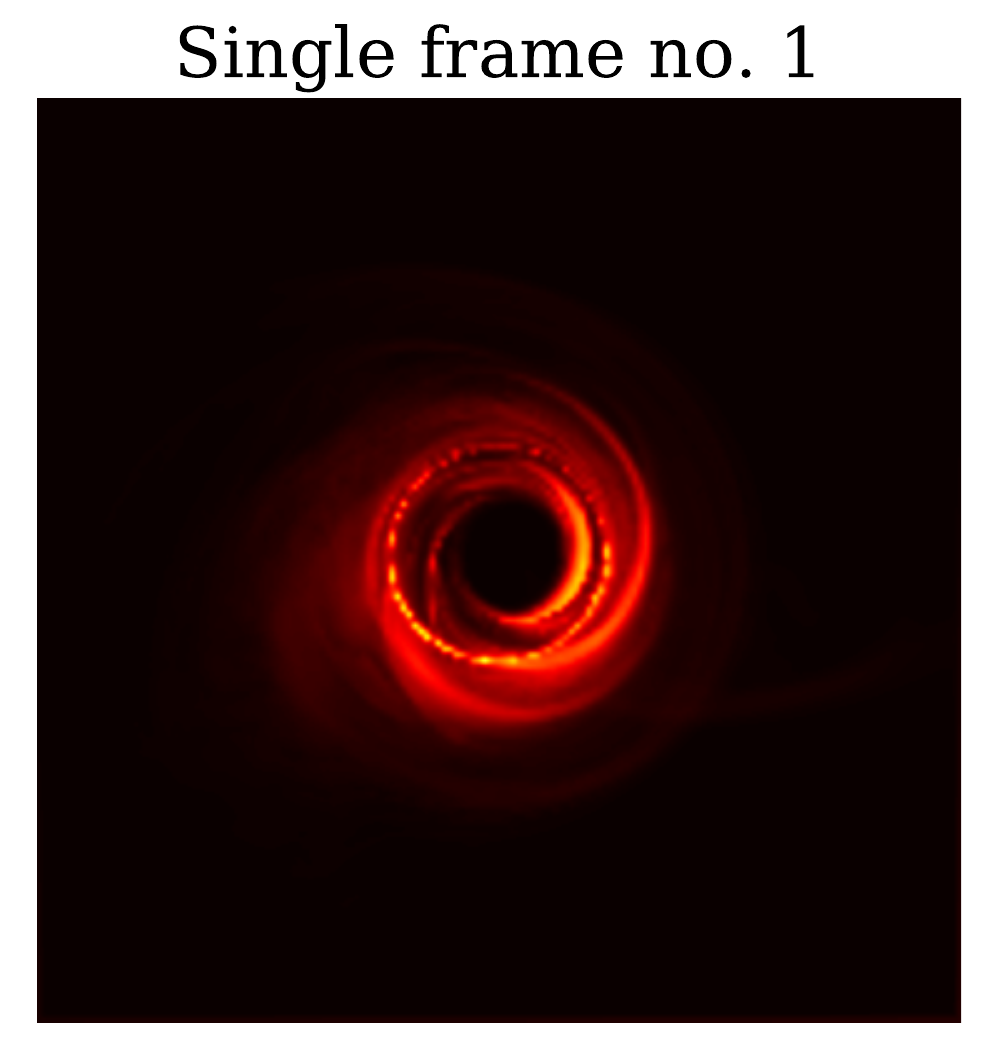}
\includegraphics[width=.16\textwidth]{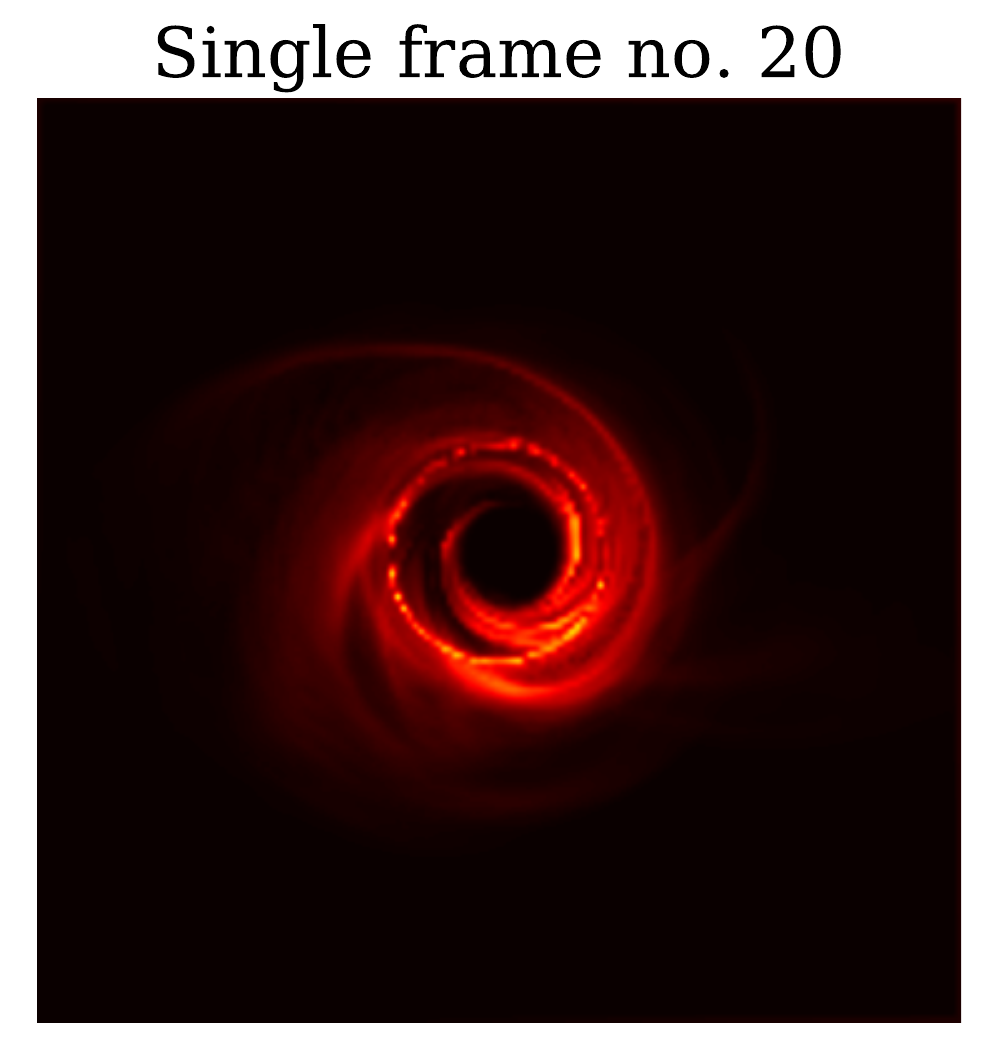}
\includegraphics[width=.16\textwidth]{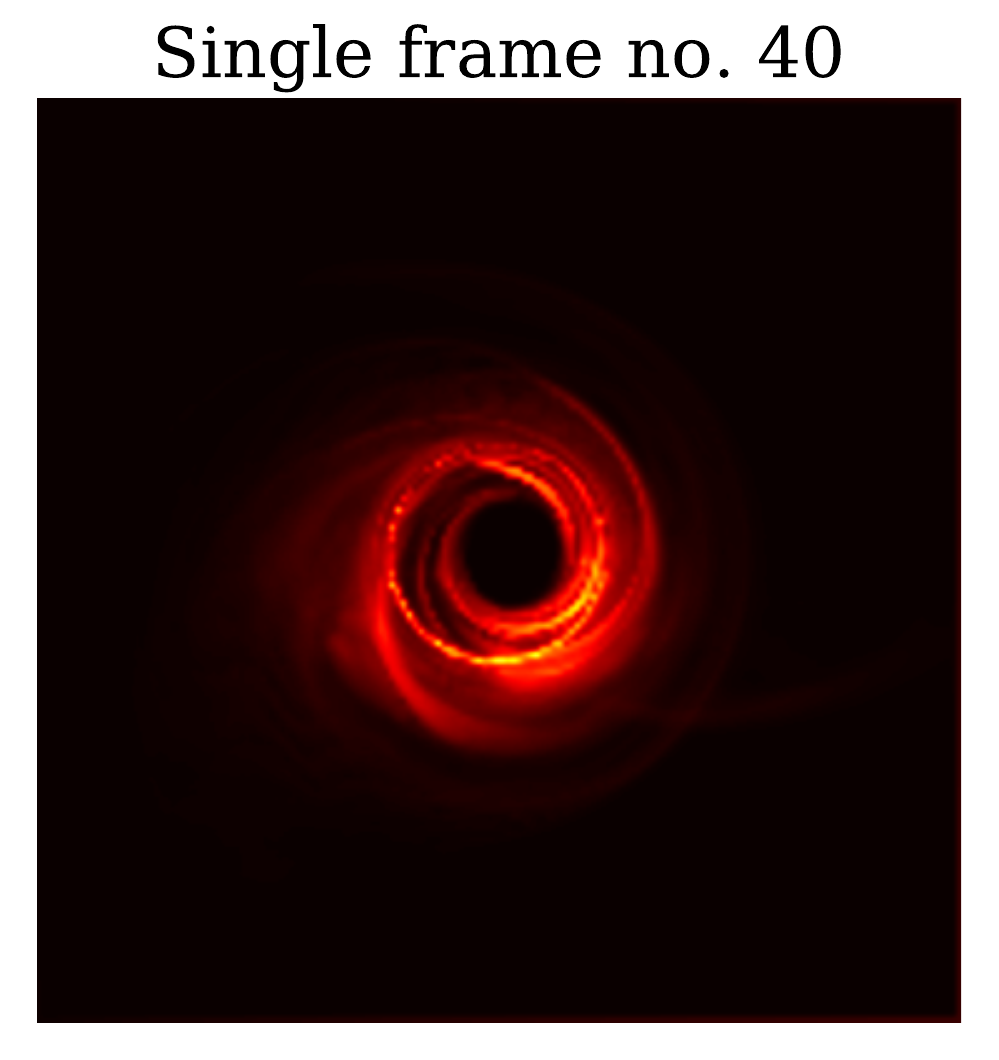}
\includegraphics[width=.16\textwidth]{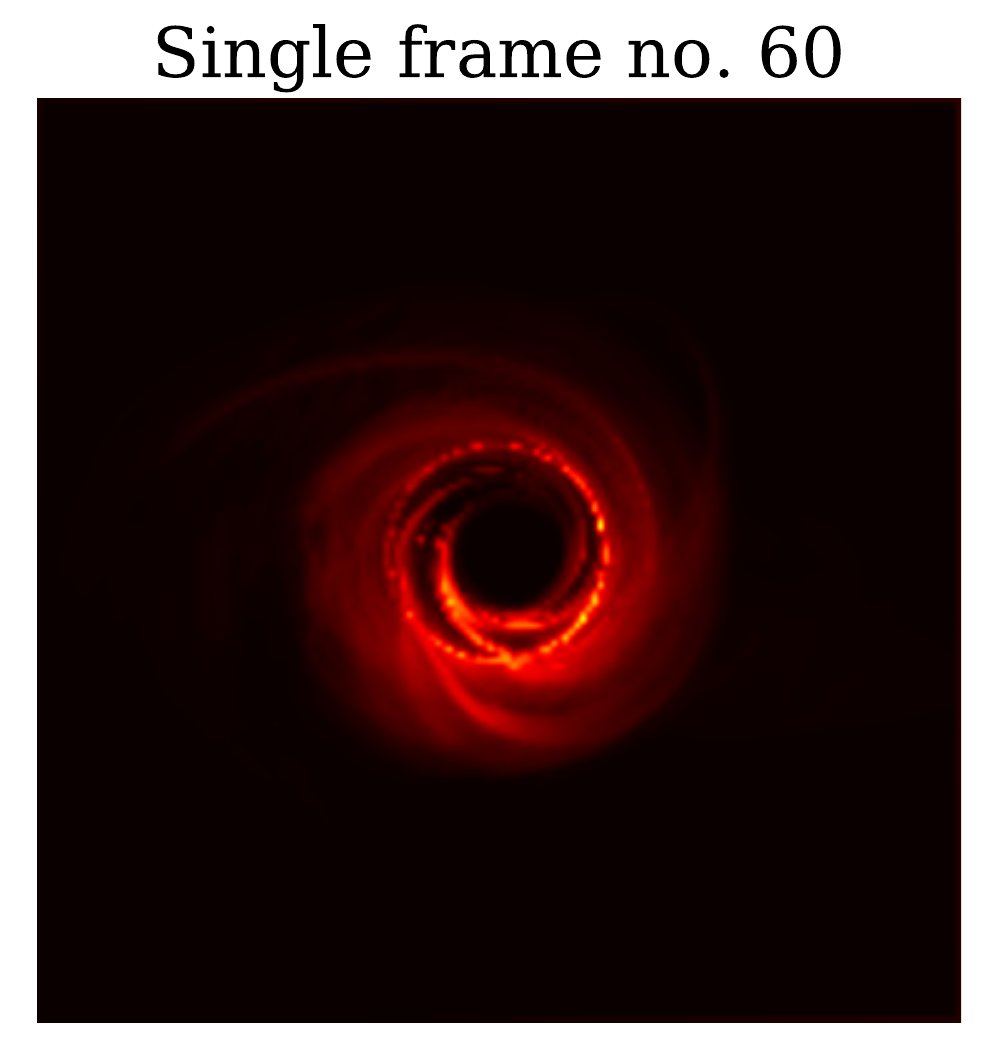}
\includegraphics[width=.16\textwidth]{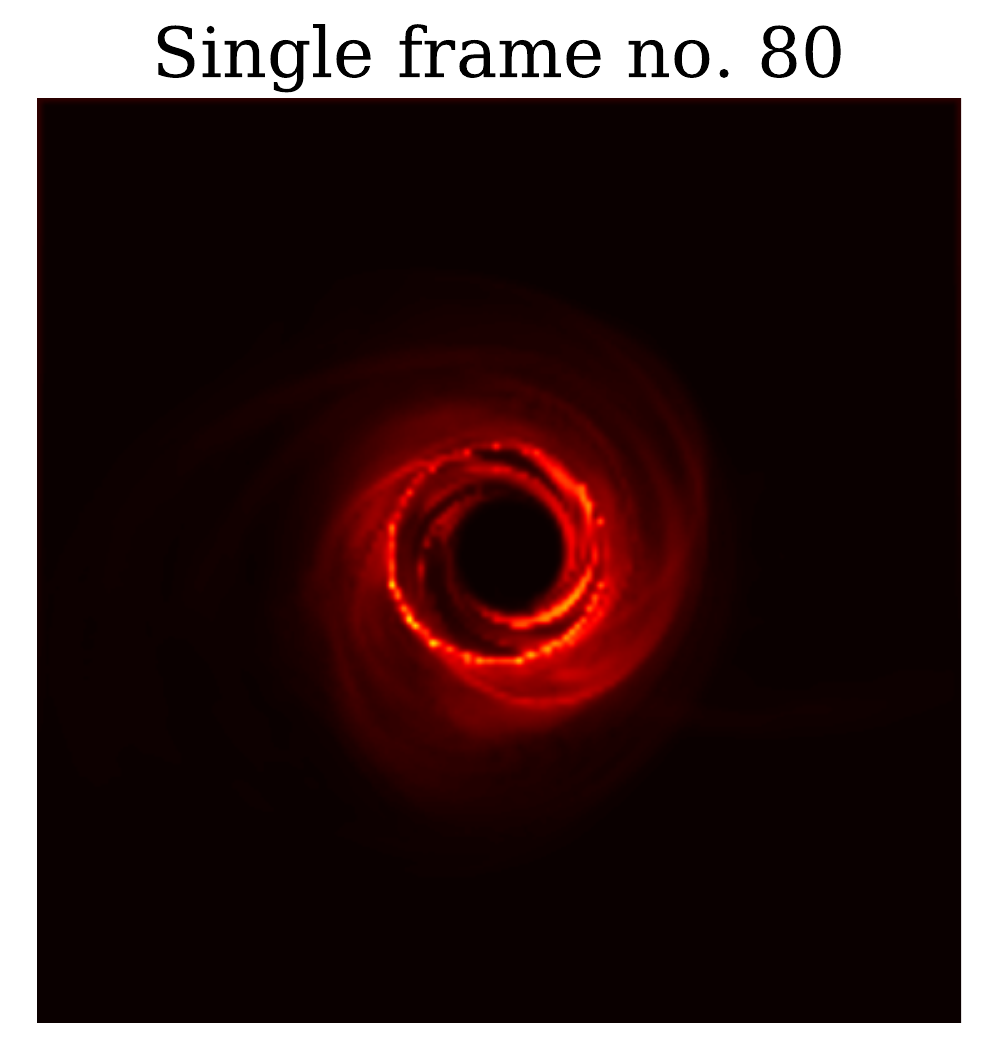}
\includegraphics[width=.16\textwidth]{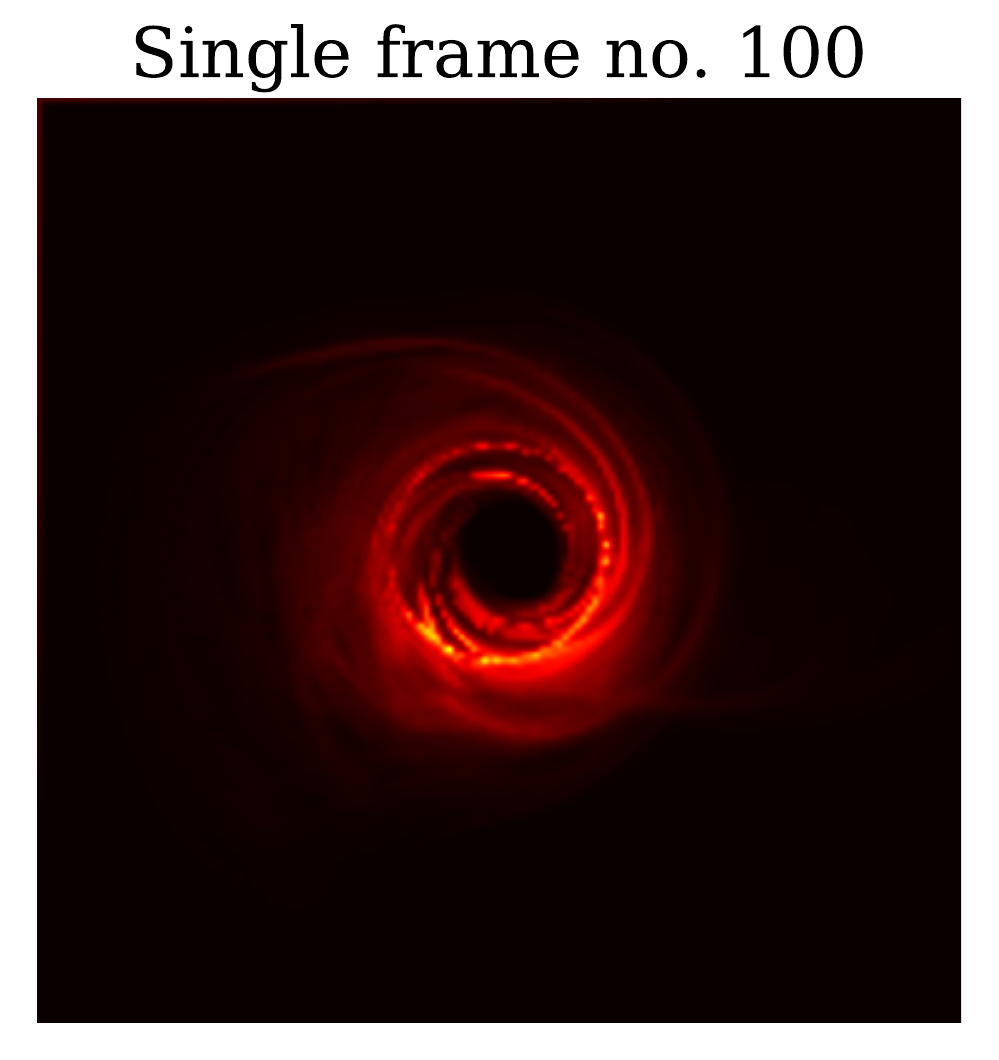} \\
\includegraphics[width=.16\textwidth]{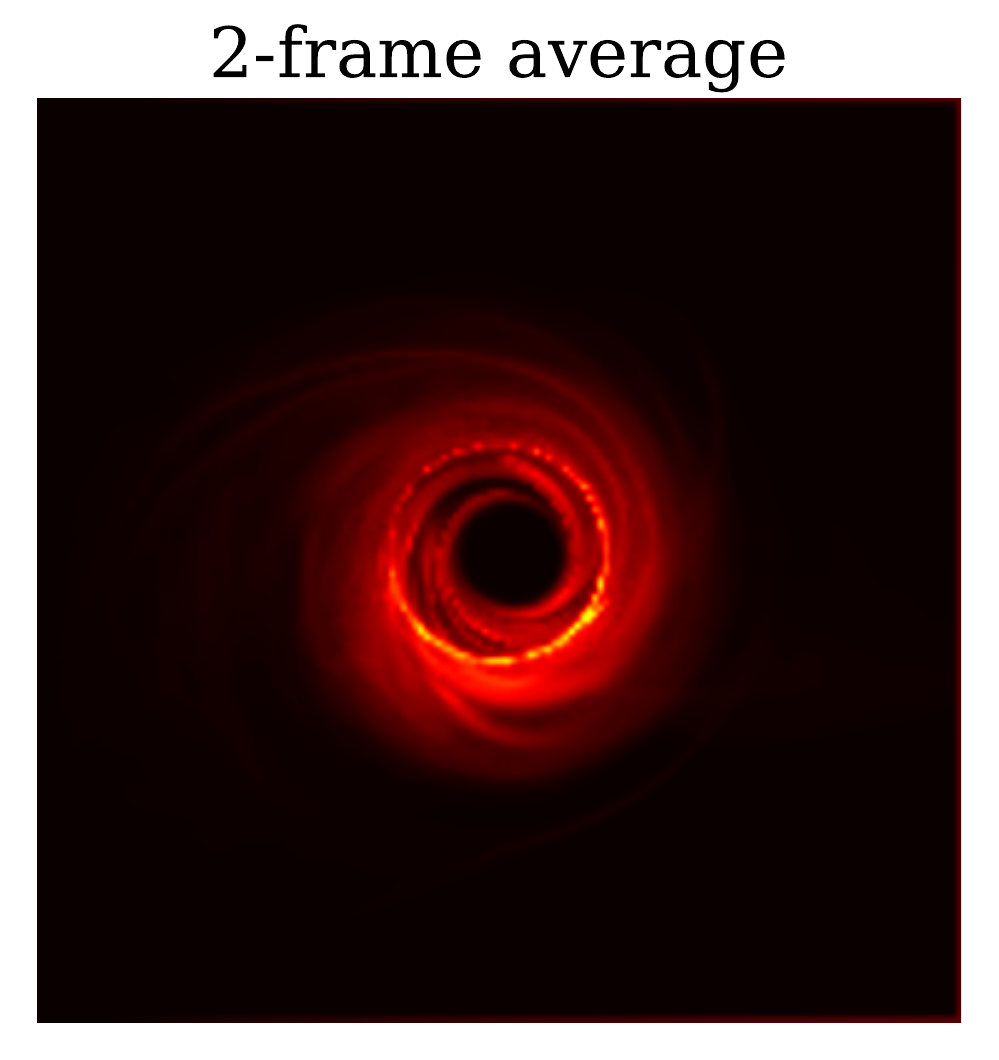}
\includegraphics[width=.16\textwidth]{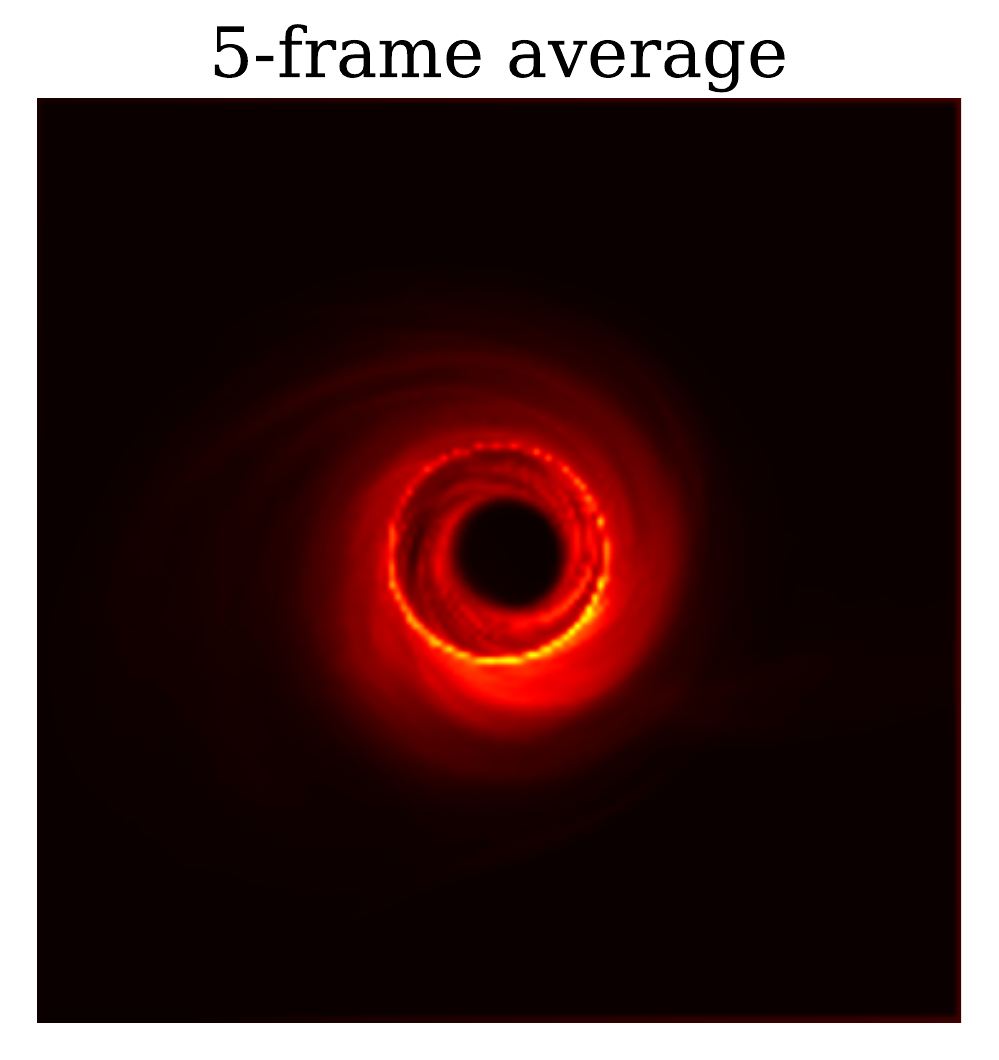}
\includegraphics[width=.16\textwidth]{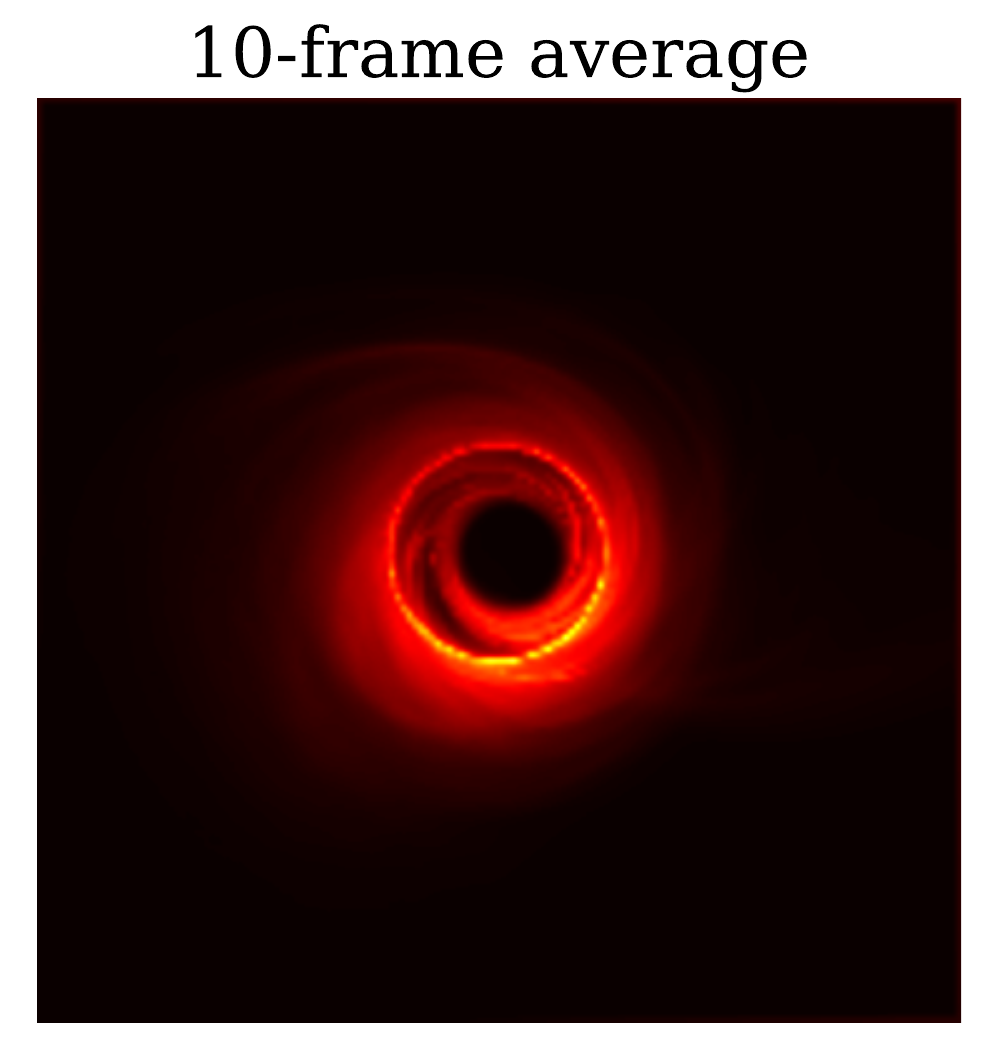}
\includegraphics[width=.16\textwidth]{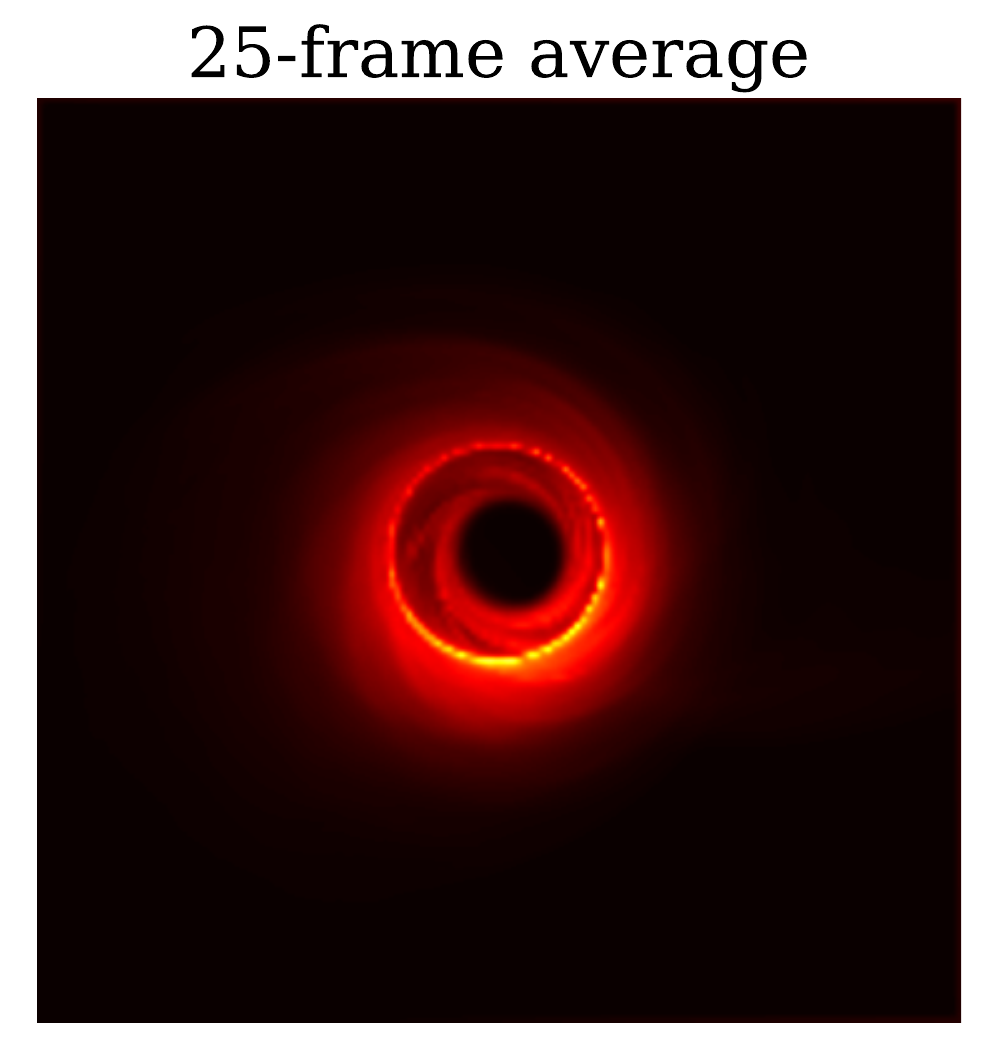}
\includegraphics[width=.16\textwidth]{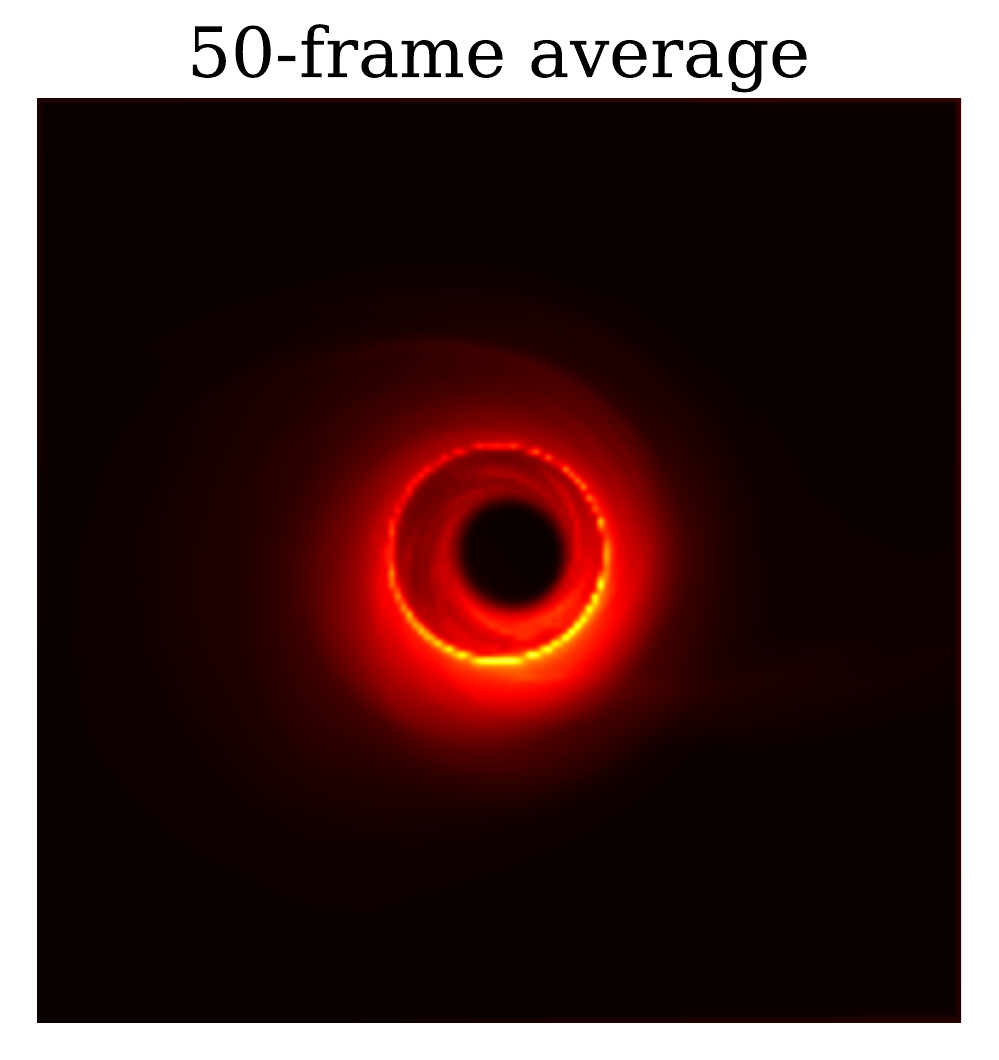}
\includegraphics[width=.16\textwidth]{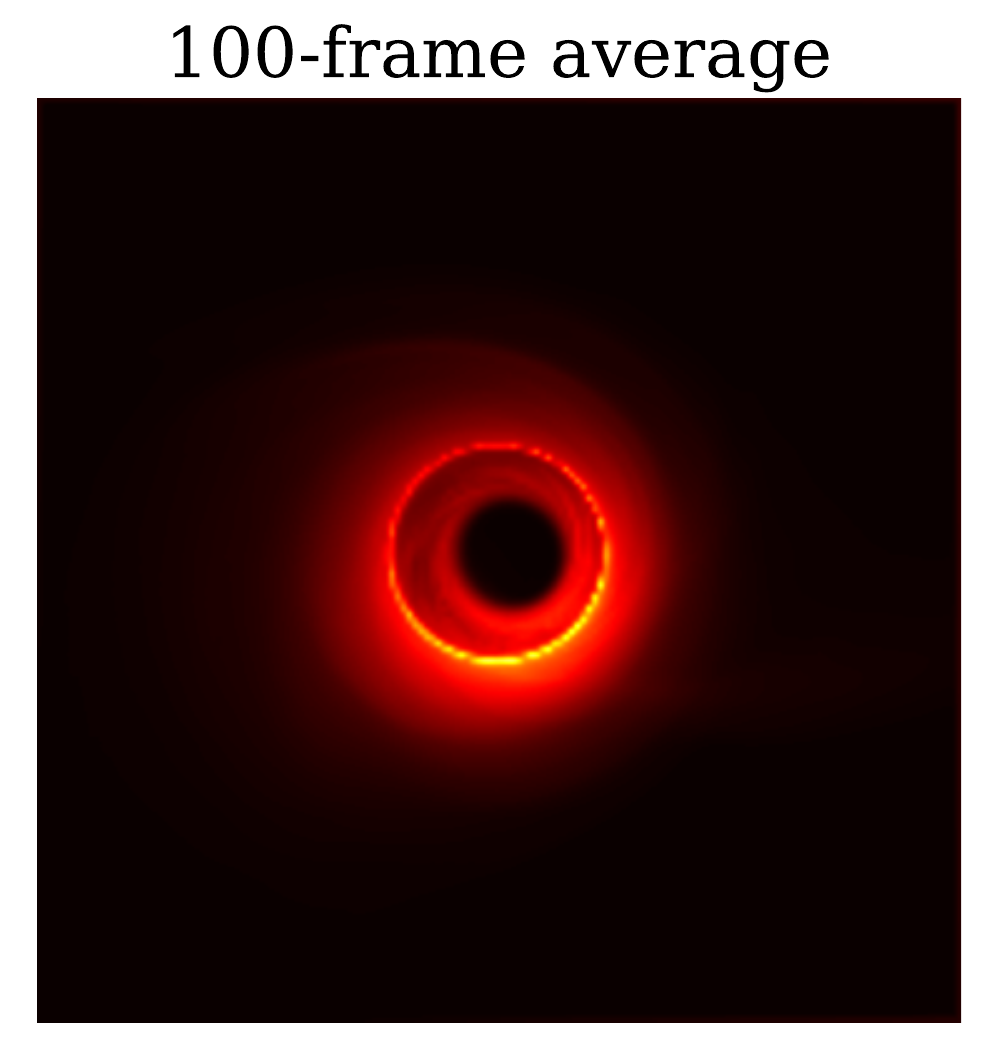}

  \caption{Upper row: ten single-frame images from the input model (SANE, $a_*=0.5$, $R_{\mathrm{high}}=80$). Lower row: example images obtained after averaging different numbers of randomly chosen frames of this model. The images are plotted on a square-root scale.}
     \label{fig:averaging_ims}
\end{figure*}

\subsection{Recovered parameters for different observing conditions}
In this section, we show how the recovered mass and sky orientation distributions and model selection by average image scoring are affected by the observation setup.

\subsubsection{Multi-epoch observations and fitting}
The synthetic data were either generated with a single observation of a single frame (designation ``1 frame''), or with ten observations from ten randomly selected frames from the same model (designation ``10 frames''). The latter case represents a situation where repeated observations of the same source take place with a fixed array, which could become an operational mode for the future EHT. The visibilities with equal $uv$-coordinates were coherently averaged after the network calibration step for the ten observations generated with independent data corruption and calibration realizations. For the EHI simulations, the source was set to remain static over the $uv$-spiral completion time of 29 days because of the limited amount of movie frames available. In reality, more realizations of the variability would be sampled within an EHI observation. Since the Fourier transform is linear, the average of ten observations of different frames corresponds to an observation of the average frame (modulo observational data corruptions). 

The effect of averaging on the visibility data is illustrated in Figure \ref{fig:averaging}, which shows comparisons between the Fast Fourier Transform (FFT) of the average image of ten frames, an EHT2021 observation of this average image, and an average of ten independent EHT2021 observations of the ten different frames. The scatter in the visibility amplitudes and closure phases reduces as the visibilities are averaged. The amplitude gain offsets between the single and averaged observation are comparable. The closure phases of a single observation have no offset to the true value, but a small offset to the FFT of the average image caused by averaging of data with gain offsets included occurs for the averaged ten-epoch observation of ten different frames. However, this offset mostly remains within the noise. On the ALMA-LMT-SMT triangle, which has the sensitive ALMA-LMT baseline and large gain variations on the SMT-LMT baseline, a systematic offset beyond the noise can be observed, but it is limited to $\sim3$ degrees. This offset is not enough to cause significant biases in the fits as all recovered mass and sky orientation distributions have the true value within $1\sigma$ (see next subsections). The variation between model frames and other observational uncertainties (e.g. the limited baseline length) are dominant over this effect.

For the average image scoring, the models were also averaged down to ten frames which are the averages of ten randomly selected model frames. In order to capture the variability between ten-frame averages, each 100-frame GRMHD movie was randomly divided into ten ten-frame averages ten times, so that a total of 100 ten-frame averages were created for each model. The data generated from the average model images should be generated in the same way as the data generated from the ground truth input model, so that the $\chi^2$ can be compared. The average image of each model was therefore observed ten times with independent data corruption and calibration realizations, and these observations were averaged as well. The fitting and average image scoring procedure was then executed exactly as described in the preceding sections. 

As the models are averaged down to ten frames, the variable small-scale turbulent structure is averaged out and the images become more dominated by the lensed photon ring and overall size and shape of the emitting region (Figure \ref{fig:averaging_ims}). The model variance is therefore reduced. Combined with the decreased scatter in the averaged data, we show below that this effect results in significantly narrower best-fit mass and sky orientation distributions and an increased ability to rule out models that do not correspond to the observed input model.

\subsubsection{Recovered mass and orientation for the input model}
\begin{figure*}[h]
\centering
\includegraphics[width=.48\textwidth]{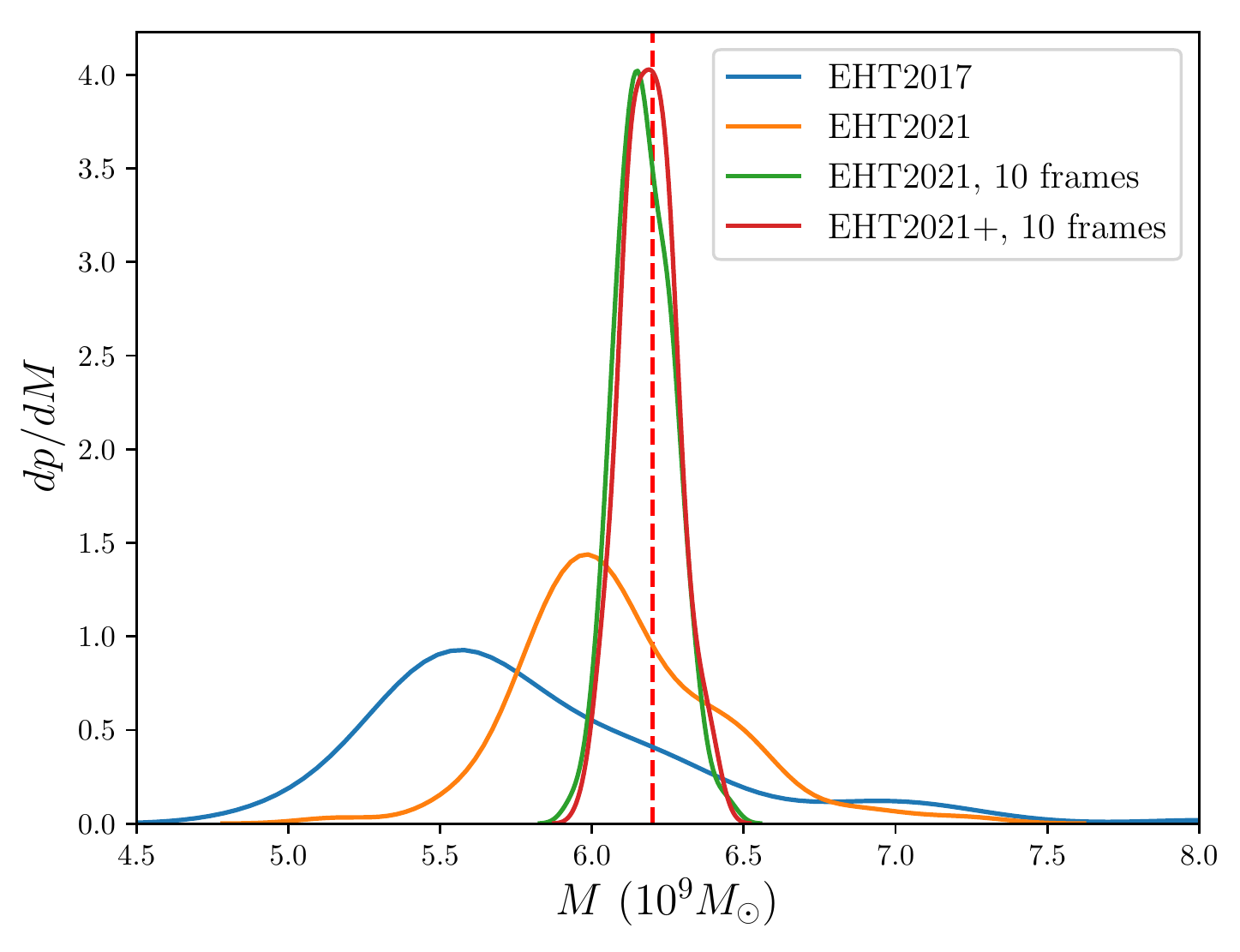}
\includegraphics[width=.48\textwidth]{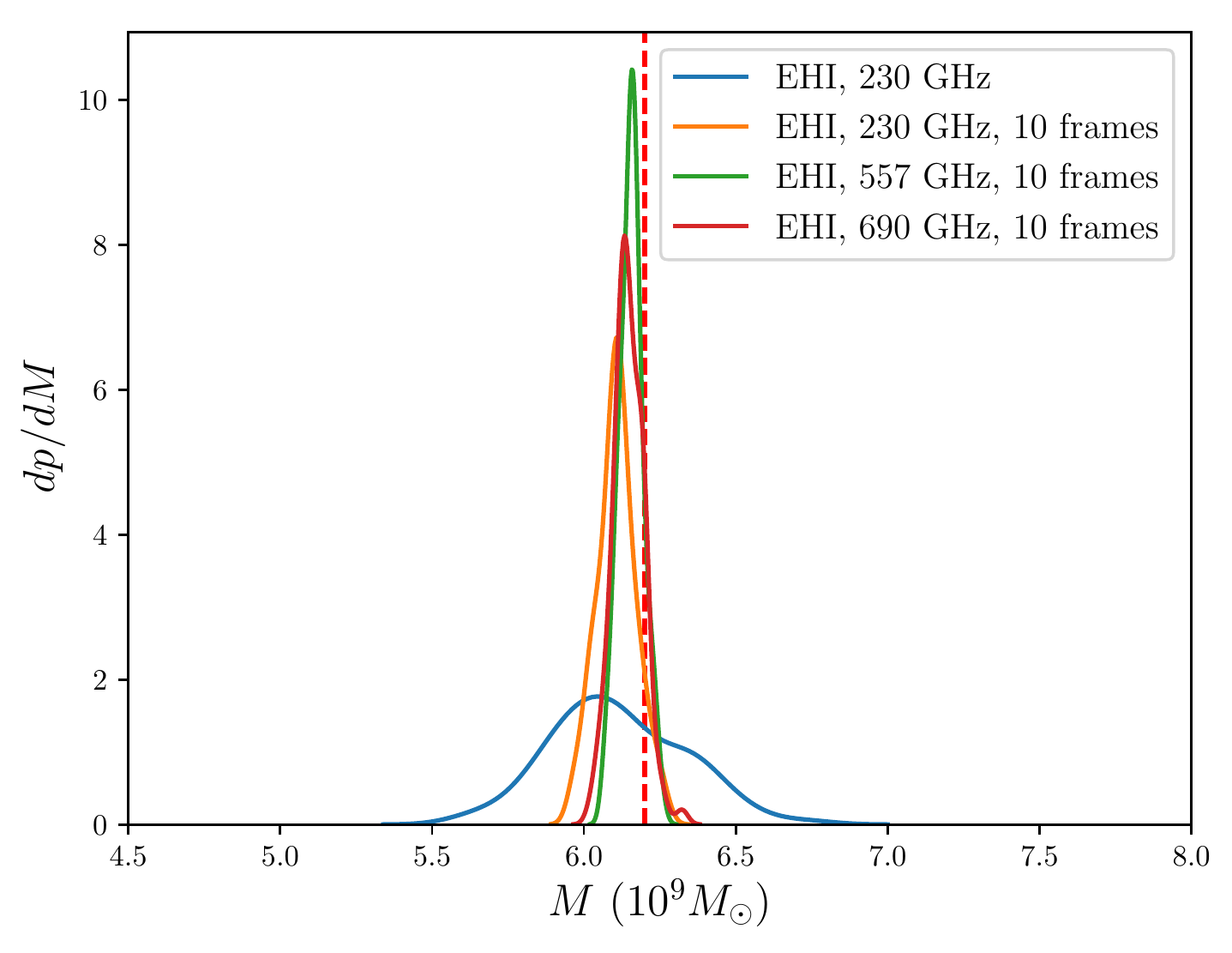} \\
\includegraphics[width=.48\textwidth]{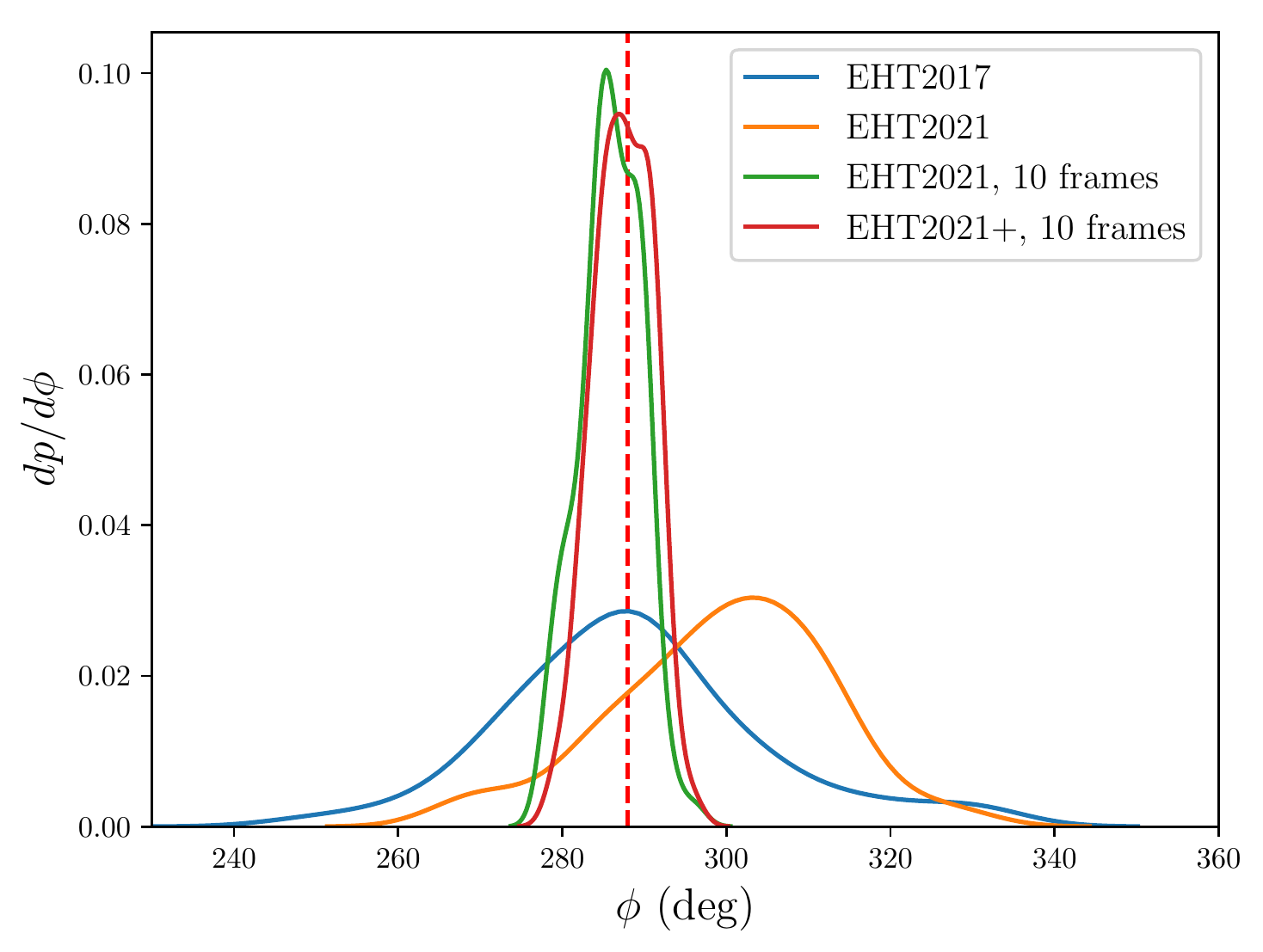}
\includegraphics[width=.48\textwidth]{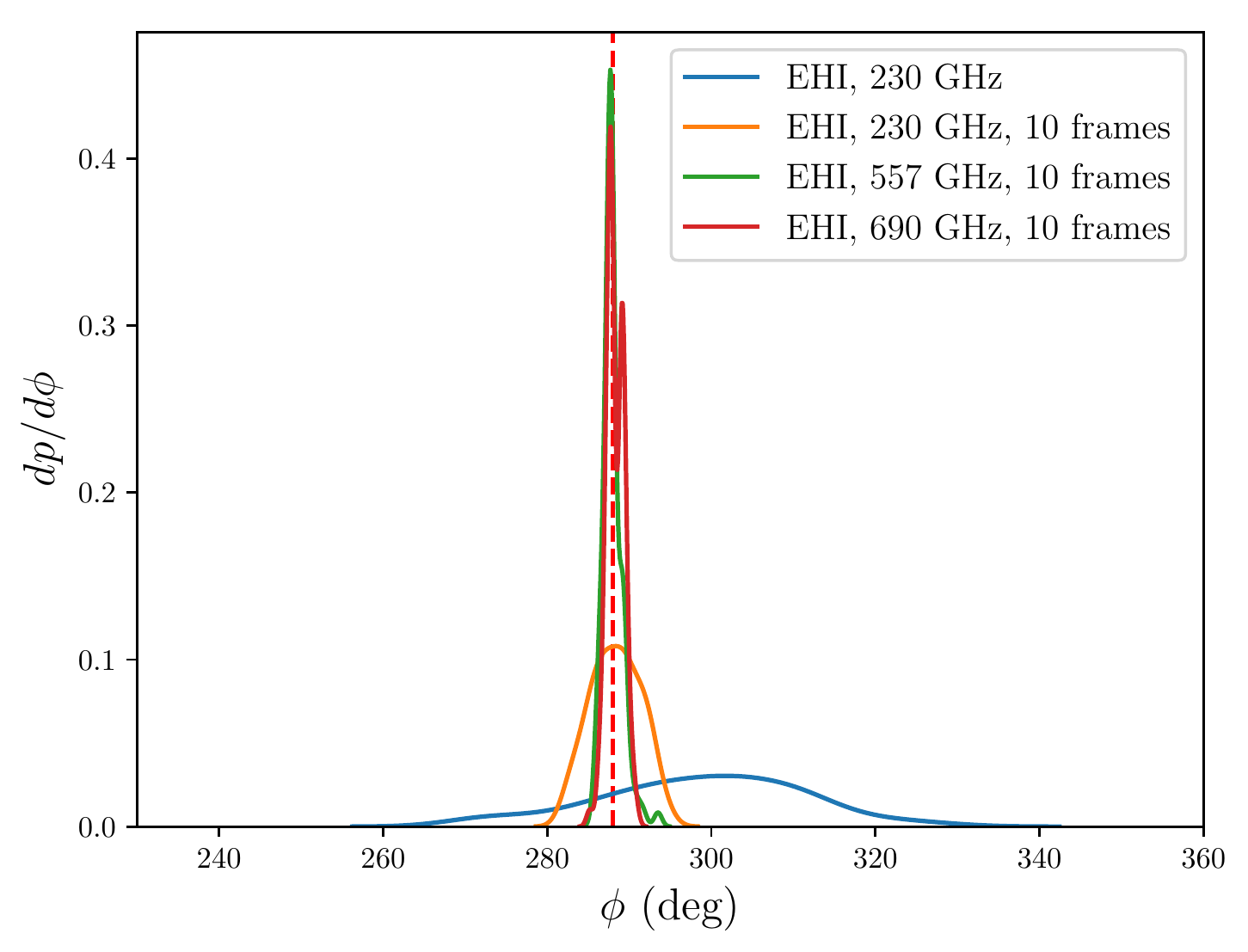}
  \caption{Best-fit mass (top) and sky orientation (bottom) distributions from fitting the frames or averages of ten frames of the input model (SANE, $a_*=0.5$, $R_{\mathrm{high}}=80$) to synthetic data generated either from frame 20 of this model or a random selection of ten frames of this model, using a ground (left) or space-based (right) array. The blue and orange line in the top left panel correspond to the distributions in the red ellipse in Figure \ref{fig:sanemassdist}, and the blue and orange line in the bottom left panel correspond to the distributions in the red ellipse in Figure \ref{fig:sanephidist}. The true input mass and sky orientations are indicated with a red dashed line.}
     \label{fig:singlemassdist}
\end{figure*}

First, we investigate the recovered mass and sky orientation distributions by fitting to the frames of the input model (SANE, $a_*=0.5$, $R_{\mathrm{high}}=80$) as shown in Figure \ref{fig:singlemassdist}. As a measure for the width of the distributions, Table \ref{tab:sigma_singlemodel} shows the standard deviations $\sigma_M$ and $\sigma_{\phi}$ for the best fit mass and orientation values, respectively, where $\sigma_{\phi}$ is the circular standard deviation accounting for wraps around $360^{\circ}$.

The recovered mass distribution clearly narrows and peaks closer to the true value as the array improves. The significant improvement as multiple observations are combined especially stands out. Repeated observations with the 2021 array even recover the mass better than a single observation with the EHI Space VLBI concept ($\sigma_M$ of $0.092\times10^9M_{\odot}$ and $0.217\times10^9M_{\odot}$, respectively), because the model variance is significantly reduced and the lensed photon ring becomes more prominent (Fig. \ref{fig:averaging_ims}), allowing for a sharper determination of the mass scaling. A potential explanation for the increase in mass estimates as the array improves is that GRMHD models generally show more emission outside the photon ring than inside, causing a tendency for frames with more large-scale emission than the input model to be scaled down, leading to an underestimate of the mass. However, it should be noted that the peak offsets are within $\sim\,1\sigma$, and could depend on the input model.

\begin{table*}[h!]
\caption{$1\sigma$ width of the mass and orientation distributions in Figure \ref{fig:singlemassdist}. The percentages are relative to the true input mass of $6.2\times10^9M_{\odot}$ and the full $360^{\circ}$ circle for $\sigma_M$ and $\sigma_{\phi}$, respectively.}
\resizebox{\textwidth}{!}{%
\begin{tabular}{l|llllllll}
   &   EHT2017   &   EHT2021   &   EHT2021   &   EHT2021+   &   EHI, 230GHz   &   EHI, 230GHz   &   EHI, 557 GHz   &   EHI, 690 GHz   \\
   &   1 frame   &   1 frame   &   10 frames   &   10 frames   &   1 frame   &   10 frames   &   10 frames   &   10 frames   \\
   \hline
$\sigma_M$ ($10^9M_{\odot}$)   & 0.534 & 0.320 & 0.092 & 0.087 & 0.217 & 0.063 & 0.039 & 0.049 \\
$\sigma_M$ (\%)   & 8.61 & 5.15 & 1.48 & 1.41 & 3.51 & 1.03 & 0.63 & 0.79 \\
$\sigma_{\phi}$ (deg)  & 15.7 & 12.9 & 3.7 & 3.5 & 12.2 & 3.0 & 1.2 & 1.0 \\
$\sigma_{\phi}$ (\%)  & 4.35 & 3.58 & 1.02 & 0.96 & 3.40 & 0.84 & 0.33 & 0.28 \\
\end{tabular}}
\label{tab:sigma_singlemodel}
\end{table*}

Adding in the six EHT2021+ stations marginally improves the mass and position angle measurements here. The precision obtained with EHT2021 coverage in combination with using multiple epochs already seems to be approaching a plateau within the maximum baseline length. The widths of the distributions are thus likely not dominated by uncertainties due to gaps in the $uv$-plane, but by the variation of the different model frames and by the maximum baseline length (resolution), which is set the Earth's size and observing frequency. We only see large improvement beyond EHT2021 here when we go to long space baselines. 

Repeated observations with the EHI further sharpen the distribution. The 230 GHz distribution peaks at slightly lower mass here, but still about $1\sigma$ from the true value.
The mass determination is less sharp for 690 GHz observations compared to 557 GHz observations ($\sigma_M$ of $0.049\times10^9M_{\odot}$ and $0.039\times10^9M_{\odot}$, respectively), which may be a consequence of the lower signal-to-noise ratio as the total flux density decreases as a function of observing frequency. 

The recovered sky orientation does not improve much as the array is improved when only single observations are considered. Going from the EHT2017 array to the EHT2021 or EHI array, the peak even moves away from the true input value, although that is not much more than a $1\sigma$ offset. This trend may reflect the variability of the source model becoming more apparent to the array as it is improved (see also Sec. \ref{sec:gentrends}). The offset disappears when multiple epochs are combined and much of the variability is averaged out. The position angle measurement improves substantially with high-frequency EHI observations compared to 230 GHz, reaching 1-degree precision. 

The distributions in Figure \ref{fig:singlemassdist} show the mass and sky orientation recovery where the synthetic data is fit to the model from which it was generated. The width of the distributions is set by the combined effects of the model variance and the ability of the array to resolve the source structure. However, as we will show in the next section and is already apparent from Figures \ref{fig:sanemassdist} and \ref{fig:sanephidist}, a substantially larger uncertainty in recovered mass and position angle results from the array's limited ability to distinguish between models with different magnetic flux, spin, and electron temperature distributions.

\begin{table*}[ht!]
\caption{GRMHD library model acceptance (ACC) and rejection (REJ) by the average image scoring procedure for different arrays. The model used for the input synthetic data generation is indicated in boldface.} 
\resizebox{0.85\textwidth}{!}{%
\begin{tabular}{lll|lllllll}
Magn.   &   $a_*$   &   $R_{\mathrm{high}}$   &   EHT2017   &   EHT2021   &   EHT2021   &   EHT2021+   &   EHI, 230GHz   &   EHI, 557 GHz   &   EHI, 690 GHz   \\
   &   &   &   1 frame   &   1 frame   &   10 frames   &   10 frames   &   10 frames   &   10 frames   &   10 frames   \\
\hline
MAD   &   -0.94   &   1   &   ACC   &   ACC   &   REJ   &   REJ   &   REJ   &   REJ   &   REJ   \\
MAD   &   -0.94   &   10   &   ACC   &   ACC   &   ACC   &   ACC   &   REJ   &   REJ   &   REJ   \\
MAD   &   -0.94   &   20   &   ACC   &   ACC   &   ACC   &   ACC   &   REJ   &   REJ   &   REJ   \\
MAD   &   -0.94   &   40   &   ACC   &   ACC   &   ACC   &   ACC   &   REJ   &   REJ   &   REJ   \\
MAD   &   -0.94   &   80   &   ACC   &   ACC   &   ACC   &   ACC   &   REJ   &   REJ   &   REJ   \\
MAD   &   -0.94   &   160   &   ACC   &   ACC   &   ACC   &   ACC   &   ACC   &   ACC   &   REJ   \\
MAD   &   -0.5   &   1   &   ACC   &   ACC   &   REJ   &   REJ   &   REJ   &   REJ   &   REJ   \\
MAD   &   -0.5   &   10   &   REJ   &   ACC   &   ACC   &   ACC   &   REJ   &   REJ   &   REJ   \\
MAD   &   -0.5   &   20   &   REJ   &   ACC   &   ACC   &   ACC   &   REJ   &   REJ   &   REJ   \\
MAD   &   -0.5   &   40   &   REJ   &   ACC   &   REJ   &   REJ   &   REJ   &   REJ   &   REJ   \\
MAD   &   -0.5   &   80   &   ACC   &   ACC   &   REJ   &   REJ   &   REJ   &   REJ   &   REJ   \\
MAD   &   -0.5   &   160   &   ACC   &   ACC   &   REJ   &   REJ   &   REJ   &   REJ   &   REJ   \\
MAD   &   0.0   &   1   &   ACC   &   ACC   &   ACC   &   ACC   &   REJ   &   REJ   &   REJ   \\
MAD   &   0.0   &   10   &   ACC   &   ACC   &   REJ   &   REJ   &   REJ   &   REJ   &   REJ   \\
MAD   &   0.0   &   20   &   ACC   &   ACC   &   REJ   &   REJ   &   REJ   &   REJ   &   REJ   \\
MAD   &   0.0   &   40   &   ACC   &   ACC   &   REJ   &   REJ   &   REJ   &   REJ   &   REJ   \\
MAD   &   0.0   &   80   &   ACC   &   ACC   &   REJ   &   REJ   &   REJ   &   REJ   &   REJ   \\
MAD   &   0.0   &   160   &   ACC   &   ACC   &   ACC   &   ACC   &   REJ   &   REJ   &   REJ   \\
MAD   &   0.5   &   1   &   ACC   &   ACC   &   REJ   &   REJ   &   REJ   &   REJ   &   REJ   \\
MAD   &   0.5   &   10   &   ACC   &   ACC   &   ACC   &   REJ   &   REJ   &   REJ   &   REJ   \\
MAD   &   0.5   &   20   &   ACC   &   ACC   &   REJ   &   REJ   &   REJ   &   REJ   &   REJ   \\
MAD   &   0.5   &   40   &   ACC   &   ACC   &   REJ   &   REJ   &   REJ   &   REJ   &   REJ   \\
MAD   &   0.5   &   80   &   ACC   &   ACC   &   REJ   &   REJ   &   REJ   &   REJ   &   REJ   \\
MAD   &   0.5   &   160   &   ACC   &   ACC   &   REJ   &   REJ   &   REJ   &   REJ   &   REJ   \\
MAD   &   0.94   &   1   &   ACC   &   ACC   &   REJ   &   REJ   &   REJ   &   REJ   &   REJ   \\
MAD   &   0.94   &   10   &   ACC   &   ACC   &   ACC   &   ACC   &   REJ   &   REJ   &   REJ   \\
MAD   &   0.94   &   20   &   ACC   &   ACC   &   ACC   &   REJ   &   REJ   &   REJ   &   REJ   \\
MAD   &   0.94   &   40   &   ACC   &   ACC   &   ACC   &   REJ   &   REJ   &   REJ   &   REJ   \\
MAD   &   0.94   &   80   &   ACC   &   ACC   &   ACC   &   ACC   &   REJ   &   REJ   &   REJ   \\
MAD   &   0.94   &   160   &   ACC   &   ACC   &   ACC   &   ACC   &   REJ   &   REJ   &   REJ   \\
SANE   &   -0.94   &   1   &   ACC   &   REJ   &   ACC   &   ACC   &   REJ   &   REJ   &   REJ   \\
SANE   &   -0.94   &   10   &   REJ   &   ACC   &   REJ   &   REJ   &   REJ   &   REJ   &   REJ   \\
SANE   &   -0.94   &   20   &   REJ   &   ACC   &   REJ   &   REJ   &   REJ   &   REJ   &   REJ   \\
SANE   &   -0.94   &   40   &   REJ   &   ACC   &   REJ   &   REJ   &   REJ   &   REJ   &   REJ   \\
SANE   &   -0.94   &   80   &   ACC   &   ACC   &   REJ   &   REJ   &   REJ   &   REJ   &   REJ   \\
SANE   &   -0.94   &   160   &   ACC   &   ACC   &   REJ   &   REJ   &   REJ   &   REJ   &   REJ   \\
SANE   &   -0.5   &   1   &   ACC   &   REJ   &   ACC   &   ACC   &   REJ   &   REJ   &   REJ   \\
SANE   &   -0.5   &   10   &   ACC   &   ACC   &   REJ   &   REJ   &   REJ   &   REJ   &   REJ   \\
SANE   &   -0.5   &   20   &   ACC   &   ACC   &   REJ   &   REJ   &   REJ   &   REJ   &   REJ   \\
SANE   &   -0.5   &   40   &   REJ   &   REJ   &   REJ   &   REJ   &   REJ   &   REJ   &   REJ   \\
SANE   &   -0.5   &   80   &   REJ   &   REJ   &   REJ   &   REJ   &   REJ   &   REJ   &   REJ   \\
SANE   &   -0.5   &   160   &   REJ   &   ACC   &   REJ   &   REJ   &   REJ   &   REJ   &   REJ   \\
SANE   &   0.0   &   1   &   ACC   &   ACC   &   ACC   &   ACC   &   REJ   &   REJ   &   REJ   \\
SANE   &   0.0   &   10   &   ACC   &   ACC   &   ACC   &   ACC   &   ACC   &   ACC   &   ACC   \\
SANE   &   0.0   &   20   &   ACC   &   ACC   &   ACC   &   ACC   &   ACC   &   ACC   &   ACC   \\
SANE   &   0.0   &   40   &   ACC   &   ACC   &   ACC   &   ACC   &   ACC   &   REJ   &   REJ   \\
SANE   &   0.0   &   80   &   ACC   &   ACC   &   ACC   &   ACC   &   REJ   &   REJ   &   REJ   \\
SANE   &   0.0   &   160   &   ACC   &   ACC   &   REJ   &   REJ   &   REJ   &   REJ   &   REJ   \\
SANE   &   0.5   &   1   &   ACC   &   ACC   &   ACC   &   ACC   &   REJ   &   REJ   &   REJ   \\
SANE   &   0.5   &   10   &   ACC   &   ACC   &   ACC   &   ACC   &   ACC   &   ACC   &   ACC   \\
SANE   &   0.5   &   20   &   ACC   &   ACC   &   ACC   &   ACC   &   ACC   &   ACC   &   ACC   \\
SANE   &   0.5   &   40   &   ACC   &   ACC   &   ACC   &   ACC   &   ACC   &   ACC   &   ACC   \\
\textbf{SANE}   &   \textbf{0.5}   &   \textbf{80}   &   \textbf{ACC}   &   \textbf{ACC}   &   \textbf{ACC}   &   \textbf{ACC}   &   \textbf{ACC}   &   \textbf{ACC}   &   \textbf{ACC}   \\
SANE   &   0.5   &   160   &   ACC   &   ACC   &   ACC   &   ACC   &   ACC   &   ACC   &   ACC   \\
SANE   &   0.97   &   1   &   ACC   &   REJ   &   REJ   &   REJ   &   REJ   &   REJ   &   REJ   \\
SANE   &   0.97   &   10   &   ACC   &   ACC   &   REJ   &   REJ   &   REJ   &   REJ   &   REJ   \\
SANE   &   0.97   &   20   &   ACC   &   ACC   &   REJ   &   REJ   &   REJ   &   REJ   &   REJ   \\
SANE   &   0.97   &   40   &   ACC   &   REJ   &   REJ   &   REJ   &   REJ   &   REJ   &   REJ   \\
SANE   &   0.97   &   80   &   ACC   &   ACC   &   REJ   &   REJ   &   REJ   &   REJ   &   REJ   \\
SANE   &   0.97   &   160   &   ACC   &   ACC   &   REJ   &   REJ   &   REJ   &   REJ   &   REJ   \\
\hline
\% ACC  &   &   &   85   &   90   &   47   &   42   &   15   &   13   &   12   \\
\end{tabular}}
\label{tab:ais}
\end{table*}

\subsubsection{Recovered library model parameters}
\label{sec:libraryparams}

\begin{figure}[tbp]
\centering
\includegraphics[width=.5\textwidth]{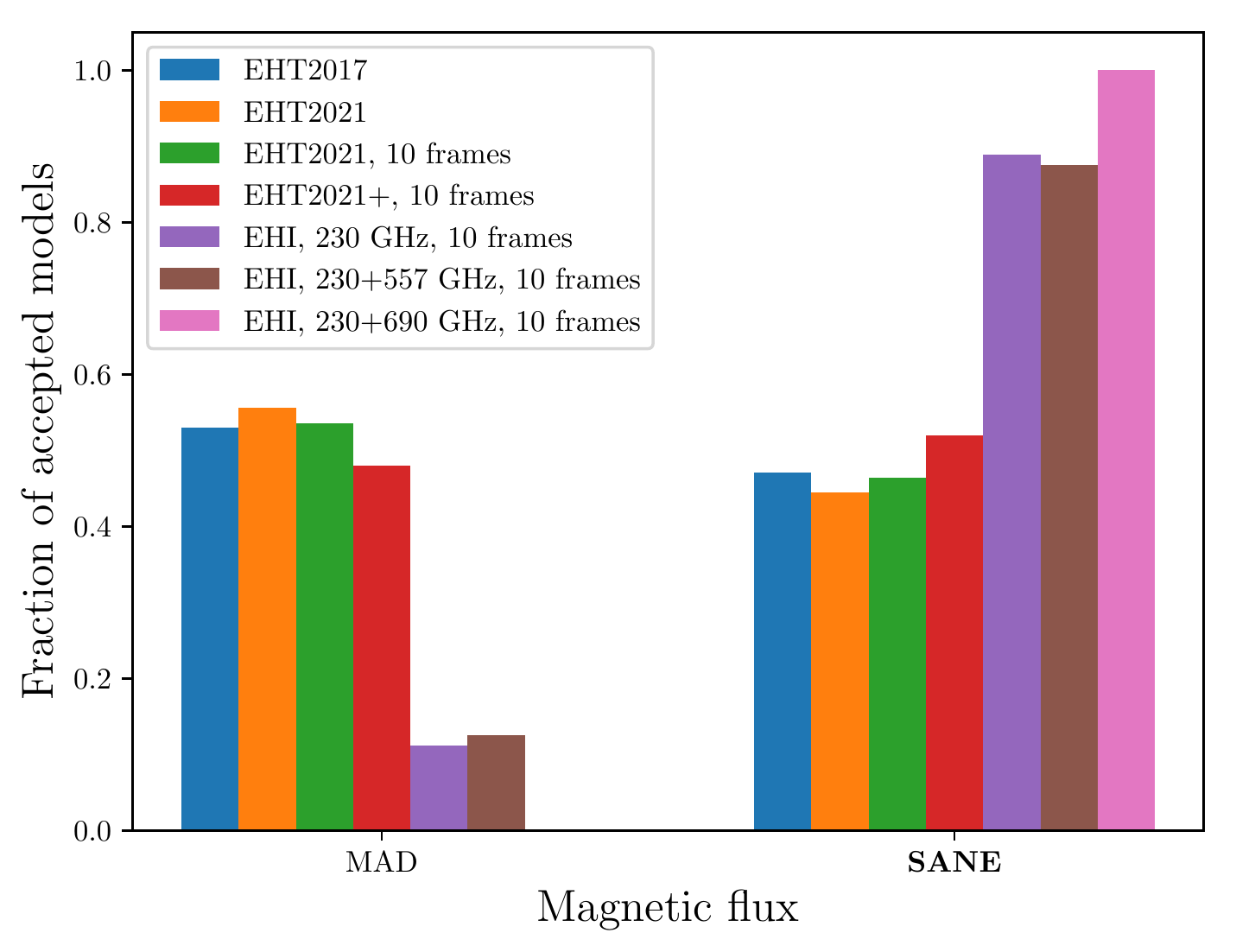}
  \caption{Distribution over the magnetic flux of the accepted models in Table \ref{tab:ais}. The true input magnetic flux (SANE) is indicated in boldface. The lacking MAD bar for the 690 GHz EHI observation indicate that none of the accepted models for this dataset were MAD models.}
     \label{fig:magndist}
\end{figure}

\begin{figure}[tbp]
\centering
\includegraphics[width=.5\textwidth]{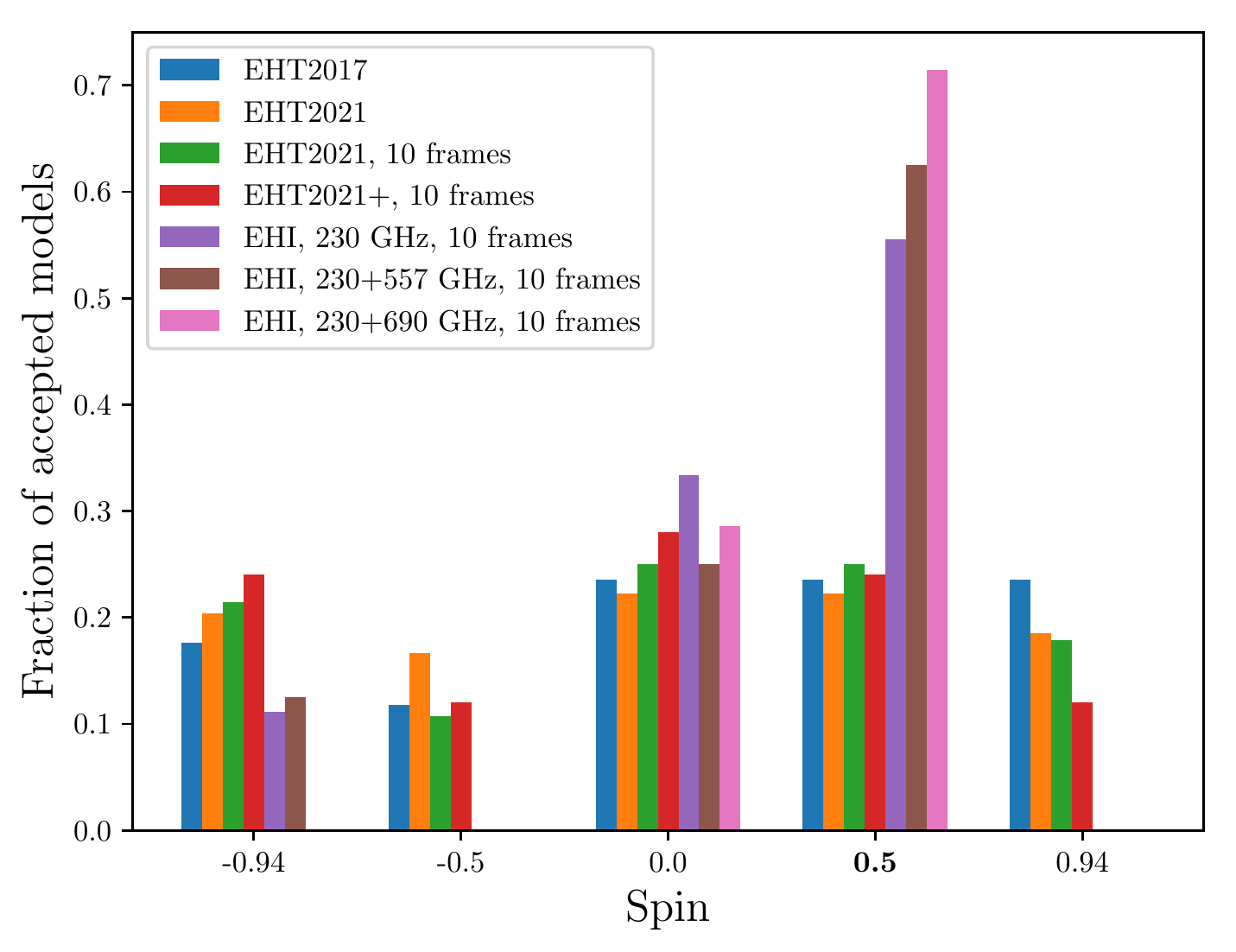}
  \caption{Distribution over black hole spin of the accepted models in Table \ref{tab:ais}. The SANE models with $a_*=0.97$ are incorporated in the $a_*=0.94$ bins. The true input spin (0.5) is indicated in boldface.}
     \label{fig:spindist}
\end{figure}

\begin{figure}[htbp]
\centering
\includegraphics[width=.5\textwidth]{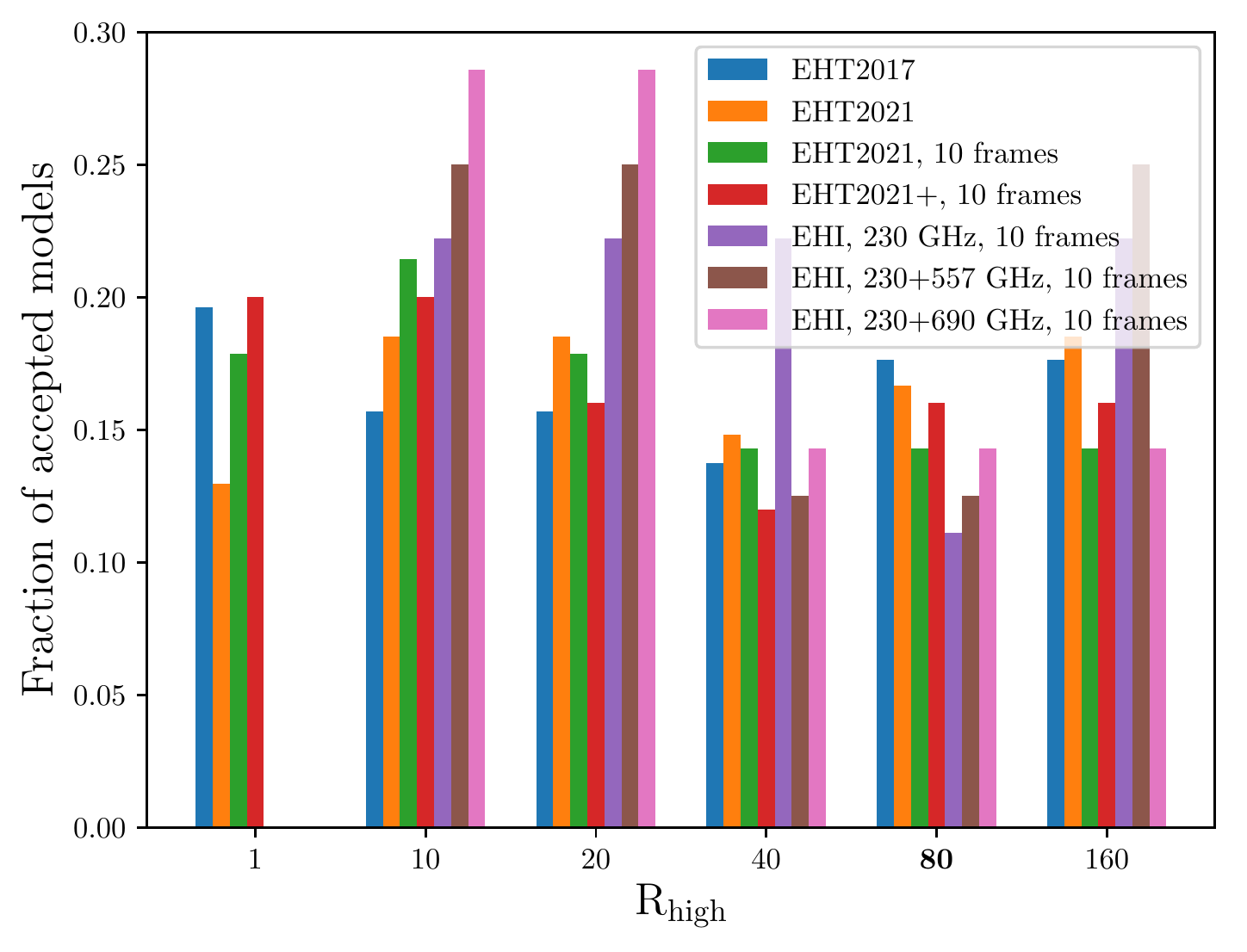}
  \caption{Distribution over $R_{\mathrm{high}}$ of the accepted models in Table \ref{tab:ais}. The true input $R_{\mathrm{high}}$ (80) is indicated in boldface.}
     \label{fig:rhighdist}
\end{figure}

In this section, we investigate how enhancements of the array could improve the ability to rule out models from the GRMHD library. Table \ref{tab:ais} shows the average image scoring results of fitting synthetic data from the SANE model with $a_*=0.5$ and $R_{\mathrm{high}}=80$ to the GRMHD models for different arrays (Sect. \ref{sec:synthdata}). As the observation configuration improves, the accepted models converge towards the parameters of the input model (bold). The percentage of accepted models generally decreases as the array improves or multiple observations are combined in the columns from left to right. An exception is the single EHT2021 observation compared to the single EHT2017 observation, where more models are accepted for the former. Like for the SANE models (see Sec. \ref{sec:gentrends}), the MAD models that were rejected for EHT2017 but accepted for EHT2021 have large model variance that becomes more apparent with the EHT2021 coverage. These models are all rejected with repeated observations and with further improvements of the array. For the EHT2021+ and high-frequency EHI observations, the model acceptance for the repeated EHT2021 observations and 230 GHz EHI observations was respectively used as a filter, resulting in the additional rejection of a few models. 

Figure \ref{fig:magndist} shows the distribution of accepted models between MAD and SANE models for each array configuration in Table \ref{tab:ais}. The magnetic flux is challenging to recover with ground-based observations: the ratio between accepted MAD and SANE models is close to 1:1. A clear preference for SANE models becomes apparent with EHI observations. For the 690 GHz EHI observations, all accepted models are SANE models, which indeed corresponds to the ground truth input model.

The black hole spin distribution (Fig. \ref{fig:spindist}) is close to flat for the single EHT2017 and EHT2021 observations, although retrograde spins are slightly disfavored compared to prograde spins. With repeated EHT2021+ observations, the $a_*=0.94$ models are disfavored more strongly. With EHI observations, the true input spin of 0.5 is strongly favored. In this case, the $a_*=-0.5$ and $a_*=0.94$ models are completely ruled out with 230 GHz observations. Only $a_*=0$ and $a_*=0.5$ models remain with high-frequency EHI observations.

\begin{figure*}[h]
\centering
\includegraphics[width=.48\textwidth]{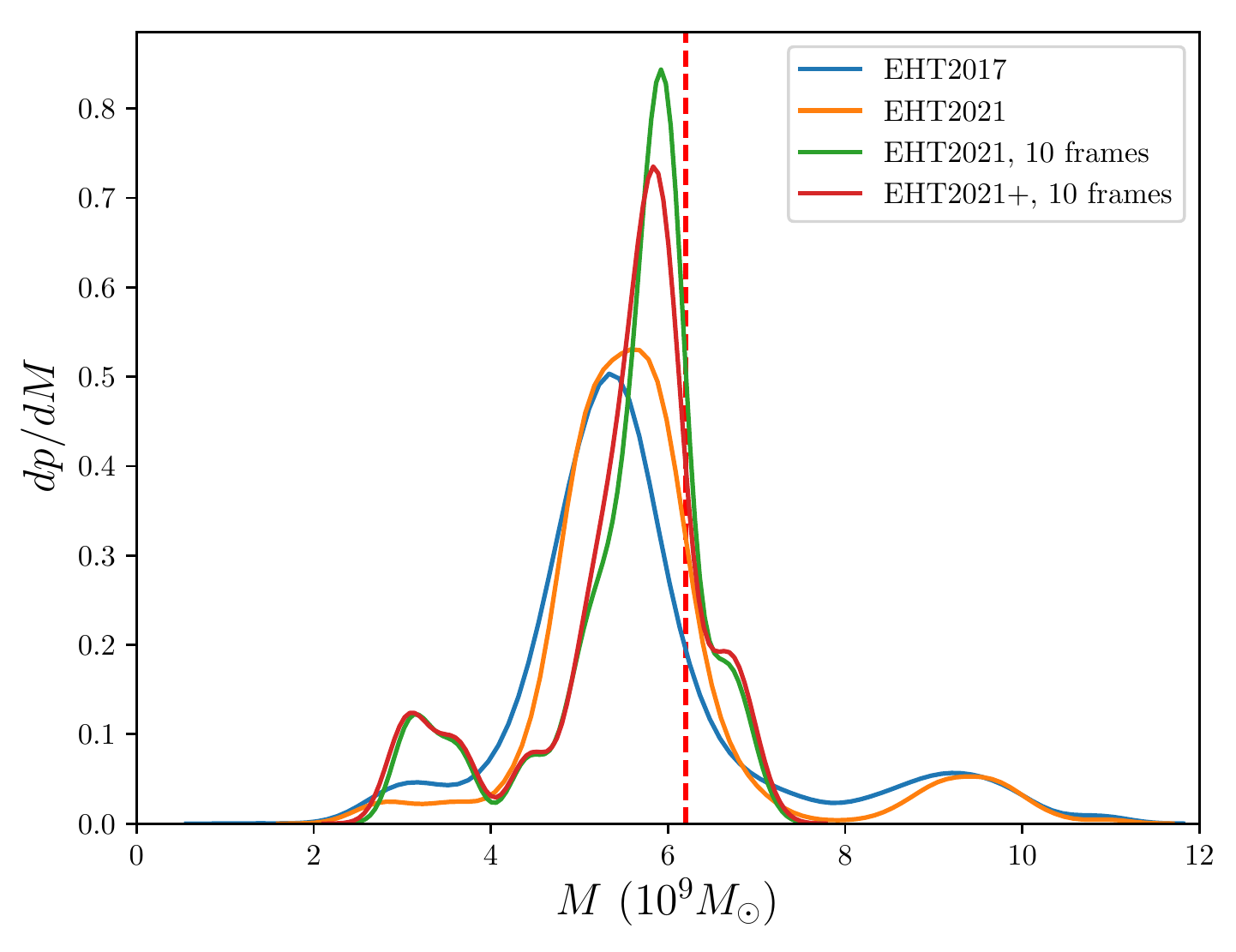}
\includegraphics[width=.48\textwidth]{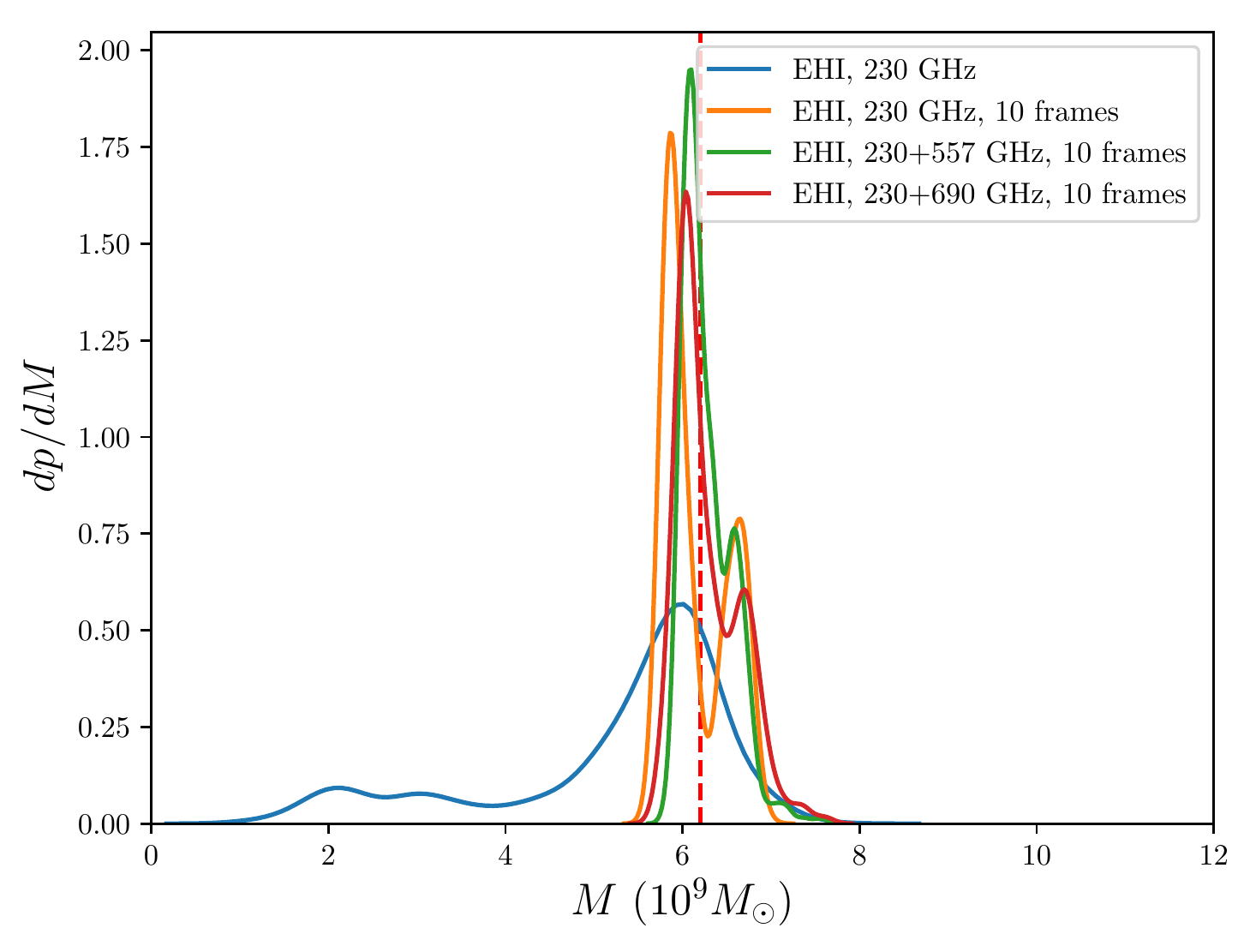} \\
\includegraphics[width=.48\textwidth]{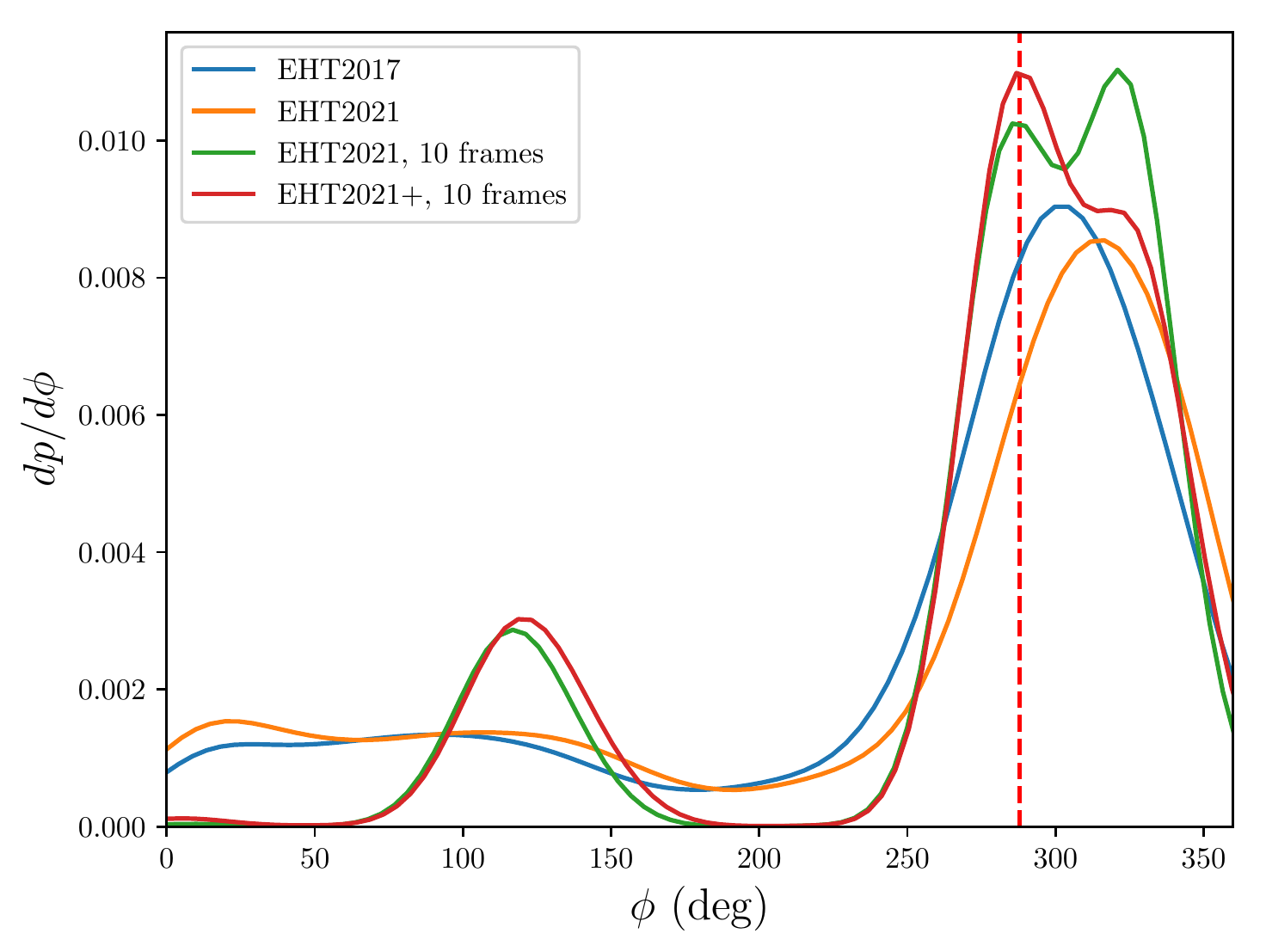}
\includegraphics[width=.48\textwidth]{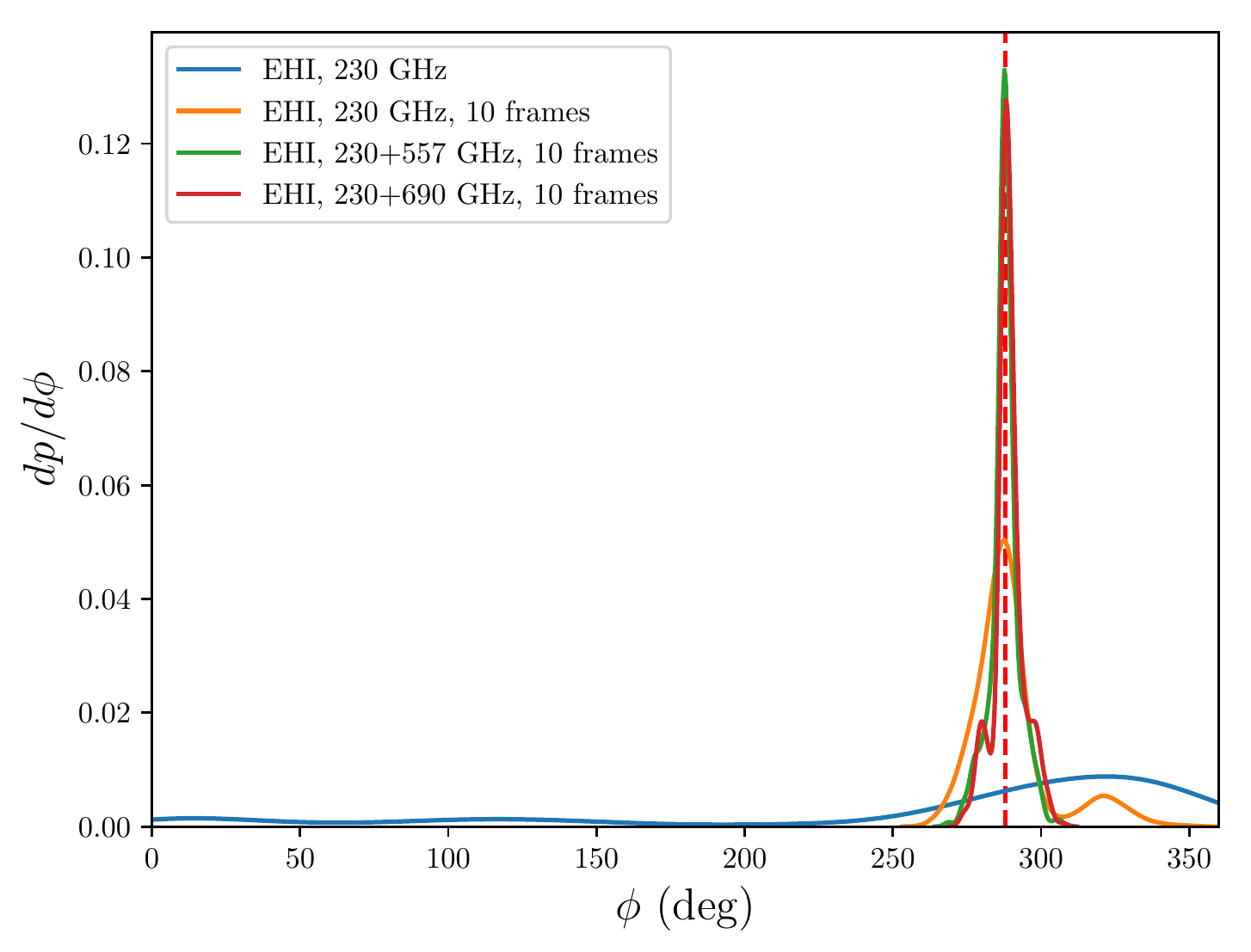}

  \caption{Best-fit mass (top) and sky orientation (bottom) distributions of the accepted models in Table \ref{tab:ais} after average image scoring for synthetic data generated with different observing configurations of ground (left) and space-based (right) arrays. The true input mass and sky orientations are indicated with a red dashed line. The sky orientations for models with retrograde spin were shifted by $180^{\circ}$.}
     \label{fig:fullmassdist}
\end{figure*}

$R_{\mathrm{high}}$ (Fig. \ref{fig:rhighdist}) is the most difficult parameter to measure. Its distribution remains relatively flat, with little changes between the observing configurations. Only EHI observations rule out the disk-dominated ($R_{\mathrm{high}}=1$) models. Low $R_{\mathrm{high}}$-values seem to be favored over the true value of 80, although it should be noted that only five to eight of the GRMHD models are accepted for the EHI observations (see Table \ref{tab:ais}). The result that $R_{\mathrm{high}}$ is more difficult to constrain than. e.g., the black hole spin can be understood from the fact that, when the magnetic flux and black hole spin are fixed, it typically does not have a large influence on the model images. This can also be seen in Figures \ref{fig:sanemassdist} and \ref{fig:sanephidist}: while varying the spin can change the fitted mass by a factor 2 or more due to, e.g., the appearance of a jet footprint and the changing plasma orbits for high spins (Fig. \ref{fig:models}), the differences between the panels with different $R_{\mathrm{high}}$ is small. An exception is the $R_{\mathrm{high}}=1$ panel. These models are disk-dominated, resulting in more emission outside the black hole shadow and a lower mass estimate (Fig. \ref{fig:models}). The $R_{\mathrm{high}}=10-160$ models all show more of the jet emission, resulting in a similar morphology. Tighter constraints on $R_{\mathrm{high}}$ may therefore require multi-wavelength observations and spectral fitting.

\begin{figure*}[h]
\centering
\includegraphics[width=.48\textwidth]{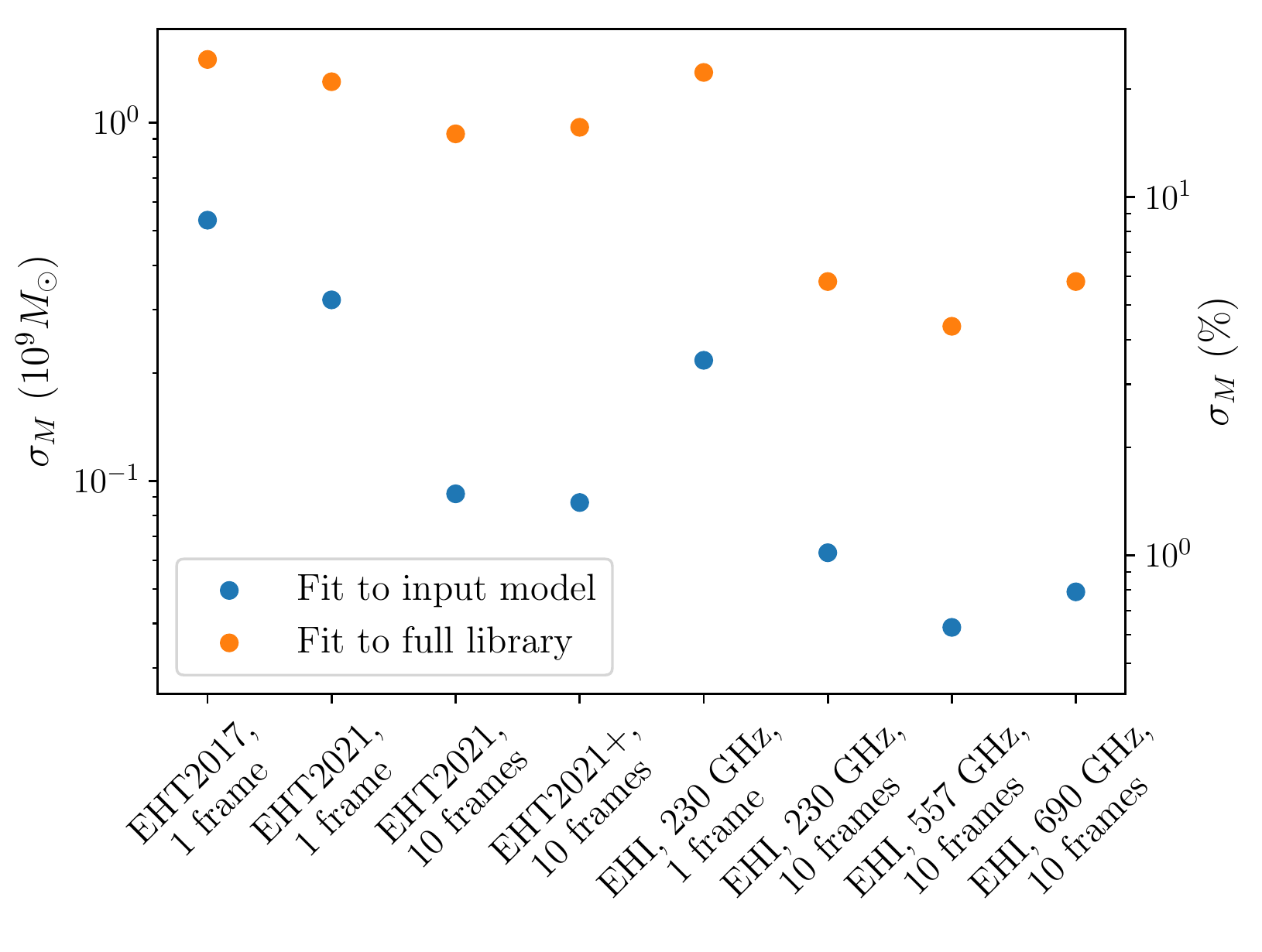}
\includegraphics[width=.48\textwidth]{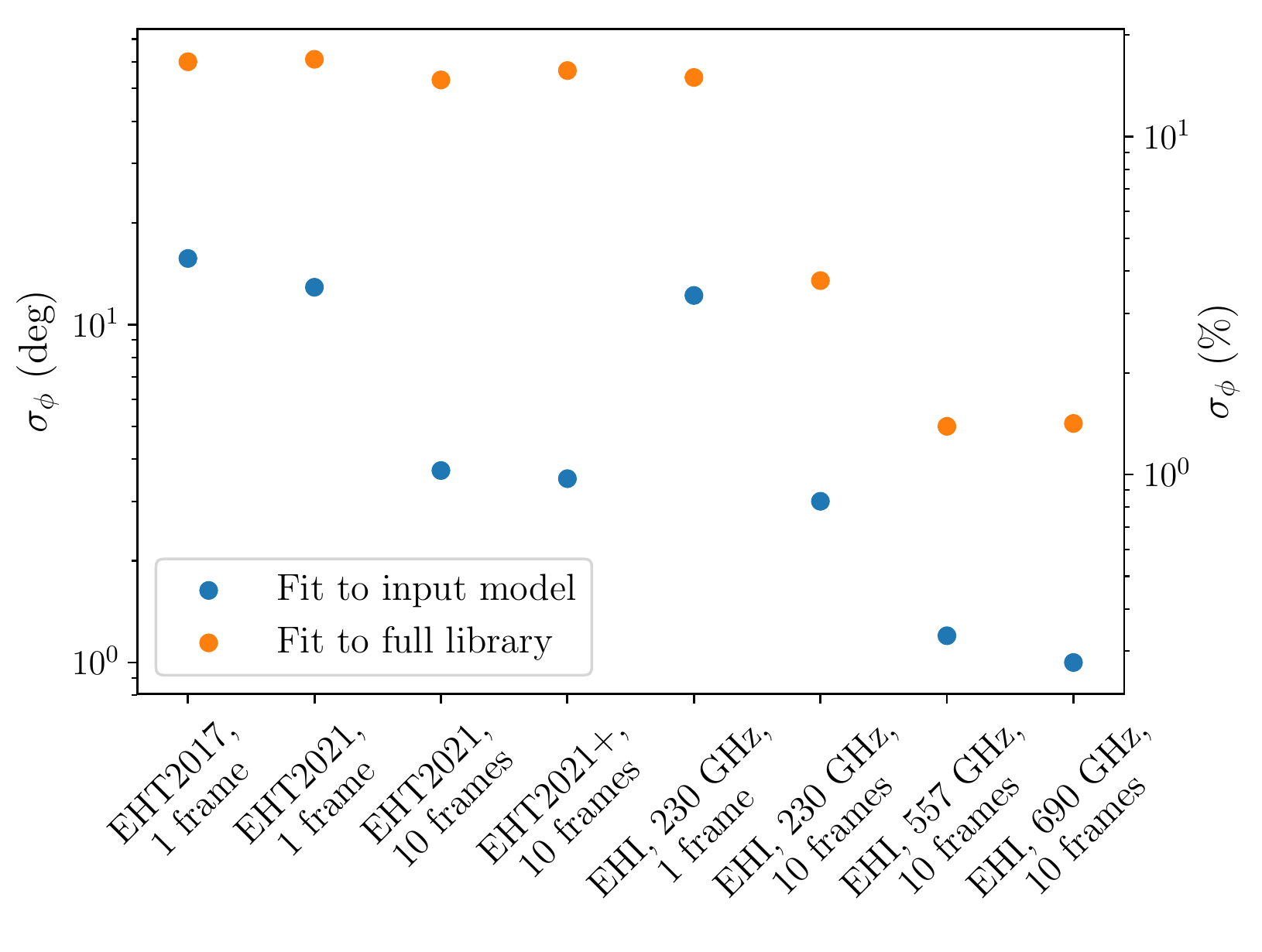}
  \caption{$\sigma_M$ (left panel) and $\sigma_{\phi}$ (right panel) values and percentages from fitting synthetic data from different observations to the input model (SANE, $a_*=0.5$, $R_{\mathrm{high}}=80$) and the full GRMHD library. The values are from Tables \ref{tab:sigma_singlemodel} and \ref{tab:sigma}. The percentages are relative to the true input mass of $6.2\times10^9M_{\odot}$ and the full $360^{\circ}$ circle for $\sigma_M$ and $\sigma_{\phi}$, respectively.}
     \label{fig:sigma}
\end{figure*}

\subsubsection{Recovered mass and orientation for the library models}
Figure \ref{fig:fullmassdist} shows the mass and sky orientation distributions for the accepted models after average image scoring (Table \ref{tab:ais}). Table \ref{tab:sigma} shows the standard deviations $\sigma_M$ and $\sigma_{\phi}$ for the best fit mass and orientation values, respectively. The $\sigma$ values from Tables \ref{tab:sigma_singlemodel} and \ref{tab:sigma} are displayed in Figure \ref{fig:sigma} as well. The distributions from fitting to the full GRMHD library are significantly broader than those from the fit to the input model only (Fig. \ref{fig:sanemassdist}), indicating that the uncertainties in recovered mass and orientation are dominated by the array's ability to rule out GRMHD models with different parameters.

The EHT2017 and EHT2021 single observation mass distributions are similar, where the EHT2021 distribution peaks slightly higher and closer to the true value due to the rejection of some SANE models with offset best-fit masses (Fig. \ref{fig:sanemassdist} and Sec. \ref{sec:gentrends}). The $\sigma_M$ of $1.50\times10^9M_{\odot}$ for the single observation with the EHT2017 array is substantially larger than the error of $0.7\times10^9M_{\odot}$ measured from the real 2017 data \citep{eht-paperI, eht-paperVI}, because that measurement was made from data on four different days using a combination of three different methods, and most models were rejected based on constraints other than those from the EHT data. Repeated observations with the EHT2021 array sharpen the distribution more significantly as, e.g., the high mass estimates for the SANE models with $a_*=0.94$ (Fig. \ref{fig:sanemassdist}) are rejected. The six EHT2021+ stations reduce the number of accepted models by $11\%$ compared to EHT2021 (Table \ref{tab:ais}), but the mass distributions are similar ($\sigma_M$ is even slightly larger for EHT2021+). The models that were additionally rejected with the ten-epoch EHT2021+ observation were thus not at the tail of the mass distribution for the ten-epoch EHT2021 observation. If the large-scale jet emission is taken into account in the fitting, EHT2021+ observations are expected to be able to rule out a larger part of the GRMHD parameter space (see also Sec. \ref{sec:summary}). The single observation with the EHI lacks S/N for a sharp peak, but repeated observations show a sharp bimodal distribution as only a few models are still accepted (Table \ref{tab:ais}). High-frequency EHI observations bring the distributions closer to the true value with a smaller width. The repeated 557 GHz observations give the lowest $\sigma_M$ of $0.27\times10^9M_{\odot}$, corresponding to an uncertainty of $4.4\%$ with respect to the true input mass.

\begin{table*}[ht!]
\caption{$1\sigma$ width of the mass and orientation distributions in Figure \ref{fig:fullmassdist}. The percentages are relative to the true input mass of $6.2\times10^9M_{\odot}$ and the full $360^{\circ}$ circle for $\sigma_M$ and $\sigma_{\phi}$, respectively.}
\resizebox{\textwidth}{!}{%
\begin{tabular}{l|llllllll}
   &   EHT2017   &   EHT2021   &   EHT2021   &   EHT2021+   &   EHI, 230GHz   &   EHI, 230GHz   &   EHI, 557 GHz   &   EHI, 690 GHz   \\
   &   1 frame   &   1 frame   &   10 frames   &   10 frames   &   1 frame   &   10 frames   &   10 frames   &   10 frames   \\
   \hline
$\sigma_M$ ($10^9M_{\odot}$)   & 1.50 & 1.30 & 0.93 & 0.97 & 1.38 & 0.36 & 0.27 & 0.36 \\
$\sigma_M$ (\%)   & 24.2 & 21.0 & 14.9 & 15.6 & 22.3 & 5.7 & 4.4 & 5.7 \\
$\sigma_{\phi}$ (deg)  & 60.0 & 61.0 & 53.0 & 56.5 & 53.9 & 13.5 & 5.0 & 5.1 \\
$\sigma_{\phi}$ (\%)  & 16.6 & 16.9 & 14.7 & 15.7& 15.0 & 3.7 & 1.4 & 1.4 \\
\end{tabular}}
\label{tab:sigma}
\end{table*}

The sky orientation distribution shows similar trends. Especially the high-frequency EHI observations substantially improve the position angle constraints from a $\sigma_\phi$ of $13.5^{\circ}$ with repeated EHI observations at 230 GHz and a $\sigma_\phi$ of $5.0^{\circ}$ and $5.1^{\circ}$ at 557 and 690 GHz, respectively.

\subsubsection{MAD input model}
The above analysis was done using a single model (SANE, $a_*=0.5$, $R_{\mathrm{high}}=80$) as input. The parameter estimates obtained for the different arrays may depend on the input model. In order to get a rough idea of this input model dependence of the parameter estimates, we here repeat the analysis performed above using a MAD, $a_*=0.5$, $R_{\mathrm{high}}=80$ input model. MAD models are generally less variable than SANE models. As the parameter estimation procedure is computationally expensive, we only focus on the single and ten-epoch EHT2021 and ten-epoch 557 GHz EHI observations here.

\begin{figure*}[h]
\centering
\includegraphics[width=.32\textwidth]{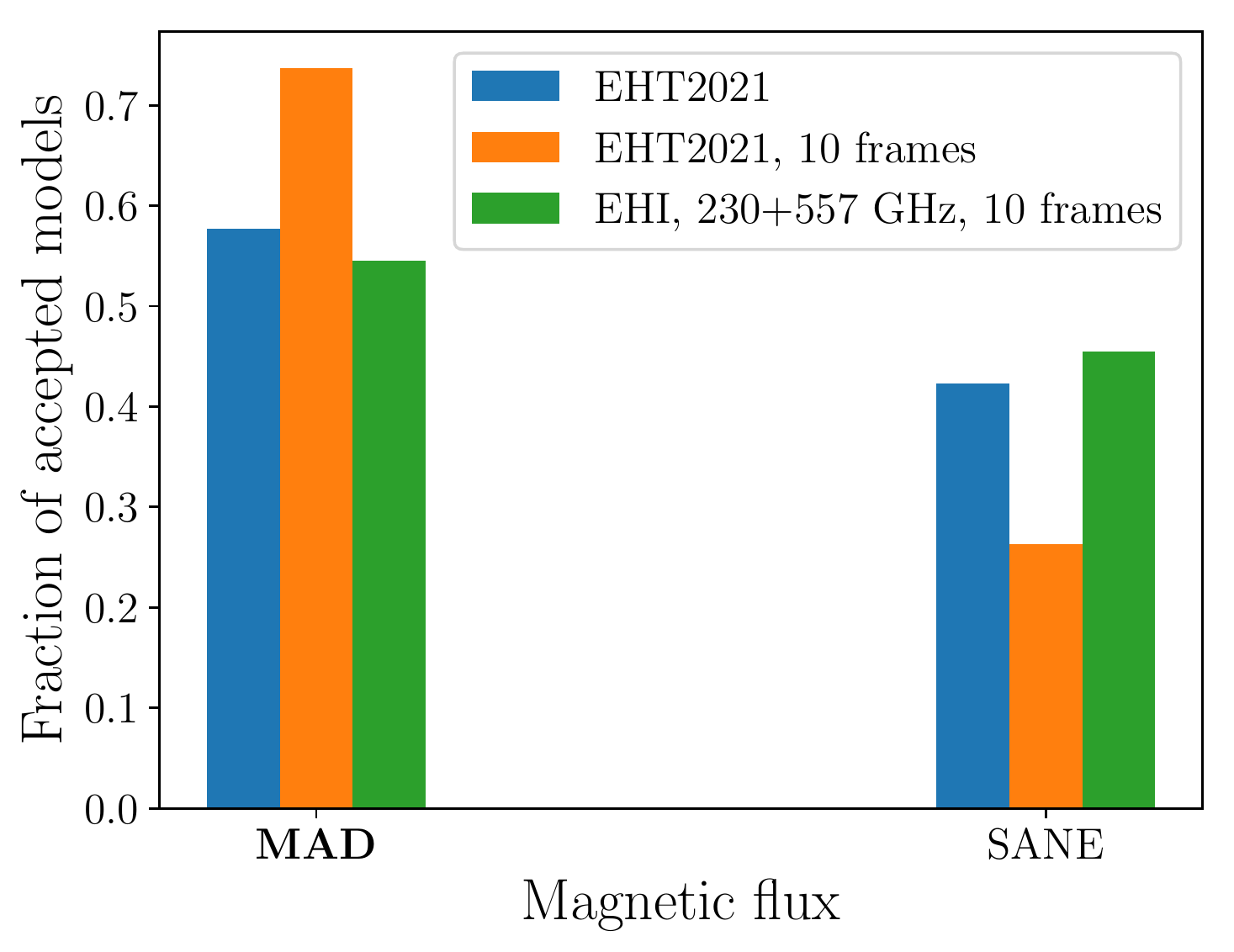}
\includegraphics[width=.32\textwidth]{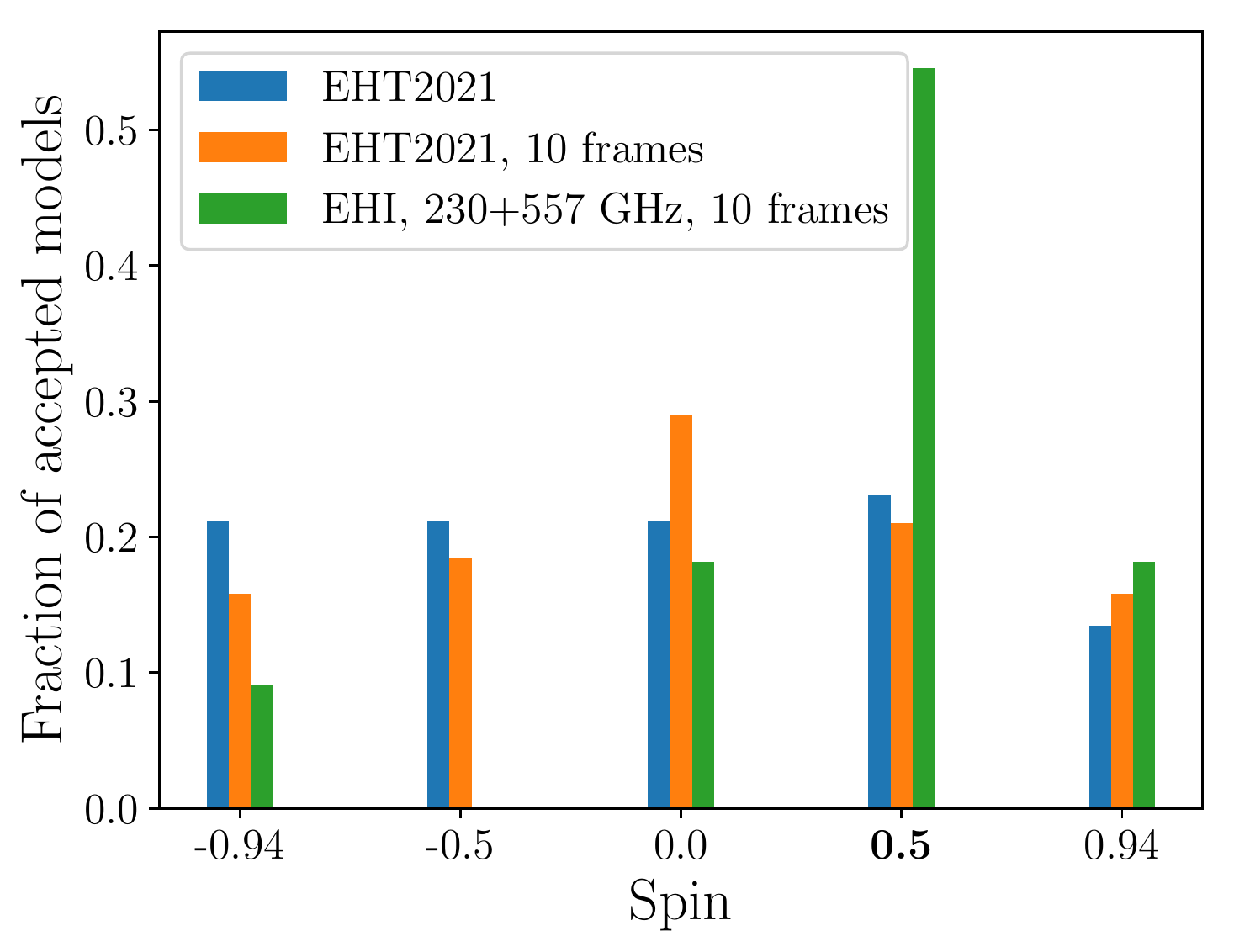}
\includegraphics[width=.32\textwidth]{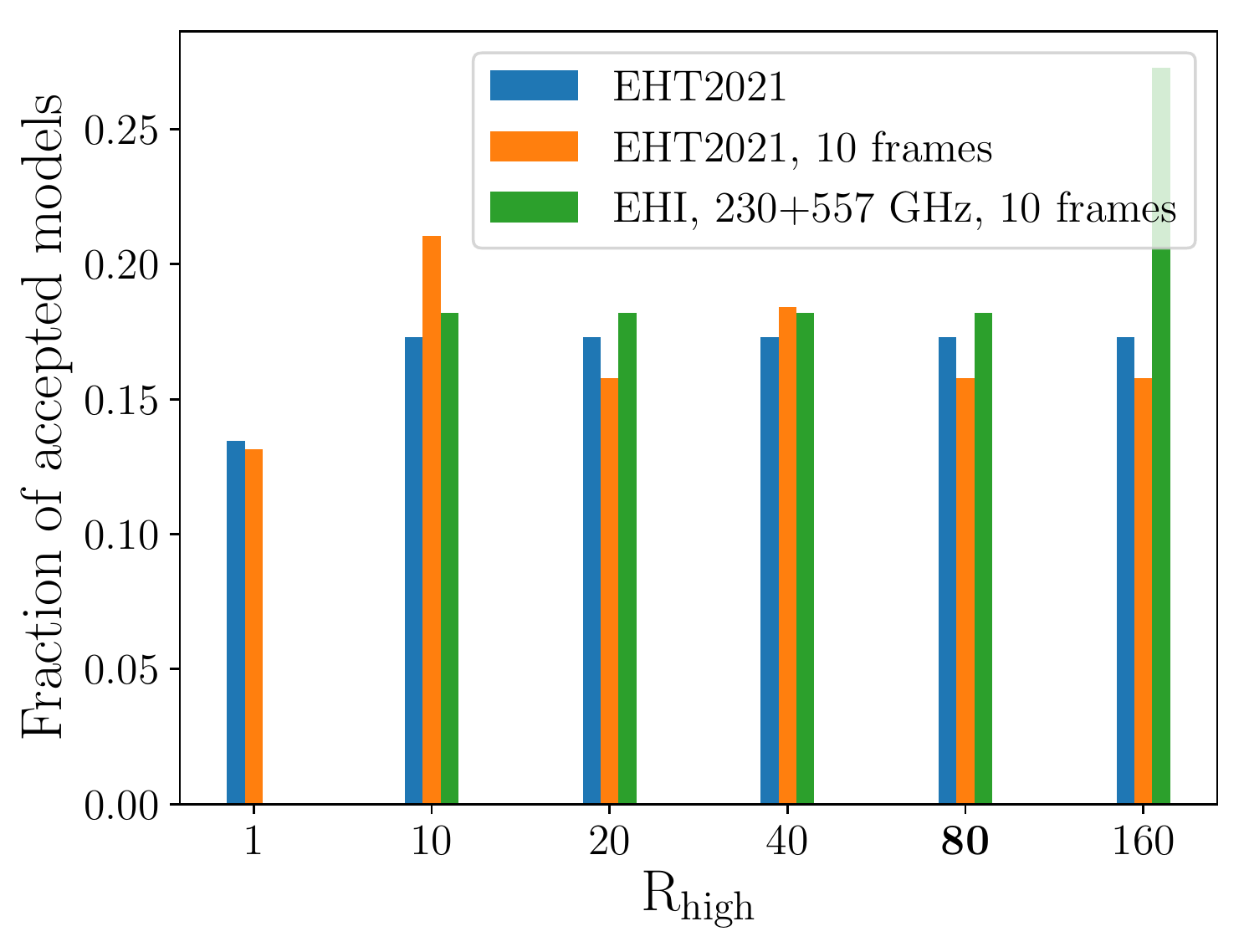}

  \caption{Left to right: same as Figures \ref{fig:magndist}, \ref{fig:spindist}, and \ref{fig:rhighdist}, respectively, but using a MAD, $a_*=0.5$, $R_{\mathrm{high}}=80$ input model.}
     \label{fig:paramdist_mad}
\end{figure*}

With this input model, the number of accepted models after the single and ten-epoch EHT2021 and ten-epoch 557 GHz EHI observations are 87\%, 63\%, and 18\%, respectively. Hence, slightly more models are accepted after repeated ground-based and space-based observations, but the overall trend is similar to using a SANE input model (\ref{tab:ais}). Figure \ref{fig:paramdist_mad} shows the distribution over magnetic flux, spin, and $R_{\mathrm{high}}$ after average image scoring. The true input magnetic flux (MAD, left panel) is favored more strongly for the ground-based array, but the distribution becomes flatter again for the space-based array as some MAD models with different spin and $R_{\mathrm{high}}$-values are rejected. This is also visible in the middle and right panels. Like for the SANE input model (Fig. \ref{fig:spindist}), the spin distribution remains relatively flat with ground-based arrays, but the true input spin of 0.5 is favored with the space-based array. Finally, the $R_{\mathrm{high}}$ distributions are also similar to those for the SANE input model (Fig. \ref{fig:rhighdist}), where the space-based array only rules out the disk-dominated ($R_{\mathrm{high}}=1$) models and the distributions are relatively flat otherwise.

\begin{figure}[h]
\centering
\includegraphics[width=.5\textwidth]{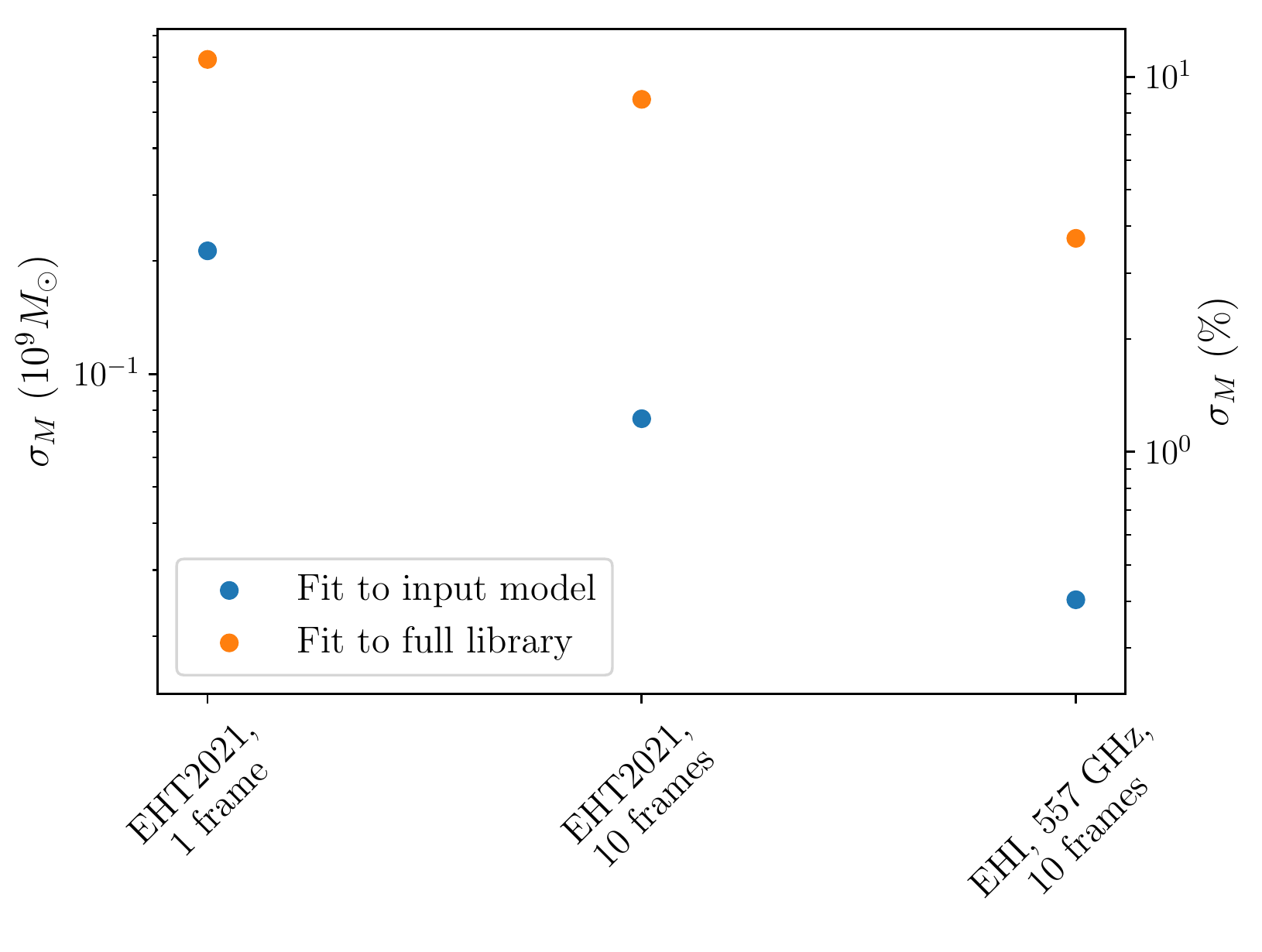}
  \caption{Same as Figure \ref{fig:fullmassdist}, left panel, but using a MAD, $a_*=0.5$, $R_{\mathrm{high}}=80$ input model.}
     \label{fig:fullmassdist_mad}
\end{figure}

Figure \ref{fig:fullmassdist_mad} shows the widths of the mass distributions ($\sigma_M$) when fitting to the input model only and to the full GRMHD library after average image scoring. Here, we see similar numbers and trends as when using the SANE input model (Fig. \ref{fig:fullmassdist}). The effect of repeated observations is a bit smaller here. The single EHT2021 observation of the MAD model already ruled out both the high-spin and disk-dominated SANE models, which caused the mass distribution to be wide when using the SANE input model. When fitting to the input model only, the single EHT2021 observation also gives a smaller $\sigma_M$ when using the MAD model as input, which could be due to weaker variability in MAD compared to SANE. The repeated EHT2021 observations give similar $\sigma_M$ for the MAD and SANE models. The high-frequency EHI observations constrain the mass to 0.4\% when fitting to the MAD input model, compared to 0.6\% when fitting to the SANE input model.

\section{Summary and outlook}
\label{sec:summary}
In this paper, we use the methods developed in \citet{eht-paperV, eht-paperVI} to investigate the potential black hole and accretion parameter estimation capabilities of the current and future Event Horizon Telescope and the Event Horizon Imager Space VLBI concept. We use synthetic data from two single input models (MAD+SANE, $a_*=0.5$, $R_{\mathrm{high}}=80$) fit to a GRMHD image library at 230, 557, and 690 GHz. We find that the percentage of models rejected from the library and the ability to measure the black hole mass and black hole spin axis sky orientation increases significantly when observations from multiple epochs are combined.  
By averaging multiple model frames and epochs, the variance within the models is reduced, leaving images of the quiescent source structure, which has a more prominent photon ring than the individual frames and can therefore be fit more precisely to the observed data. 

When the array is extended or a Space VLBI array is employed, the parameter constraints improve as well. Multi-epoch observations with the EHT2021 array improve the mass constraint by $38\%$ compared to a single EHT2017 observation. The Event Horizon Imager, with potential maximum baseline lengths of 19 to 57 G$\lambda$ depending on frequency, significantly reduces the GRMHD model parameter space: it strongly favors models with the correct input magnetic flux, rules out retrograde and strongly prograde spin models when using our SANE input model, and rules out disk-dominated ($R_{\mathrm{high}}=1$) models when using both input models. $R_{\mathrm{high}}$ is difficult to constrain further: once the emission is jet-dominated, the results are only weakly dependent on the microphysics of particle heating. The 1$\sigma$ black hole mass constraint is 4 to 5\% for EHI observations at 557 GHz. This constraint is still limited by the array's ability to rule out different GRMHD models: when only the input model is fit to the data, it is 0.4 to 0.6\%. The mass and sky orientation constraints from fitting to the input model only are about an order of magnitude better than those from fitting to the full GRMHD library for all synthetic observations (Fig. \ref{fig:sigma}). With, e.g., multi-wavelength observations, lightcurve analysis, or polarization measurements and modeling, the library of acceptable models could be significantly reduced beyond those ruled out by high-frequency VLBI observations only, leading to better constraints. Multi-wavelength constraints were used to rule out 32 of the 60 library models in \citet{eht-paperV} in addition to the 9 that were ruled out based on the EHT data only. With the 2017 M87 polarization data, all SANE models in the EHT GRMHD library could be ruled out \citep{eht-papervii, eht-paperviii}.

Assuming that the mass estimation results are, at least in order of magnitude, applicable to observations of Sgr A* as well, they provide an estimate of the precision with which general relativity can be tested. This precision is limited by the $\sim\,0.5\%$ mass measurement uncertainty from fitting the input model to the ten-epoch 557 GHz synthetic EHI data. This uncertainty is comparable to the current uncertainty in the Sgr A* mass measurements by \citet{Do2019} and \citet{Gravity2019}. If, in the coming decades, we indeed manage to independently constrain the GRMHD plasma parameters, and potentially get a spin measurement from further Gravity observations of stellar orbits and infrared flares in the Galactic Center \citep[e.g.][]{Eisenhauer2011} or from timing measurements of a pulsar in the Galactic Center \citep{Liu2012, Goddi2016}, this means a potential null hypothesis test of the Kerr metric \citep{Psaltis2015, Johannsen2016} at sub-percent level with high-frequency Space VLBI. Additionally, such a measurement could potentially strongly constrain non-Kerr alternatives such as boson stars \citep{Olivares2020}, dilaton black holes \citep{Mizuno2018}, and axion models \citep{Chen2020}. 

All parameter estimates presented in this paper were derived from two representative input models. They could be investigated for more input models for a more complete picture, as different input models could result in different constraints on the fitted parameters. For both the EHT2021+ array and the EHI, only one out of many potential array configurations was simulated here. The simulations give a rough idea of the parameter estimation potential, but the framework used in this paper can now be used to optimize these future array concepts to provide the best possible parameter constraints. For the EHT2021+ array, the amount and locations of new antennas can be varied, and simulations at 345 GHz could be considered as well. For the EHI, the system noise is a limiting factor, and it could be varied to formulate the system requirements more robustly. Combined arrays with space and ground stations \citep{Palumbo2019, Fish2020, Andrianov2021} could be investigated as well. 

Analyses like these are not only useful for investigating the science potential of future VLBI arrays, but they can also be used to optimize observing strategies with current arrays. A future study on the effects of weather, antenna pointing offsets, station dropouts, and scheduling on the parameter estimations could help formulate a set of trigger requirements for the array.

The GRMHD model library is limited in that it has only a few discrete parameter values for the magnetization, black hole spin, and electron temperature distributions. Apart from increasing the number of discrete values, one could think about ways to interpolate between these using, e.g., machine learning techniques \citep{Gucht2020, Lin2020}, or fit to semi-analytic models \citep[e.g.][]{Broderick2016}. The variability within the GRMHD models was found to be an important limitation for constraining black hole parameters, as attested by, e.g., the small difference in recovered parameters between the EHT2021 and EHT2021+ arrays. The analysis pipeline may be extended to include a characterization of the source variability as part of the model selection process \citep[e.g.][]{Kim2016, Roelofs2017, Medeiros2017, Medeiros2018, Johnson2017, Bouman2018, Wielgus2020}, which could improve the constraining power beyond the averaging method introduced here. EHT expansions are expected to make the large-scale jet visible in reconstructed images of the black hole shadow due to an increased dynamic range \citep{Doeleman2019, RoelofsJanssen2020, Raymond2020}. This connection between event-horizon scales and the extended jet has not been taken into account in the parameter estimation framework used here, as the GRMHD library images have a small field of view (160 $\mu$as). With the development of GRMHD simulations that have the ability to connect large \citep[e.g.][]{Fromm2017, Fromm2018, Fromm2019, Liska2018, Chatterjee2019} and small scales at different wavelengths and of an extended fitting framework, the constraining power is expected to improve especially for EHT extensions and space arrays. For a mass measurement, feature extraction techniques such as a ring fit \citep{eht-paperIV, eht-paperVI} may be used, potentially in combination with fitting the more extended (variable) structure \citep{Broderick2020hybrid}. Models and analysis techniques for Sgr A* and polarization could be considered as well. These possible avenues for further simulation and fitting framework development make that the parameter constraints presented in this paper should not be interpreted as set limits on the constraining power of the considered arrays. Rather, they show what is achievable with the current state of the art.

While the analysis can indeed be extended, this work already provides some directions into the future. Repeated observations with a modestly extended array (EHT2021) already improve black hole and accretion parameter estimates significantly. Repeated observations seem more important than large array expansions, at least within the current fitting framework. Order of magnitude improvements become possible with a small Space VLBI array, and even stronger constraints may be attainable with a larger and more sensitive Space VLBI array, involving more and larger dishes and longer baselines. Multi-wavelength and possibly also polarization data can help constraining the model parameter space, allowing precise mass measurements and hence tests of general relativity. In the long term, we have the machinery and tools to make high-precision tests of the physics and astrophysics of black holes a reality.

\begin{acknowledgements}
This work is supported by the ERC Synergy Grant “BlackHoleCam: Imaging the Event Horizon of Black Holes” (Grant 610058). C.M.F. is supported by the Black Hole Initiative at Harvard University, which is supported by a grant from the John Templeton Foundation. JD is supported by NASA grant NNX17AL82G and a Joint Columbia/Flatiron Postdoctoral Fellowship. Research at the Flatiron Institute is supported by the Simons Foundation. Z.~Y.~is supported by a Leverhulme Trust Early Career Fellowship and a UKRI Stephen Hawking Fellowship.
The GRMHD and ray-tracing simulations were performed on GOETHE at the CSC-Frankfurt and Iboga at ITP Frankfurt. The \texttt{GENA} pipeline was developed primarily by C.M.F.
We thank Alexander Raymond for providing a list of potential EHT2021+ stations that are suitable for observations of M87. We thank Kotaro Moriyama, Dominic Pesce, and Sheperd Doeleman for useful comments and discussion on this work.
This work has made use of NASA's astrophysics data system (ADS), the Numpy \citep{numpy}, Scipy \citep{scipy}, and Astropy \citep{astropy1,astropy2} libraries, and the KERN software bundle \citep{molenaar2018kern}.

\end{acknowledgements}

\bibliographystyle{aa} 
\bibliography{bibliography}

\begin{appendix}
\section{Black hole mass scaling}
\label{app:mass_scaling}
During the optimization process we allow the black hole mass $M$ to vary and we re-scale the pixel size of the images accordingly. The obtained range of black hole masses within our work is varying between about $2\times10^9\,M_{\odot}$ and $12\times10^9\,M_{\odot}$ (see Figs. \ref{fig:sanemassdist} and \ref{fig:fullmassdist}). In order to test the applicability and the uncertainties introduced by the black hole mass scaling, we performed the radiative transfer calculations for our reference model (SANE, $a_*=0.5$, $R_{\rm high}=80$, $M=6.2\times10^9\,M_{\odot}$) using several  black hole masses (2,3,..,12$\times10^9\,M_{\odot}$) and compared them to the re-scaled reference model. The comparison is computed in the image plane via the mean square error (MSE) and in the Fourier plane using the visibility amplitude (VA).

\begin{figure*}[htbp]
\centering
\includegraphics[width=\textwidth]{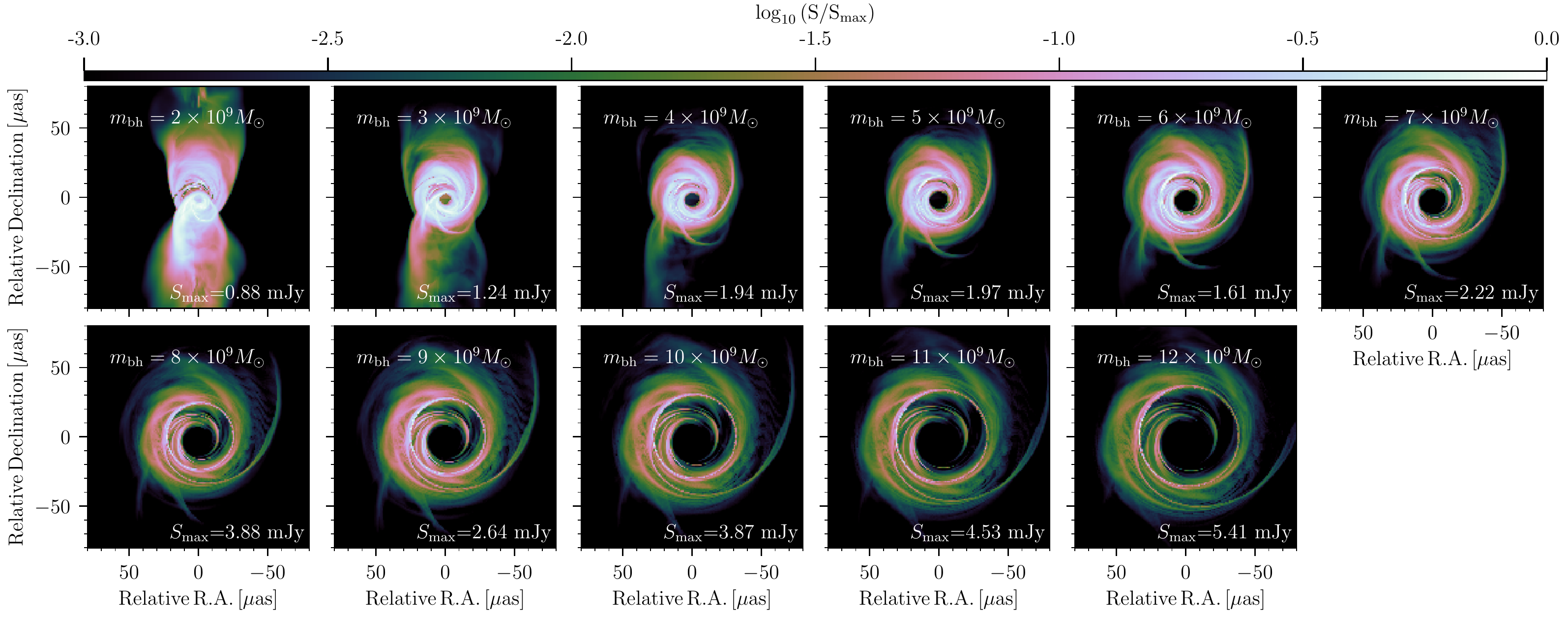}
  \caption{Reference model (SANE, $a_*=0.5$, $R_{\rm high}=80$, $i=163^\circ$) using 
 different black hole mass during the radiative transfer calculations, with masses of 2$\times10^9\,M_{\odot}$ to 12$\times10^9\,M_{\odot}$ increasing from left to right. The used black hole mass is indicated at the top of each panel.}
     \label{fig:varymassGRRT}
\end{figure*}

During the radiative transfer calculations, we fixed the field of view to 160\,$\mu$as, used a resolution of 1 pixel per $\mu$as and iterated the mass accretion rate until a total flux of 0.8\,Jy was obtained. In Figure \ref{fig:varymassGRRT} we show the GRRT images for several different black hole masses using $R_\mathrm{high}=80$. The GRRT images using a black hole mass $>4\times10^9\,M_{\odot}$ show very similar structures: a prominent photon ring and the foot point of the jet as inner smaller ring. While we decrease the black hole mass keeping the field of view fixed to 160\,$\mu$as more jet structure moves into the field of view (see for example the first panel for $M=2\times 10^9\,M_{\odot}$). More importantly, for smaller black hole masses we need to increase the mass accretion rate to obtain a total flux density of 0.8\,Jy. Connected to the increase of the accretion rate, the opacity of the plasma orbiting the black hole increases. The density in the disk is typically higher than in the jet, and thus the disk turns optically thick and the emitting regions are shifted into the jet.

\begin{figure*}[htbp]
\centering
\includegraphics[width=\textwidth]{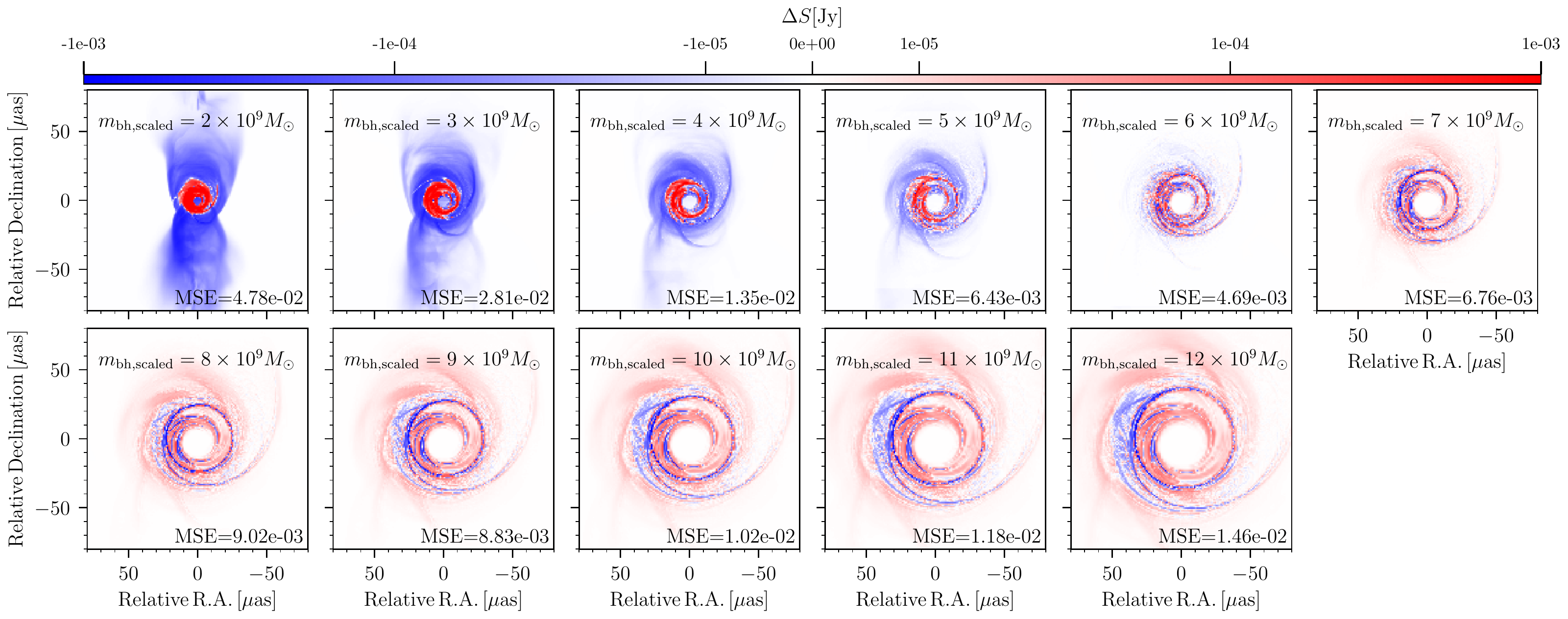}
  \caption{Pixel differences between the re-scaled reference model (SANE, $a_*=0.5$, $i=163^\circ$, using $M=6.2\times10^9\,M_{\odot}$) and images using  
 different black hole mass during the radiative transfer calculations from left to right 2$\times10^9\,M_{\odot}$ to 12$\times10^9\,M_{\odot}$.}
     \label{fig:diffmassGRRT}
\end{figure*}

\begin{figure}[htbp]
\centering
\includegraphics[width=0.48\textwidth]{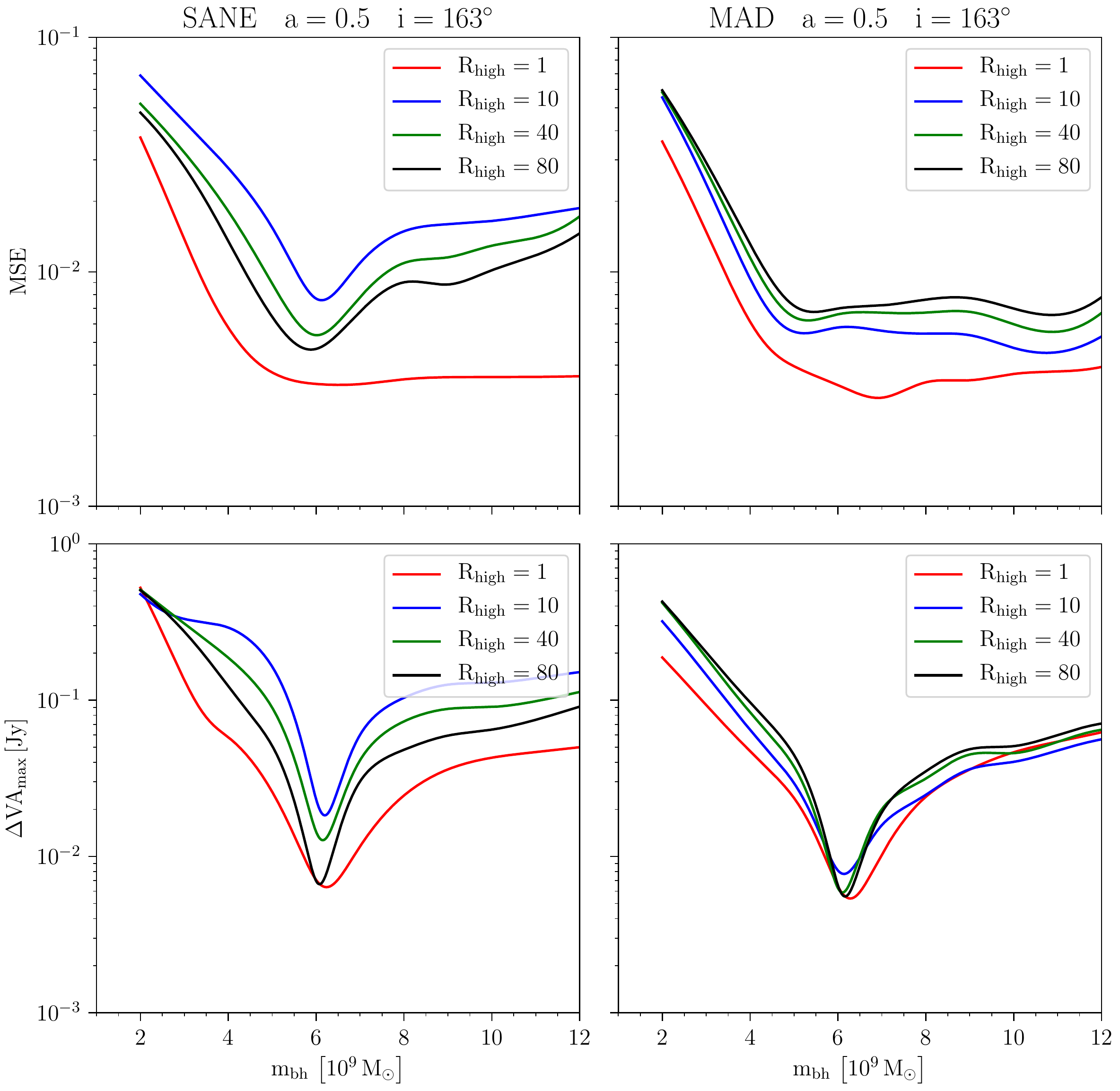}
  \caption{Result of the mass scaling analysis for SANE (left) and MAD (right) model with $a_*=0.5$ for different $R_{\rm high}$ values. The top panels show the MSE and the bottom ones the maximal difference in the visibility amplitude.}
     \label{fig:resmasstest}
\end{figure}

 In the next step, we re-scale our standard SANE model and compare it to SANE model computed with different black hole masses. The result of this analysis can be seen in Figure \ref{fig:diffmassGRRT}. For $M<4\times10^9\,M_{\odot}$ the largest difference is coming from the jet and secondly from the optically thick regions in the disk. For larger black hole masses, the differences are minor and mainly occur from slightly better resolved arcs in the highest mass cases. The MSE between the images computed from the different black hole masses and re-scaled one reflects the above mentioned behaviour and is increasing faster for the lower masses than for the higher masses. In Figure \ref{fig:resmasstest} we present the evolution of the MSE (top) and the largest differences in the visibility amplitude (VA) using the EHT2017 coverage (bottom) for a SANE $a_*=0.5$ (left) and a MAD $a_*=0.5$ model (right) with different values for $R_{\rm high}$. Both models show the same behaviour: up-scaling the images to larger black hole masses than used in the radiative transfer introduces smaller errors than down-scaling the images to smaller black hole masses\footnote{Assuming a fixed field of view and constant flux of 0.8\,Jy}. The comparison across the accretion model (SANE or MAD) shows that re-scaling MAD models introduces smaller errors than re-scaling SANE models. A possible explanation for this behaviour could be the less variable and more compact emission structure seen in the MAD models. 
 
 Inspecting the images (Figures \ref{fig:varymassGRRT} and \ref{fig:diffmassGRRT}), the structure starts to change significantly around $M=4\times10^9\,M_{\odot}$. From the bottom panels of Figure \ref{fig:resmasstest}, the VA error is then limited to about 0.1\,Jy. Up-scaling the black hole mass does not introduce VA errors larger than 0.1\,Jy (see bottom panels in Fig. \ref{fig:resmasstest}). Based on this analysis, images using a black hole mass of $6.2\times10^9\,M_{\odot}$ during the radiative transfer should not be down-scaled to black hole masses smaller than $4\times10^9\,M_{\odot}$, while up-scaling them leads to less serious issues. Since for this work the GRMHD models were ray-traced at $6.2\times10^9\,M_{\odot}$ and the input synthetic data were also generated from a $6.2\times10^9\,M_{\odot}$ model, the mass scaling did not lead to serious errors in this work. Some library models had a best-fit mass below $4\times10^9\,M_{\odot}$ (Fig. \ref{fig:sanemassdist}), but most of these were rejected by the average image scoring procedure, so that the final mass distributions  (Fig. \ref{fig:fullmassdist}) hardly extend below $4\times10^9\,M_{\odot}$. When fitting to real data, it is recommended to perform a similar analysis as presented in this appendix, and redo the ray-tracing if a significant fraction of the fitted masses is in the problematic regions.
 
\end{appendix}

\end{document}